# Review on Superconducting Materials

**Roland Hott[1], Reinhold Kleiner[2], Thomas Wolf[1] & Gertrud Zwicknagl[3]**

[1] Karlsruher Institut für Technologie, Institut für Festkörperphysik,
P. O. Box 3640, 76021 Karlsruhe, GERMANY

[2] Physikalisches Institut and Center for Quantum Science in LISA$^+$,
Universität Tübingen,
72076 Tübingen, GERMANY

[3] Institut für Mathematische Physik, Technische Universität Braunschweig,
38106 Braunschweig, GERMANY

## 1. INTRODUCTION

Superconductivity with its flow of electric current without friction amounts to a realization of the old human dream of perpetual motion. The ratio of resistance between the normal-conducting and the superconducting ("SC") state has been tested to exceed $10^{14}$, i.e., it is at least as large as between a usual insulator and silver as the best normal-conducting material [1]. As the progress of cooling technique gave access to lower and lower temperatures, superconductivity established as common low-temperature instability of most, possibly all metallic systems (see Fig. 1).

The discovery of superconductivity in 1911 came as the result of straightforward research to investigate the microscopic source of the electrical resistance of metals: Studies on alloys and temperature dependent measurements had evidenced that this could be decreased by reducing the density of impurity atoms as well as by lowering temperature. Mercury with its exceptionally low melting and boiling temperatures offered the best perspectives with respect to very low impurity levels. Heike Kamerlingh Onnes had built at the University of Leiden a unique cryogenic facility where he was able to achieve substantially lower temperature than any other laboratory. This led him in 1911 to his famous experiment to see how low one can go concerning electrical resistance. The observed sudden jump to apparently zero resistivity below 4 K came nevertheless as a big surprise. Kamerlingh Onnes immediately recognized it as indication of a new state of matter. In 1933, the discovery of Walter Meißner and Robert Ochsenfeld that magnetic fields are expelled from a superconductor, demonstrated that superconductivity is actually a true thermodynamical state since in contrast to the situation for a merely perfect conductor this expulsion is independent of the experimental history [1].

Fritz and Heinz London developed two years later equations that were capable of describing the interaction of superconductors with electromagnetic fields in terms of a SC current density which expells a magnetic field from the SC bulk deeper than a penetration depth $\lambda$. Fritz London realized that this mechanism could have a quantum-mechanical origin provided that the SC state was described by a "rigid" wavefunction which stays essentially unchanged by the presence of an applied magnetic field. Later on, he refined this idea to the concept of superconductivity as a macroscopic quantum state.

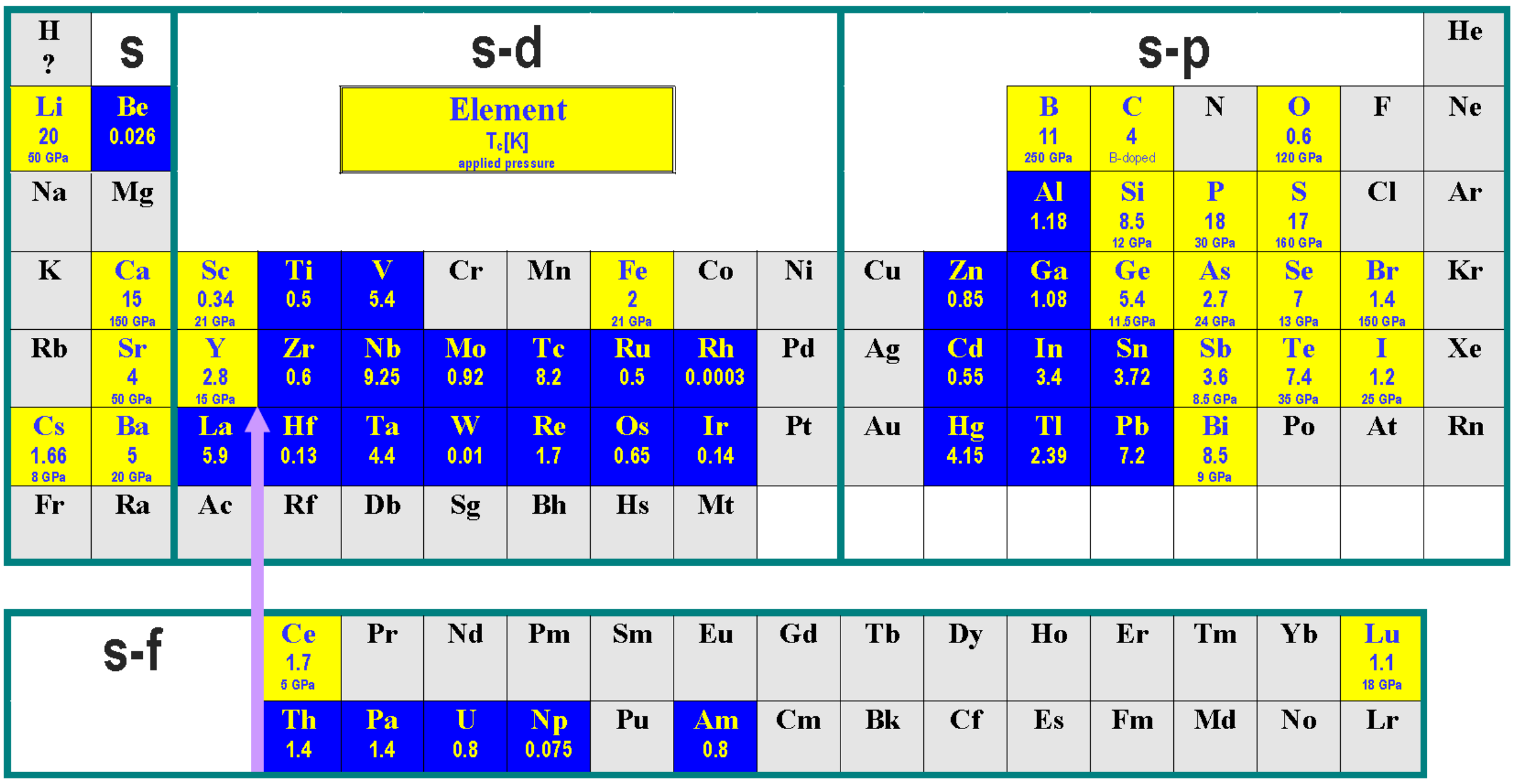

**Fig. 1** Periodic table with the distribution and $T_c$ [K] of the chemical elements for which superconductivity has been observed with or without application of pressure [1,2,3,4].

Soon after mercury, Kamerlingh Onnes expanded the list of SC materials to include tin (3.7 K) and lead (7.2 K). After the discovery of superconductivity in thallium (2.4 K) and indium (3.4 K) again in Leiden [5], Meißner successfully continued the search through the periodic table finding in 1928 tantalum (4.2 K), 1929 thorium (1.4 K) and 1930 titanium (0.4 K), vanadium (5.3 K) and niobium, the element with the highest critical temperature $T_c$ = 9.2 K [6]. The extension to binary alloys and compounds in 1928 by de Haas and Voogd was fruitful bringing in SbSn, $Sb_2Sn$, $Cu_3Sn$ and $Bi_5Tl_3$ [7]. $Bi_5Tl_3$ and shortly afterwards a Pb-Bi eutectic alloy established first examples of critical magnetic field values in the Tesla range.

This had revived the hope for electromagnets where high persistent currents in SC coils would generate strong permanent magnetic fields as already envisioned by Kamerlingh Onnes. In the first generation of SC materials ("*type-I*") superconductivity was easily suppressed by magnetic fields: The self-field generated by the injected current prevented high-field as well as high-current applications. The first step towards this goal was the discovery of *type-II* superconductors where the magnetic penetration depth $\lambda$ exceeds the SC coherence length $\xi$, a materials parameter that Pippard had introduced in 1950 to describe the nonlocal character of the SC state and which was later recognized as the average size of an electron pair. $\lambda > \xi$ enables coexistence of superconductivity and magnetic fields which are allowed here to penetrate into the SC bulk in the quantized form of vortices of the size $\xi$. Thus the material can benefit energetically from SC condensation energy as well as from the energetically favorable magnetic field penetration. This coexistence enables the survival of superconductivity even in strong magnetic fields, at least up to a certain critical field $B_{c2}$ where the average distance between vortices becomes smaller than $\xi$ and the SC state no longer survives the "vortex swiss cheesing". The last ingredient required for technically applicable "*hard*" superconductors was the discovery and engineering of pinning centers which fix penetrated magnetic flux and prevent a Lorentz force driven flow through the superconductor that otherwise would generate power dissipation.

After 1930, SC materials research was more or less in a hiatus until Bernd T. Matthias and John K. Hulm started in the early 1950s a extensive systematic search that delivered a number of new compounds with high $T_c$ well above 10 K as well as technically attractive $B_{c2}$ well above 10 T. The most prominent examples of this crop were NbTi ($T_c$ = 9.2 K) and the A15 materials, $A_3B$ type intermetallic compounds, the materials basis for todays SC wire industry. Having generated some 3000 different alloys in the 1950s and 1960s Matthias became a recognized wizard of the SC materials science. He condensed his huge practical knowledge into "rules" of how to prepare "good" superconductors: High crystal symmetry, high density of electronic states at the Fermi level, no oxygen, no magnetism, no insulators! [8]

In spite of his inofficial sixth rule "Stay away from theorists!", in 1957 the *BCS theory* [9] brought up a desperately awaited breakthrough of theoretical solid state physics, a microscopic explanation of superconductivity. Due to the atomic bomb development, single isotope SC samples, e.g., mercury, had become available where the dependence of $T_c$ on the isotope mass pointed to a phonon-driven mechanism. John Bardeen realized that phonons introduce an attractive interaction between electrons close to the Fermi surface ("overscreening"). The key idea of BCS theory is that even a tiny attractive interaction between these electrons leads to the formation of bound electron pair states ("Cooper pairs"). The electrons in these paired states are no longer obliged to obey the Fermi-Dirac statistics, which enforces the electrons to occupy high kinetic energy single particle states due to the Pauli principle. The energy gain of the SC state with respect to the normal state does not result from the small binding energy of the pairs but it is the condensation energy of the pairs merging into a macroscopic quantum state. This results in an energy gap for electron excitations into single particle states. Although the BCS theory was derived from the physical idea of attractive electron-phonon coupling, the model-based weak pair coupling theory as its mathematical essence is well applicable to other pairing mechanisms. BCS theory had an impact not only on solid state physics but also on elementary particle physics where it was further developed to the idea of the Higgs mechanism of elementary particle mass generation (origins, impact and current state of the BCS theory are reviewed in [10]).

In 1979, in violation of another Matthias rule, superconductivity was discovered in the magnetic material $CeCu_2Si_2$ as the first representative of a new class of *"heavy-fermion" superconductors* [11] where magnetism is suspected to be the mechanism responsible for the Cooper pairing: In these intermetallic compounds the electronic degrees of freedom which are responsible for superconductivity are directly linked with magnetic moments of partially filled f-shells of lanthanide or actinide atoms. Superconductivity below a typical $T_c \sim 1$ K seems to arise here from the delicate balance between the localized magnetic moments trying to imprint their magnetic signature on the shielding conduction electrons, and the conduction electrons which try to screen these magnetic moments by spin flipping, e. g., via Kondo effect [12].

The search for *organic superconductors* had been boosted in the 1960s by the idea that conductive polymer chains with polarizable molecular groups may provide for electrons running along the polymer chains a highly effective Cooper pair coupling by means of an energy exchange via localized excitons [13]. Since the first discovery of an organic superconductor in 1980 [14] remarkably high critical temperatures $T_c > 10$ K have been achieved [15]. However, the origin of superconductivity has turned out to be certainly far from the suggested excitonic mechanism. The electric conduction stems here from π-electrons in stacked aromatic rings which form one- or two-dimensional delocalized electron systems. This restriction of the effective dimensionality and strong Coulomb repulsion effects push the systems towards metal-insulator, magnetic and SC transitions (for a recent review see [16]).

The verdict of the Mermin-Wagner theorem [17] that a long-range order can not exist in two dimensions at finite temperature due to the influence of fluctuations has long been believed to restrict superconductivity to the physical dimension d = 3. The *cuprate "high-temperature superconductors" ("HTS")* discovered in 1986 [18] proved that the limiting case $d = 2+\varepsilon$ ($\varepsilon \rightarrow 0$), i.e., a scenario of basically two-dimensional CuO layer-oriented superconductivity with slight SC coupling to neighbouring CuO layers, can even be enormously beneficial for SC long-range order (the evolution of HTS is reviewed in [19]). The problem with the theoretical description of cuprate HTS within BCS theory and its extensions [20,21] is not the high $T_c$ of up to 138 K under normal pressure [22], far above the pre-HTS record of 23 K [23]: According to the McMillan-Rowell formula [24], a commonly used $T_c$ approximation for the BCS superconductors, this HTS $T_c$ range is readily accessible with reasonable materials parameters [25]. The real problem is that, in contrast to the "deep" Fermi sea of quasi-free electrons in classical metals where the Cooper-pair condensed electrons amount only to a small part of the valence electron system ($k_B T_c << E_{\mathrm{Fermi}}$), in these layered cuprate compounds there is only a "shallow" reservoir of charge carriers ($k_B T_c \sim E_{\mathrm{Fermi}}$) which first have to be introduced in the insulating antiferromagnetic ("AF") stoichiometric parent compound by appropriate doping. Moreover, these charge carriers turn out to be complex physical objects in which strong Coulomb correlations connect their internal charge and spin degrees of freedom and for which Landau's quasiparticle concept of a description in terms of quasi-free electrons "dressed" by screening charge clouds no longer applies [26]. The normal charge transport based on these carriers corresponds to a "bad/strange metal" behaviour with a quite high electrical resistance which increases with temperature in a linear way without saturation up to the melting temperature, in stark contrast to the well-known saturating scattering behaviour of quasi-particles in a normal metal. Accordingly, the BCS recipe to express the SC wavefunction in terms of simple normal-metal single particle states does not work for cuprate HTS since the macroscopic many-particle wavefunction is thoroughly changing in the superconductive transition. Moreover, the strong connection of charge and spin degrees of freedom in the normal-conducting state paves the way for a plethora of other forms of broken symmetry, e.g. charge and spin density waves, which may compete or cooperate with superconductivity [26]. A satisfactory theoretical description has not yet been achieved [29,30] but the SC instability in cuprate HTS, as well as in the structurally and chemically related layered ruthenate [27] and cobaltate [28] compounds, is believed to stem predominantly from a magnetic and not from a phononic interaction as in the case of the classical metallic superconductors where magnetism plays only the role of an alternative, intrinsically antagonistic long range order instability.

Fullerenes ($C_{60}$, $C_{70}$, ... ; doping with metal atoms leads to *fullerides*) were discovered in 1985 and attracted much attention as a third modification of elementary carbon. The superconductivity in $C_{60}$ introduced by doping and intercalation of alkali-metal atoms, with $T_c$ values up to 33 K at normal pressure [31], well above the pre-HTS record of 23 K [23] followed soon as another surprise. In spite of this high $T_c$, the explanation of superconductivity seems to be well within reach of conventional BCS theory based on intramolecular phonons [32].

*Borides* had been investigated systematically with respect to high-$T_c$ superconductivity already in the 1950s. The rationale was the BCS $T_c$-formula [24] where a high characteristic phonon frequency, as provided by the light boron atoms, was predicted to be particularly helpful with respect to a high $T_c$. In the 1990s, the borocarbide superconductors RE $Ni_2B_2C$ (RE: Rare Earth element) with $T_c$ up to 16.5 K [33] fulfilled this promise at least halfway. However, phonons are here only one of the candidates of contributing superconductivity mechanisms. For Rare Earth borocarbides there is additional magnetism due to localized $RE^{3+}$ 4f-electrons which is weakly interacting with superconductivity associated with the 3d-electrons of the $Ni_2B_2$ layers. A huge surprise came in 2001 with the discovery of superconductivity up to $T_c = 40$ K in $MgB_2$, a compound which had been well-known since the 1950s and which was in 2001 already commercially available in quantities up to metric tons [34]. As for the fullerides, in spite of the high $T_c$ a phononic mechanism is here highly plausible: Coulomb correlation effects do not play a role since the conduction electron system involves neither inner atomic shells nor a reduction of the effective dimensionality. A new feature, multiband superconductivity, i.e., the coherent coupling of the Cooper-pair instabilities of several Fermi surfaces [35,36] turned out to be essential for the theoretical description of the SC properties [37].

This multiband mechanism plays an even more dominant role in *iron-based superconductors* [38] where the scattering between the Fermi surfaces of up to five Fe-derived electronic bands is apparently the origin of a complicated magnetic superconductive coupling mechanism [39,40]. In contrast to the cuprate HTS the parent compounds are here metals with a spin-density wave long-range magnetic order. LaFePO, the first representative of this SC materials family was discovered in 2006 [41] but did not gain much attention due to its comparatively low $T_c = 5$ K. $T_c = 26$ K, however, observed two years later for the next *iron pnictide* compound LaFeAsO [42] started a “gold rush” resembling the cuprate HTS situation twenty years before: $T_c$ was immediately pushed up to 55 K [43]. The *iron chalcogenide* FeSe [44] with a bulk-$T_c = 8$ K attracted only little attention as a structural oddity representing the irreducible SC basis structure of a combined iron-based superconductor family until recently $T_c$ up to 109 K was achieved in a FeSe monolayer on a $SrTiO_3$ substrate [45,46] It is still under debate whether this high $T_c$ is an intrinsic FeSe bulk property or due to interface superconductivity of the quasi 2d electron gas at a polar interface as it was first observed for ultrathin $LaAlO_3$ films on $SrTiO_3$ [47, 48].

Finally, the longstanding quest for *room-temperature superconductivity* has met with success: The observed superconductivity in hydrogen sulfide up to a temperature of 203 K [49] is able to fulfill this promise, at least in a room in Antartica at a very high pressure ~100 GPa! As a success of conventional phonon-based first-principles theory this experimental result had even been predicted [50] and a posteriori reproduced with amazing accuracy in more refined calculations [51]. In these experiments with sulfur-hydride and recently also with phosphorous-hydride with $T_c$ up to ~100 K [52] the high pressure is obviously required in order to destabilize the starting compound. As a proof of principle these results revive hopes for high-temperature superconductivity mediated by high-energy hydrogen-based phonons [53]. Nevertheless, they can not be easily translated into a recipe for the preparation of technically useful materials where chemical pressure stabilizes superconductivity even under regular ambience conditions. This caveat holds true even more for the induction of transient superconductivity up to a temperature of 300 K under intense infrared or terahertz frequency pulse illumination as recently reported for the strongly underdoped cuprate HTS $YBa_2Cu_3O_{7-\delta}$ („YBCO“) [54, 55] and for similar experiments on the cuprate HTS $La_{1.675}Eu_{0.2}Sr_{0.125}CuO_4$ and $La_{1.875}Ba_{0.125}CuO_4$ [56, 57] as well as on $K_3C_{60}$ [58]. Even accepting the spectroscopic evidence as experimental criterion for the presence of superconductivity as a thermodynamic bulk state, it is on the one hand clear that the involved dynamical process can not be directly mimicked by any materials design. On the other hand, the details of the concluded mechanism of an antagonistic interplay between superconductivity and charge (density wave) order [55, 57] are still under debate: Recent experiments on $TiSe_2$ as model substance for a study of this interplay under well-defined electric-field controlled doping conditions demonstrate a spatial texturing of superconductivity indicating a complicated local phase separation scenario [59].

As for *practical applications*, NbTi wire for coils generating magnetic fields up to 9 T in magnetic resonance imaging (MRI) and high-energy physics particle accelerator systems plays today the most important commercial role by far of all SC materials [60]. Wire fabrication from a brittle material such as $Nb_3Sn$ is substantially more difficult [61], but has nevertheless reached a level of 20 t per year, the wires being used in coil systems where magnetic field levels above 10 T, up to 20 T are required. The demand for the international fusion project ITER boosted this annual production rate, e.g. with 97 t for the recently completed European share of cable materials for the toroidal field coils [62]. Metal tapes coated with the cuprate HTS YBCO or REBCO (RE-$Ba_2Cu_3O_{7-\delta}$) are nowadays available as conductor materials from several commercial suppliers [63]. High production cost still limits their use today to demonstration projects where either magnetic field level substantially above 20 T [64] or operation temperatures far above liquid helium ("LHe") temperature are required. 35.4 T has been achieved by means of a REBCO coil at 4.2 K in a 31.2 T background field [65]. 26.4 T by an all REBCO magnet represent the highest field generated by an all superconducting magnet system until present [66]. With respect to magnets and cables operated in the temperature range up to ~ 20 K, $MgB_2$ wire may be a substantially cheaper competitor [44,61,67]. Nb-coated rf cavities are now introduced worldwide in high-energy physics particle accelerators to boost the beam energies distinctly beyond the hitherto possible level [68]. Nb thin film technology in combination with Nb/$AlO_x$/Nb Josephson junctions [69] is at present the workhorse for LHe operated electronic circuitry, e.g., in commercial SQUID magnetic field sensors. Workable YBCO thin films have been developed but their use is still limited to niche applications [70].

Quantum computers promise to open up new capabilities in the fields of optimization and simulation that are not possible using today's computers. Superconducting circuits offer the best prospects to achieve this goal via scalable solid state technology [71]. In a more remote future, *topological superconductors* may offer here new possibilities: Topological insulators are new states of quantum matter, characterized by a full insulating gap in the bulk and gapless edge or surface states which are protected by time-reversal symmetry [72]. By direct analogy between the superconducting gap and the bandgap in insulators this concept can be generalized to time-reversal invariant topological superconductors and superfluids. Gapless surface states enable here the realization of Majorana fermions [73]. Localized Majorana fermions obey non-Abelian exchange statistics, making them interesting building blocks for topological quantum computing [74]. Recently, an electronic circuit has been demonstrated to provide a suitable platform for topological superconductivity and Majorana physics [75].

## 1. CUPRATE HIGH-TEMPERATURE SUPERCONDUCTORS

Cuprate High-Temperature Superconductors ("HTS") have played an outstanding role in the scientific and technological development of superconductors. Except for semiconductors, no other class of materials has been investigated so thoroughly by thousands and thousands of researchers worldwide. The plethora of preparational degrees of freedom, the inherent tendency towards inhomogeneities and defects did not allow easy progress in the preparation of these materials. The very short SC coherence lengths of the order of the dimensions of the crystallographic unit cell were on one hand a high hurdle. On the other hand, this SC link of nano-scale microstructure and macroscopic transport properties provided minute monitoring of remnant obstructive defects which could then be tackled by further materials optimization. For cuprate HTS, many materials problems have been solved or at least thoroughly discussed which had not been even realized as a problem for other superconductors before. Due to this exemplifying character, the chemistry and physics of cuprate HTS will be discussed here in more detail.

The structural element of HTS compounds related to the location of mobile charge carriers are stacks of a certain number n = 1, 2, 3, ... of $CuO_2$ layers "glued" on top of each other by means of intermediate Ca layers (see Fig. 2b) [77,78,79,80]. Counterpart of these "*active blocks*" of $(CuO_2/Ca/)_{n-1}CuO_2$ stacks are "*charge reservoir blocks*" $EO/(AO_x)_m/EO$ with m = 1, 2 monolayers of a quite arbitrary oxide $AO_x$ (A = Bi [78], Pb [79], Tl [78], Hg [78], Au [81], Cu [78], Ca [82], B [79], Al [79], Ga [79]; see Table 1) "wrapped" on each side by a monolayer of alkaline earth oxide EO with E = Ba, Sr (see Fig. 2b). The HTS structure results from alternating stacking of these two block units. The choice of BaO or SrO as "wrapping" layer is not arbitrary but depends on the involved $AO_x$ since it has to provide a good spatial adjustment of the $CuO_2$ to the $AO_x$ layers.

The general chemical formula $\mathbf{A_m}E_\mathbf{2}Ca_\mathbf{n-1}Cu_\mathbf{n}O_{2n+m+2+y}$ (see Fig. 2b) is conveniently abbreviated as **A-m2(n-1)n** [80] (e. g., $Bi_2Sr_2Ca_2Cu_3O_{10}$: Bi-2223) neglecting the indication of the alkaline earth element (see Tab.1). The family of all n = 1, 2, 3, ... representatives with common $AO_x$ are often referred to as "A-HTS", e. g., Bi-HTS. The most prominent compound $\mathbf{Y}Ba_2\mathbf{Cu}_3\mathbf{O}_7$ (see Fig. 2a), the first HTS discovered with a critical temperature $T_c$ above the boiling point of liquid nitrogen [83], is traditionally abbreviated as "**YBCO**" or "**Y-123**" ($\mathbf{Y_1}Ba_\mathbf{2}Cu_\mathbf{3}O_{7-\delta}$). It also fits into the general HTS classification scheme as a modification of Cu-1212 where Ca is completely substituted by Y. This substitution introduces extra negative charge in the $CuO_2$ layers due to the higher valence of Y (+3) compared to Ca (+2). The HTS compounds $REBa_2Cu_3O_{7-\delta}$ ("RBCO", "RE-123") where RE can be La [78] or any rare earth element [84] except for Ce or Tb [85] can be regarded as a generalization of this substitution scheme. The lanthanide contraction of the RE ions provides here an experimental handle on the distance between the two $CuO_2$ layers of the active block of the doped Cu-1212 compound [86,87,88]. $\mathbf{Y}_1Ba_2Cu_4O_8$ ("**Y-124**") [89] is the m = 2 counterpart Cu-2212 of YBCO.

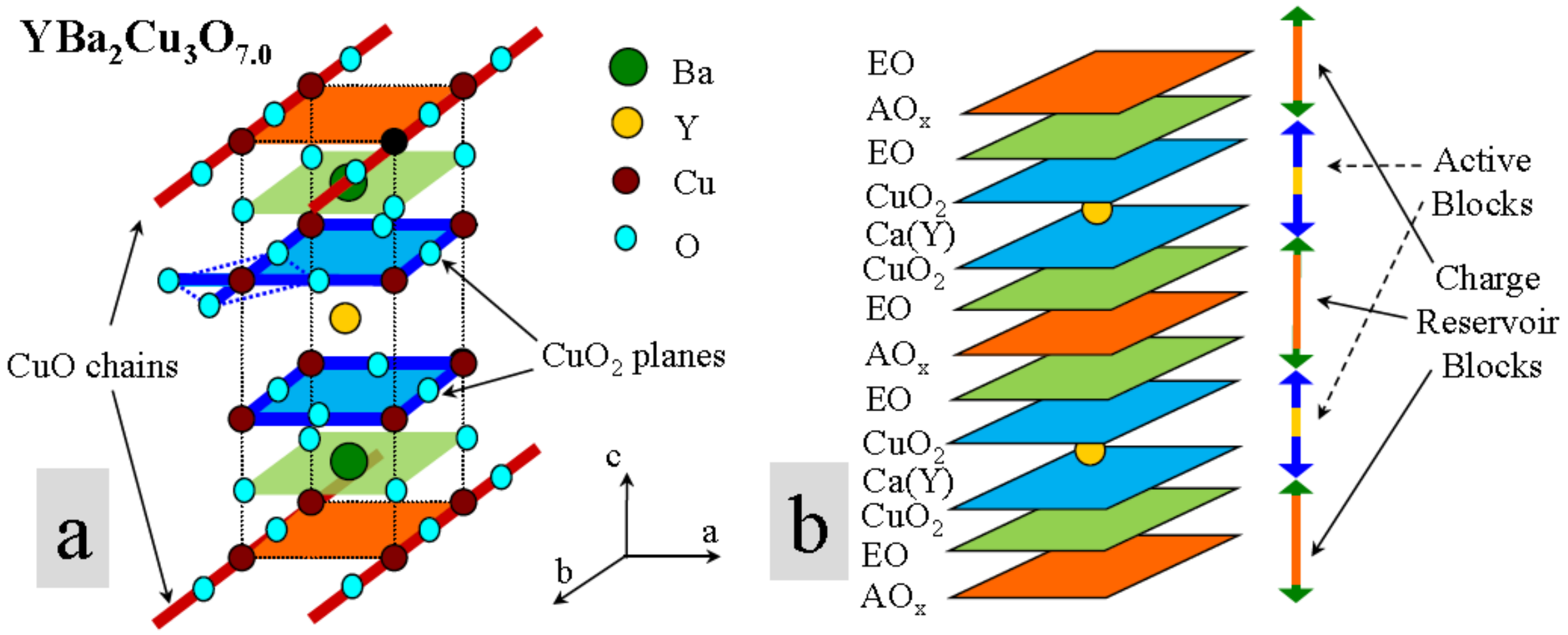

**Fig. 2** a) Crystal structure of $YBa_2Cu_3O_7$ ("YBCO"). The presence of the CuO chains introduces an orthorhombic distortion of the unit cell (a = 0.382 nm, b = 0.389 nm, c = 1.167 nm [76]).
b) General structure of a cuprate HTS A-m2(n-1)n ($A_mE_2Ca_{n-1}Cu_nO_{2n+m+2+y}$) for m = 1. For m = 0 or m = 2 the missing (additional) $AO_x$ layer per unit cell leads to a (a/2, b/2, 0) "side step" of the unit cells adjoining in c-axis direction.

The "**214**" HTS compounds $E_2Cu_1O_4$ (s. Tab1), e. g., $La_{2-x}Sr_xCuO_4$ ("LSCO") or $Nd_{2-x}Ce_xCuO_4$ ("NCCO") are a bit exotic in this ordering scheme but may also be represented here as "0201" with m = 0, n = 1 and $E_2 = La_{2-x}Sr_x$ and $E_2 = Nd_{2-x}Ce_x$, respectively.

Further interesting chemical modifications of the basic HTS compositions are the introduction of fluorine [115,116] or chlorine [117] as more electronegative substituents of oxygen. For Hg-1223, $T_c$ = 135 K can thus be raised to 138 K [22], the highest $T_c$ reported by now under normal pressure conditions (164 K at 30 GPa [118]).

The preparation of an HTS layer structure based on $(CuO_2/Ca/)_{n-1}CuO_2$ stacks with a well-defined number $n$ of $CuO_2$ layers (see Fig. 2) introduces another challenge: As the formation enthalpy of a compound, e. g., with a single $CuO_2$ layer ($n$ = 1), differs only little from that of the ($n$ = 2)-compound with two adjacent $CuO_2$ layers, these materials tend to form polytypes [89]. Experimentally, for all HTS families A-m2(n-1)n the optimized $T_c$ is found to increase from n = 1 to n = 3, 4 and to decrease again for higher n (s. Tab.1). It is still unclear whether this $T_c$ maximum is an intrinsic HTS property since the synthesis of higher-n members of the HTS families turns out to be more and more complicated [119,120,121,122]. In particular, there is at present no preparation technique that allows to adjust here a sufficiently high oxygen content or to provide otherwise sufficient electronic doping that would allow to clarify the possible range of $T_c$ optimization for higher n HTS A-m2(n-1)n [119,122]. Hence the question is still open if such an optimized $T_c$ may eventually continue to increase towards higher n, possibly up to $T_c \sim$ 200 K [123].

| HTS Family | Stochiometry | Notation | Compounds | Highest $T_c$ |
|---|---|---|---|---|
| Bi-HTS | $Bi_mSr_2Ca_{n-1}Cu_nO_{2n+m+2}$ | Bi-m2(n-1)n, BSCCO | Bi-1212 | 102 K [90] |
| | m = 1, 2 | | Bi-2201 | 34 K [91] |
| | n = 1, 2, 3 . . . | | Bi-2212 | 96 K [92] |
| | | | Bi-2223 | 110 K [78] |
| | | | Bi-2234 | 110 K [93] |
| Pb-HTS | $Pb_mSr_2Ca_{n-1}Cu_nO_{2n+m+2}$ | Pb-m2(n-1)n | Pb-1212 | 70 K [94] |
| | | | Pb-1223 | 122 K [95] |
| Tl-HTS | $Tl_mBa_2Ca_{n-1}Cu_nO_{2n+m+2}$ | Tl-m2(n-1)n, TBCCO | Tl-1201 | 50 K [78] |
| | m = 1, 2 | | Tl-1212 | 82 K [78] |
| | n = 1, 2, 3 . . . | | Tl-1223 | 133 K [96] |
| | | | Tl-1234 | 127 K [97] |
| | | | Tl-2201 | 90 K [78] |
| | | | Tl-2212 | 110 K [78] |
| | | | Tl-2223 | 128 K [98] |
| | | | Tl-2234 | 119 K [99] |
| Hg-HTS | $Hg_mBa_2Ca_{n-1}Cu_nO_{2n+m+2}$ | Hg-m2(n-1)n, HBCCO | Hg-1201 | 97 K [78] |
| | m = 1, 2 | | Hg-1212 | 128 K [78] |
| | n = 1, 2, 3 . . . | | Hg-1223 | 135 K [100] |
| | | | Hg-1234 | 127 K [100] |
| | | | Hg-1245 | 110 K [100] |
| | | | Hg-1256 | 107 K [100] |
| | | | Hg-2212 | 44 K [101] |
| | | | Hg-2223 | 45 K [102] |
| | | | Hg-2234 | 114 K [102] |
| Au-HTS | $Au_mBa_2Ca_{n-1}Cu_nO_{2n+m+2}$ | Au-m2(n-1)n | Au-1212 | 82 K [81] |
| 123-HTS | $REBa_2Cu_3O_{7-\delta}$ | RE-123, RBCO | Y-123, YBCO | 92 K [86] |
| | RE = Y, La, Pr, Nd, Sm, | | Nd-123, NBCO | 96 K [86] |
| | Eu, Gd, Tb, Dy, Ho, | | Gd-123 | 94 K [103] |
| | Er, Tm, Yb, Lu | | Er-123 | 92 K [**104**] |
| | | | Yb-123 | 89 K [84] |
| Cu-HTS | $Cu_mBa_2Ca_{n-1}Cu_nO_{2n+m+2}$ | Cu-m2(n-1)n | Cu-1223 | 60 K [78] |
| | m = 1, 2 | | Cu-1234 | 117 K [105] |
| | n = 1, 2, 3 . . . | | Cu-2223 | 67 K [78] |
| | | | Cu-2234 | 113 K [78] |
| | | | Cu-2245 | < 110 K [78] |
| Ru-HTS | $RuSr_2GdCu_2O_8$ | Ru-1212 | Ru-1212 | 72 K [106] |
| B-HTS | $B_mSr_2Ca_{n-1}Cu_nO_{2n+m+2}$ | B-m2(n-1)n | B-1223 | 75 K [107] |
| | | | B-1234 | 110 K [107] |
| | | | B-1245 | 85 K [107] |
| 214-HTS | $E_2CuO_4$ | LSCO | $La_{2-x}Sr_xCuO_4$ | 51 K [108] |
| | | “0201“ | $Sr_2CuO_4$ | 25 (75)K [109] |
| | | *Electron-Doped HTS* | $La_{2-x}Ce_xCuO_4$ | 28 K [110] |
| | | PCCO | $Pr_{2-x}Ce_xCuO_4$ | 24 K [111] |
| | | NCCO | $Nd_{2-x}Ce_xCuO_4$ | 24 K [111] |
| | | | $Sm_{2-x}Ce_xCuO_4$ | 22 K [112] |
| | | | $Eu_{2-x}Ce_xCuO_4$ | 23 K [112] |
| | $Ba_2Ca_{n-1}Cu_nO_{2n+2}$ | “02(n-1)n“ | “0212“ | 90K [113] |
| | | | “0223“ | 120K [113] |
| | | | “0234“ | 105K [113] |
| | | | “0245“ | 90K [113] |
| *Infinite-Layer* HTS | $ECuO_2$ | *Electron-Doped I. L.* | $Sr_{1-x}La_xCuO_2$ | 43 K [114] |

Table 1 Classification and reported optimized $T_c$ values of cuprate HTS compounds.

Another well-investigated experimental $T_c$ trend is the slight increase of the optimized $T_c$ of the RE-123 HTS with increasing distance between the two $CuO_2$ layers in the active block. It has been explained in terms of a higher effective charge transfer to the $CuO_2$ layers [84,86]. For RE-123 HTS with larger RE ions (La, Pr, Nd,), sufficient oxygenation with respect to $T_c$ optimization becomes increasingly difficult. As a further complication, these larger RE ions are comparable in size with the Ba ions which favors RE substition on Ba lattice sites [124]. This cation disorder leads to substantial $T_c$ degradation [125]. This disorder effect, oxygen deficiency and / or impurities had been suggested as reasons for the non-appearance or quick disappearance of superconductivity in Pr-123 [126] and Tb-123 as the only non-SC members of the 123-HTS family. Pr-123 samples with $T_c \sim 80$ K at normal pressure [127] and 105 K at 9.3 GPa [128] as measured immediately after preparation lost their SC properties within a few days.

In 214-HTS without BaO or SrO “wrapping” layers around the $CuO_2$ layers (see Tab. 1), $T_c$ seems to be particularly sensitive with respect to oxygen disorder [129]. The electron doped 214-HTS such as $Nd_{2-x}Ce_xCuO_4$ ("NCCO") have here the additional complication of a different oxygen sublattice where oxygen ions on interstitial lattice positions in the $Nd_{2-x}Ce_xO_2$ layer (which are yet for hole doped 214-HTS the regular oxygen positions in the non-$CuO_2$ layer!) tend to suppress $T_c$ [130].

With respect to the $T_c$ dependence of the A-m2(n-1)n HTS families on the cation A of the charge reservoir blocks, there is an increase moving in the periodic table from Bi to Hg (see Tab. 1). However, continuing to Au, the reported $T_c$ is already substantially lower. This $T_c$ trend seems to be related to the chemical nature of the A-O bonds in the $AO_x$ layers [101].

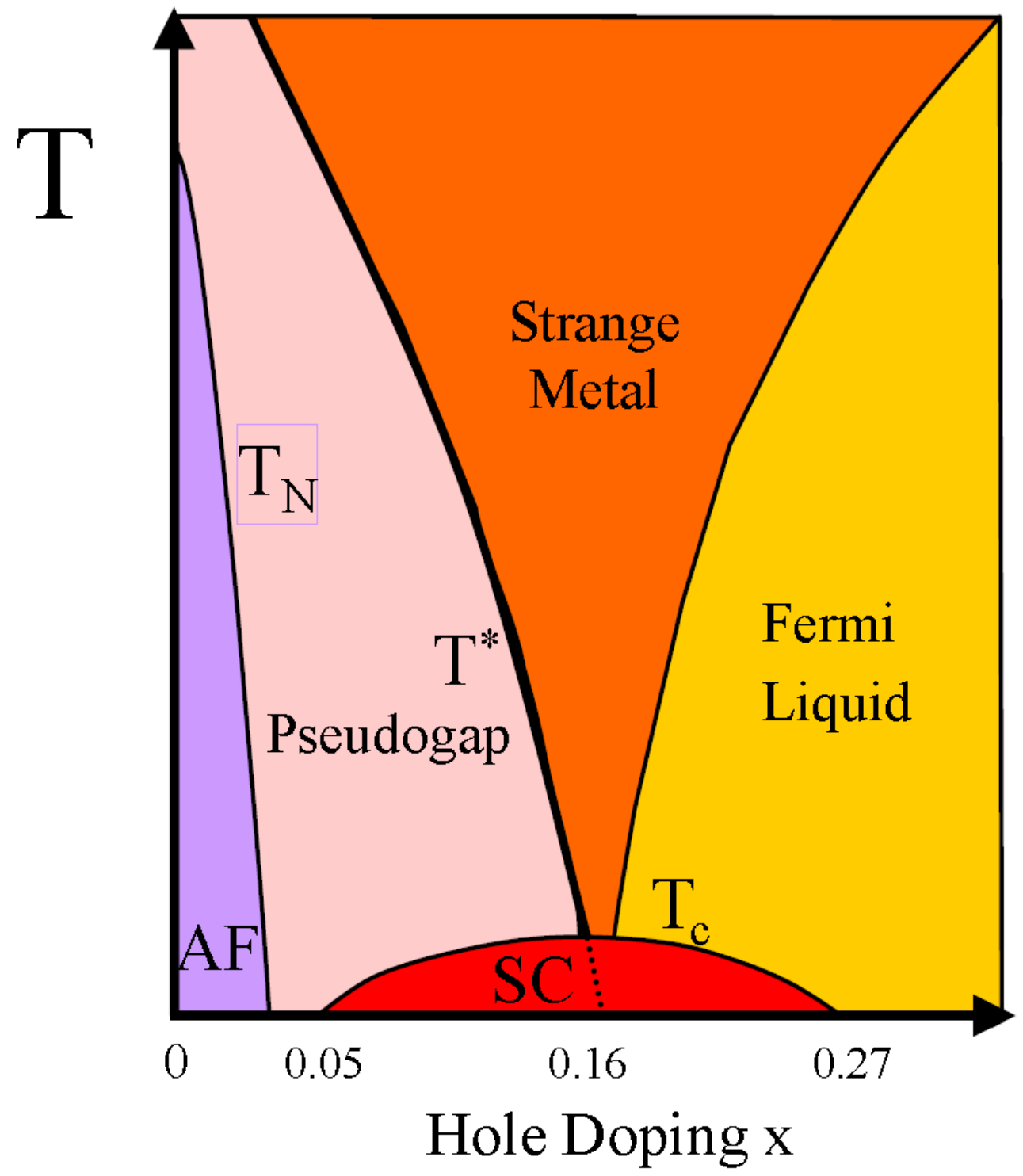

**Fig. 3** Schematic HTS temperature-doping phase diagram with the interplay of antiferromagnetism (“AF”) and superconductivity (“SC”) [135,137,138,139].

For a constant doping, buckling of the $CuO_2$ layer is observed to decrease $T_c$ [131]. Such deviations from the simple tetragonal crystal structure are found in most of the HTS compounds as a chemical consequence of the enforced $AO_x$ layer arrangement in the cuprate HTS structure. The record $T_c$ values for single $CuO_2$ layer HTS compounds of Tl-2201 ($T_c$ = 90 K)[132,133] and Hg-1201 ($T_c$ = 90 K)[134] with simple tetragonal crystal structure indicate that undistorted flat $CuO_2$ layers provide optimum superconductivity.

The SC critical temperature values $T_c$ reported in the preceeding paragraphs refer to the maximum values obtained for individually optimized doping either by variation of the oxygen content or by suitable substitution of cations. The following scenario applies to *hole-* [135] as well as to *electron-doping* of all cuprate HTS (see Fig.3) [136]: The undoped compounds are antiferromagnetic insulators up to a critical temperature $T_N$ well above 300 K, with alternating spin orientations of the hole states that are localized around the Cu atoms in the $CuO_2$ layers. Adding charge carriers by doping relaxes the restrictions of spin alignment due to the interaction of these additional spin-1/2-particles with the spin lattice. $T_N$ decreases and the insulator turns into a “bad metal”. At low temperature, however, the electric transport shows a dramatic change within a small doping range from an insulating to a SC behavior [140]. For $La_{2-x}Sr_xCuO_4$ this happens at a critical hole concentration x = 0.05 in the $CuO_2$ planes (see Fig. 2). On stronger doping, superconductivity can be observed up to an increasingly higher critical temperature $T_c$ until the maximum $T_c$ is achieved for "optimal doping" (x ≈ 0.16 for $La_{2-x}Sr_xCuO_4$). On further doping, $T_c$ decreases again until finally (x ≥ 0.27 for $La_{2-x}Sr_xCuO_4$) only normal conducting behavior is observed.

The rationale that the phenomenon of superconductivity in HTS can be conceptually reduced to the physics of the $CuO_2$ layers [141] has evolved to a more and more 2-dimensional view in terms of $CuO_2$ *planes.* The superconductive coupling between these planes within a given $(CuO_2/Ca/)_{n-1}CuO_2$ stack ("interplane coupling") is much weaker than the intraplane coupling, but still much stronger than the superconductive coupling between the $(CuO_2/Ca/)_{n-1}CuO_2$ stacks which can be described as Josephson coupling (see Fig. 4).

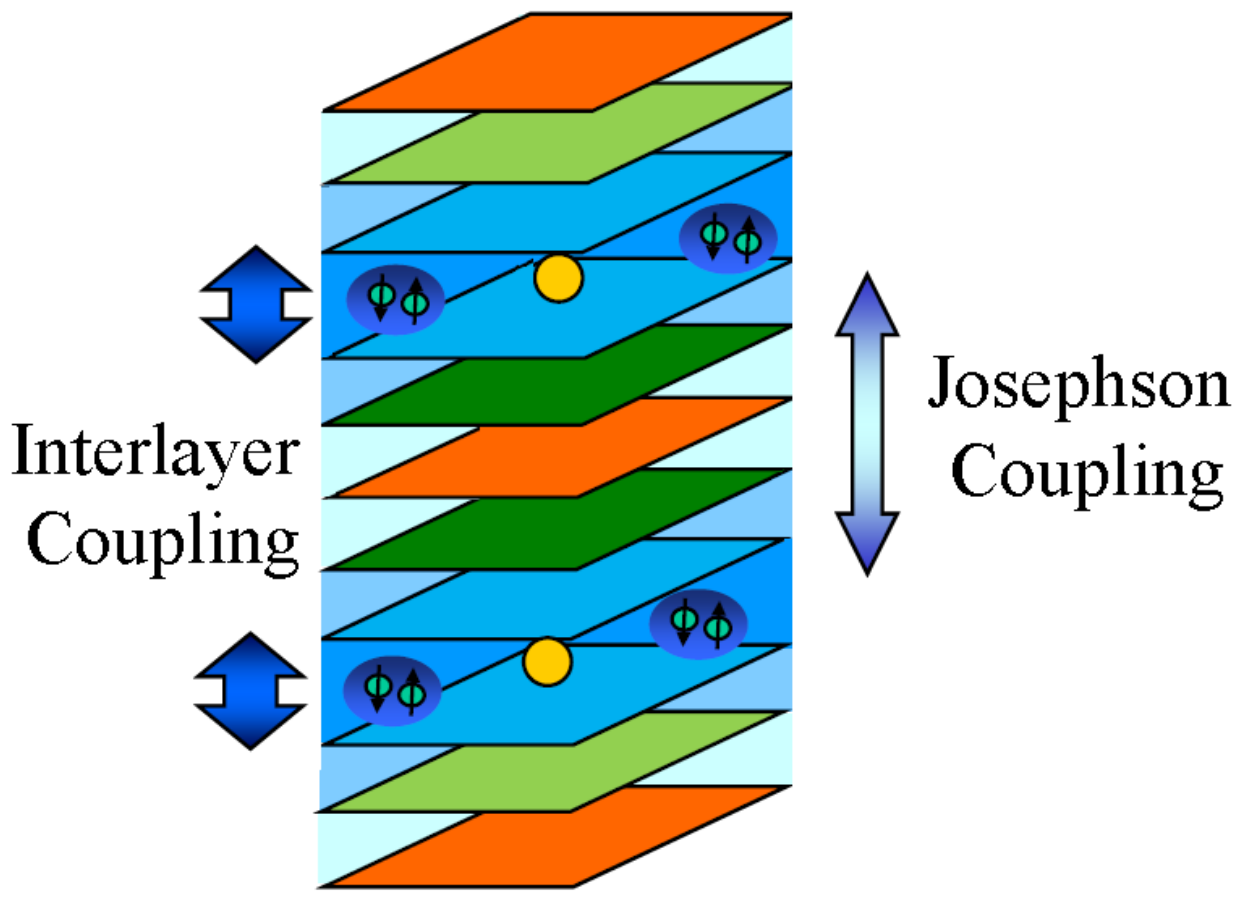

**Fig. 4** Hierarchy of the superconductive coupling in cuprate HTS.

The charge reservoir blocks $EO/(AO_x)_m/EO$ play in this idealized theoretical picture only a passive role providing the doping charge as well as the "storage space" for extra oxygen ions and cations introduced by additional doping. However, the huge pressure dependence of $T_c$ [118] in combination with the large quantitative variation of this effect for the various A-m2(n-1)n HTS families points to a more active role where the cations change not only their valency but also their transmission behavior for the interstack tunneling of Cooper pairs [142].

This becomes most evident for the Cu-HTS family, in particular for YBCO or the RE-123 HTS where the $AO_x$ layer is formed by 1-d CuO chain structures (see Fig. 2). There is experimental evidence that these CuO chains become SC, probably via proximity effect. The intercalation of superconductive CuO chain layers in-between the $CuO_2/Ca/CuO_2$ bilayer stacks is most likely the origin of the strong Josephson coupling between these bilayer stacks. This explains the remarkable reduction of the superconductive anisotropy in c-axis direction compared to all other HTS families. Moreover, in contrast to the usually isotropic SC behavior within the a-b-plane, the CuO chains seem to introduce a substantially higher SC gap in b- compared to the a-direction [143]. This particular SC anisotropy renders YBCO and the RE-123 HTS exceptional among the cuprate HTS.

HTS are extreme type-II superconductors [144] with $\lambda > 100$ nm and $\xi \sim 1$ nm. The quasi-2-dimensional nature of superconductivity in HTS leads to a pronounced anisotropy of the SC properties with much higher supercurrents along the $CuO_2$ planes than in the perpendicular direction [145,146] and a corresponding anisotropy of $\lambda$, e.g., $\lambda_{ab} = 750$ nm and $\lambda_c = 150$ nm in YBCO [147] (the indices refer to the repective orientation of the magnetic field). Material imperfections of the dimension of the coherence length which are required as pinning centers preventing the flux flow of magnetic vortices are easily encountered in HTS due to their small coherence lengths, e. g., for optimally doped YBCO $\xi_{ab} = 1.6$ nm, $\xi_c = 0.3$ nm for $T \rightarrow 0$ K [148] which are already comparable to the lattice parameters (YBCO: $a = 0.382$ nm, $b = 0.389$ nm, $c = 1.167$ nm [76]). The high $T_c$ in combination with the small value of coherence volume $(\xi_{ab})^2 \xi_c \sim 1$ nm$^3$ allows large thermally induced magnetic fluctuations in the SC phase at temperature close to $T_c$, an effect which could completely neglected in classical superconductors [7]. Moreover, since technical superconductor materials consist of a network of connected grains, already small imperfections at the grain boundaries with spatial extensions of the order of the coherence length lead to a substantial weakening of the SC connection of the grains and thus to a “weak-link behavior” of the transport properties.

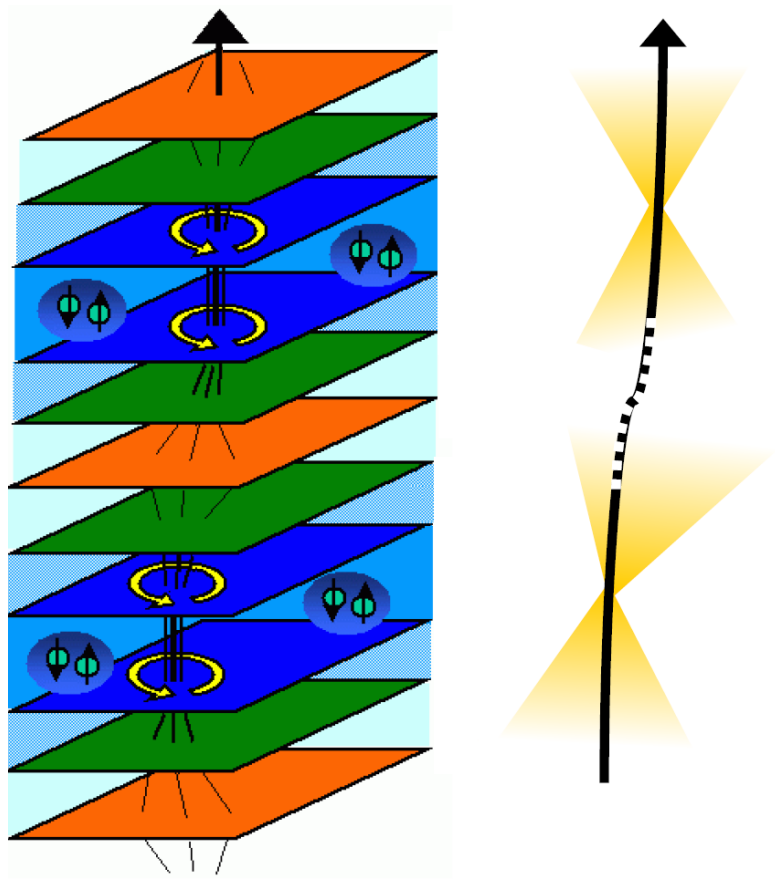

**Fig. 5** Quasi-disintegration of magnetic vortex lines into “pancake” vortices due to weak SC interlayer coupling and magnetic field overlap of neighboring vortices [153].

Obviously, this effect has to be avoided in technical conductor materials [149]. On the other hand it has also been widely exploited for the fabrication of HTS Josephson junctions [150].

The low $\xi_c$, i.e. the weak superconductive coupling between the $(CuO_2/Ca/)_{n-1}CuO_2$ stacks may lead for c-axis transport to an intrinsic Josephson effect within the unit cell even for perfectly single-crystalline materials [151]. Nowadays, the effect is used to generate electrmagnetic radiation in the THz range [152]. If the thickness of the charge reservoir blocks $EO/(AO_x)_m/EO$ in-between these stacks is larger than $\xi_c$ vortices are here no longer well-defined due to the low Cooper pair density (see Fig. 5). This leads to a quasi-disintegration of the vortices into stacks of "*pancake vortices*" which are much more flexible entities than the continuous quasi-rigid vortex lines in conventional superconductors and require therefore individual pinning centers. The extent of this quasi-disintegration is different for the various HTS compounds since $\xi_c$ is on the order of the thickness of a single oxide layers, e.g., $d_{TlOx}$ = 0.2 nm for the Tl-HTS [147]. Hence the number of layers in the charge reservoir blocks $EO/(AO_x)_m/EO$ makes a significant difference with respect to the pinning properties and thus to their supercurrents in magnetic fields. This is one of the reasons why YBCO ("Cu-**1**212") has a higher supercurrent capability in magnetic fields than the Bi-HTS Bi-**2**212 and Bi-**2**223 which for manufacturing reasons have been for a long time the most prominent HTS conductor materials. In addition, in the Cu-HTS family the $AO_x$ layer is formed by CuO chains (see Fig. 2) which apparently become SC via proximity effect. This leads here to the smallest superconductive anisotropy among all HTS families [154].

The effects described in the preceding two paragraphs combine to reduce the *irreversibility field* $B_{irr}[T]$, the tolerable limit for magnetic fields with respect to SC transport, in cuprate HTS substantially below the thermodynamical critical field $B_{c2}[T]$, a distinction which was more or less only of academic interest in the case of classical superconductor.

Beside these intrinsic obstacles for the transport of supercurrent in single-crystalline HTS materials there are additional hurdles since HTS materials are not a homogeneous continuum but rather a network of linked grains (see Fig. 6). The mechanism of crystal growth is such that all material that cannot be fitted into the lattice structure of the growing grains is pushed forward into the growth front with the consequence that in the end all remnants of secondary phases and impurities are concentrated at the boundaries between the grains. Such barriers impede the current transport even if they consist of only a few atomic layers and have to be avoided by careful control of the growth process, in particular of the composition of the offered material.

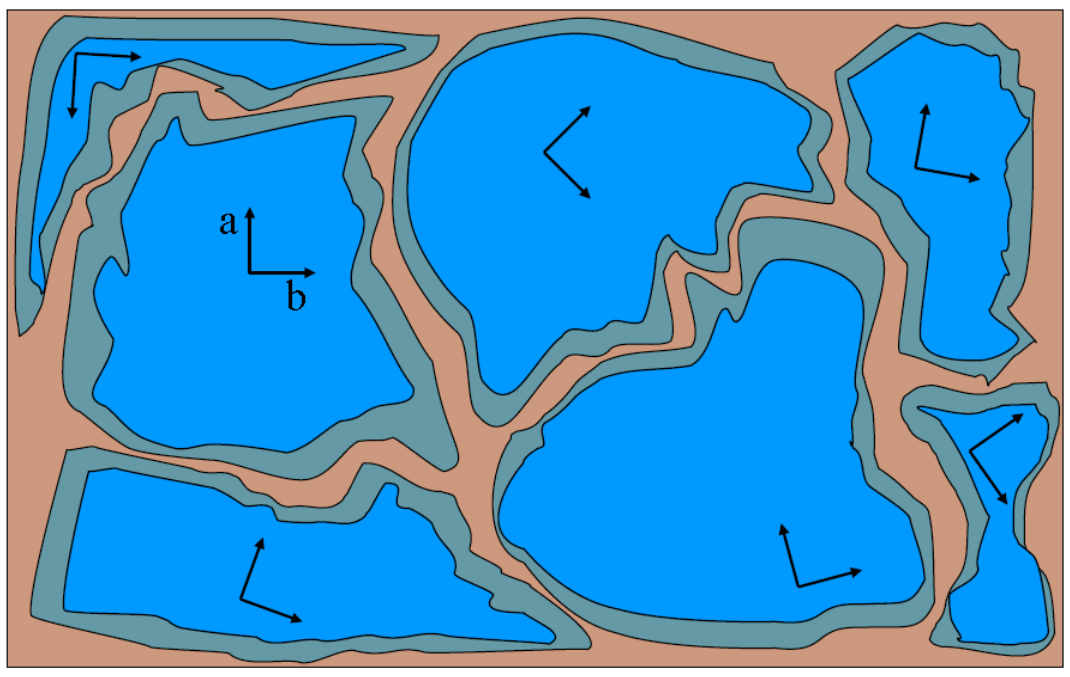

**Fig. 6** Schematics of the HTS microstructure: Differently oriented single crystal grains are separated by regions filled with secondary phases. In addition, oxygen depletion and thus $T_c$ reduction may occur at grain boundaries

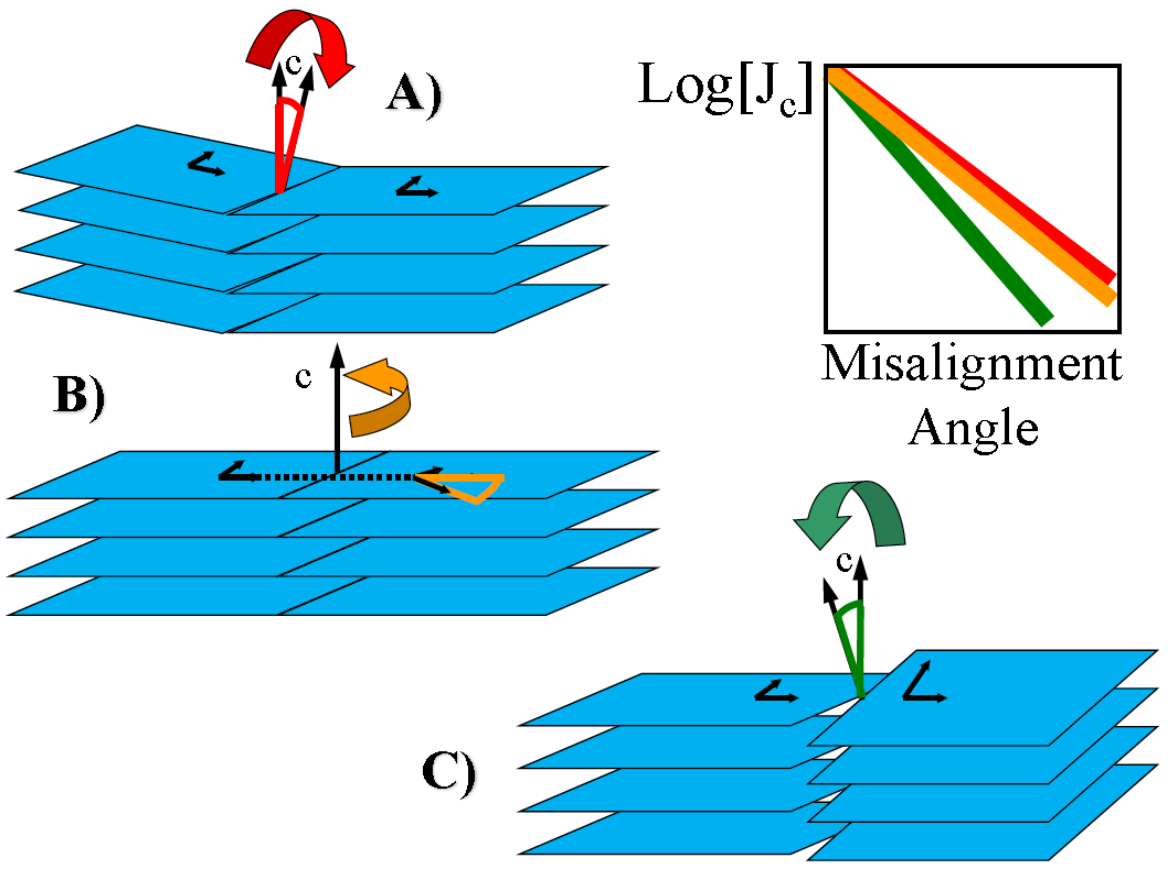

**Fig. 7** Basic grain boundary geometries and experimentally observed $J_c$ reduction $J_c \sim e^{-\alpha/\alpha_0}$ as function of the misalignment angle $\alpha$: $\alpha_0 \approx 5°$ for A) and B), $\alpha_0 \approx 3°$ for C) independent of temperature [156].

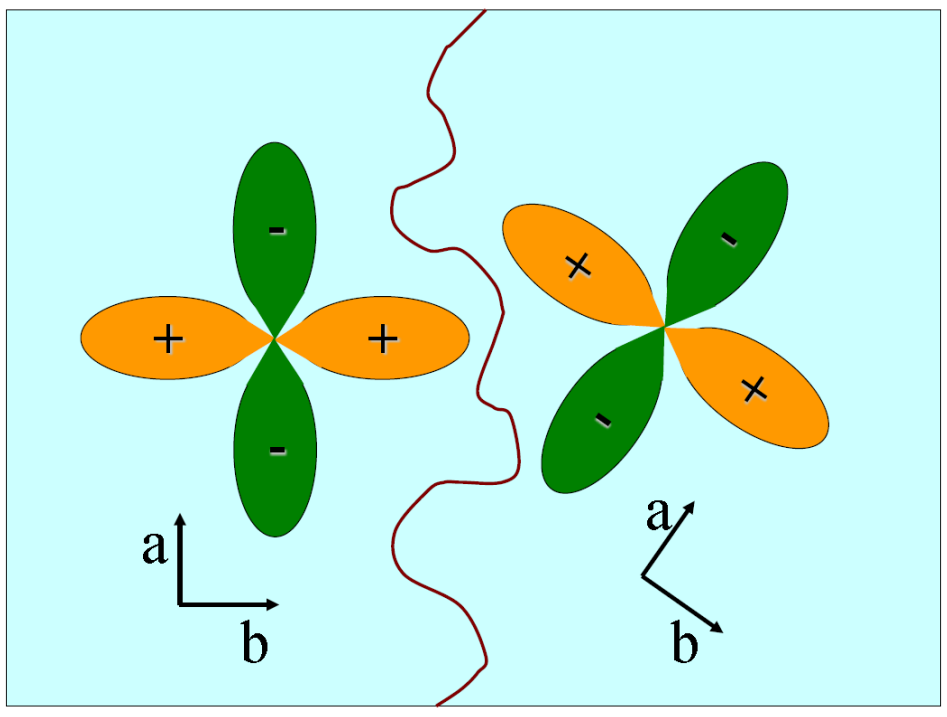

**Fig. 8** Schematics of a boundary between HTS grains. Misorientation of the SC d-wave order parameter leads to partial cancellation of the supercurrents modified by the faceting of the grain boundaries.

Another obstacle for supercurrents in HTS (which is detrimental for transport currents, but also enables the fabrication of HTS Josephson junctions) is misalignment of the grains: Exponential degradation of the supercurrent transport is observed as a function of the misalignment angle (see Fig. 7). One of the reasons for this behavior is the d-symmetry of the SC order parameter (see Fig. 8) [155]. However, the $J_c$ reduction as a function of the misalignment angle $\alpha$ turns out to be much larger than what is expected from d-wave symmetry alone [156,157]. This extra $J_c$ degradation as well as the change of the current-voltage characteristics of the transport behavior [158] were believed to arise from structural defects such as dislocations [159] and deviations from stoichiometry, in particular the loss of oxygen at the grain surfaces [160] and the concomitant local degradation of the SC properties due to the decreased doping (see Fig. 3) [161]. A recent microscopic modeling identified the build-up of charge inhomogeneities as the dominant mechanism for the suppression of the supercurrent [162].

Due to all of these limitations, practical application of cuprate HTS materials has turned out to be restricted to perfectly aligned single-crystalline materials such as epitaxial films and well-textured bulk material without *weak-link behavior*, the drastic reduction of critical currents already in low magnetic fields that results from the combination of the described effects.

Bi-HTS *(oxide) powder in tube ("(O)PIT")* tape conductors seem to be the only exception since they constantly paved their way from the first short samples in 1989 [163] to present large cable projects [164,165]. The tricky idea behind this wire preparation technique can be seen from the solution of the problem how to knot a cigarette: Wrap it in aluminum foil and then go ahead with your mechanical deformation! For Bi-HTS powder this principle works with Ag tubes as well: Bi-2212 or Bi-2223 powder or respective precursor powder is filled in Ag tubes which are subject to several mechanical deformation steps of drawing and rolling and intermediate annealing for the development of the SC Bi-HTS phase. The oxygen permeability of Ag allows for sufficient subsequent oxygenation. The two neighbouring BiO layers in the atomic Bi-2212 or Bi-2223 structure are only weakly bound and lead to graphite-like mechanical properties which allow an easy sliding or splitting of the grains along these layers (that is the chemical reason why for YBCO and Tl-1223 with only a single intermediate oxide layer OPIT wire fabrication did not meet with success). The resulting plate-like Bi-HTS grains in the filaments become aligned during the mechanical deformation steps of drawing and rolling within a few degrees. This process has been optimizied and results nowadays in conductor material with SC currents that are sufficiently high for cable applications. Unavoidably, microcracks occur in the Bi-HTS filaments, but are fortunately short-circuited by the Ag matrix which stays in close contact to Bi-HTS filaments. Nevertheless, this results in the inclusion of short resistive current paths which are reflected in the current-voltage characteristics that do not allow persistent current operation. The cost of Ag as the only possible tube material is an additional handicap. YBCO-coated metal bands with epitaxial single-crystalline YBCO grain alignment offer the prospect of overcoming both problems within foreseeable future, but are nowadays due to the required complicated fabrication process still quite expensive as well. Anyhow, even though not all cuprate HTS dreams have come true, the broad commercial interest in applications enabled by these two conductor materials will now definitively establish at least one of these cuprate HTS within the next decade as new important technical conductor material.

## 2. OTHER OXIDE SUPERCONDUCTORS

The discovery of superconductivity in the *bismuthate* $BaPb_{1-x}Bi_xO_3$ in 1975 with a (at this time) rather high $T_c$ ~ 13 K for x ~ 0.25 [166,167] raised great interest in the mechanism of superconductivity in this (at this time) quite exotic oxide compound with a low density of states at the Fermi level. The cuprate HTS soon chased away that exotic touch in spite of the rise of $T_c$ to > 30 K in $Ba_{1-x}K_xBiO_3$ ("BKBO"; x ~ 0.35) in the middle of the HTS bonanza days [168]: Tunneling showed clean gap structures consistent with weak-to-moderate coupling BCS theory [169]. In the parent compound $BaBiO_3$, a 3D charge-density wave ("CDW") arrangement of $Bi^{(4-\delta)+}O_6$ and $Bi^{(4+\delta)+}O_6$ octahedra ( $|\delta| << 1$ ) creates a gap at the Fermi level and leads to an insulating electric behavior. K or Pb doping suppresses this CDW by means of the random occupation of the A position with Ba and K or Pb ions in a simple pseudocubic $ABO_3$ solid solution structure [170,171,172]. Furthermore, this doping introduces hole carriers and finally results in a metal-insulator transition at a critical doping level $x = x_c \sim 0.35$. The maximum $T_c$ occurs for slightly higher doping. On further doping $T_c$ rapidly decreases and finally disappears at the K solubility limit x ~ 0.65. The SC pairing mechanism is apparently related to the structural and concomitant electronic 3D CDW instability.

The extensive search for other SC transition metal oxides following the discovery of the cuprate HTS came in 1994 across strontium *ruthenate* ($Sr_2RuO_4$), a layered perovskite with an almost identical crystal structure as the cuprate HTS $La_{2-x}Sr_xCuO_4$ ("LSCO"), albeit only with a $T_c$ ~ 1.5 K [27]. In both materials the conduction electrons stem from partially filled d-bands (of the Ru or Cu ions, respectively) that are strongly hybridized with oxygen p-orbitals. In contrast to the nearly filled Cu 3d-shell in cuprate HTS with only one hole state, in $Sr_2RuO_4$, in the formal oxidation state of the ruthenium ion $Ru^{4+}$ four electrons are left in the 4d-shell. The closely related ferromagnetic material $SrRuO_3$ shows the inherent tendency of $Ru^{4+}$ towards ferromagnetism. Hence, in analogy with the cuprate HTS, where on doping the AF ground state of the parent compounds seems to "dissolve" in spin-singlet Cooper pairs in a d-wave orbital channel, it was suggested that the superconductivity in $Sr_2RuO_4$ is brought about by spin-triplet pairing where the Ru ions "release" parallel-spin, i.e., triplet Cooper pairs in p-wave or even higher odd order angular orbital channels.

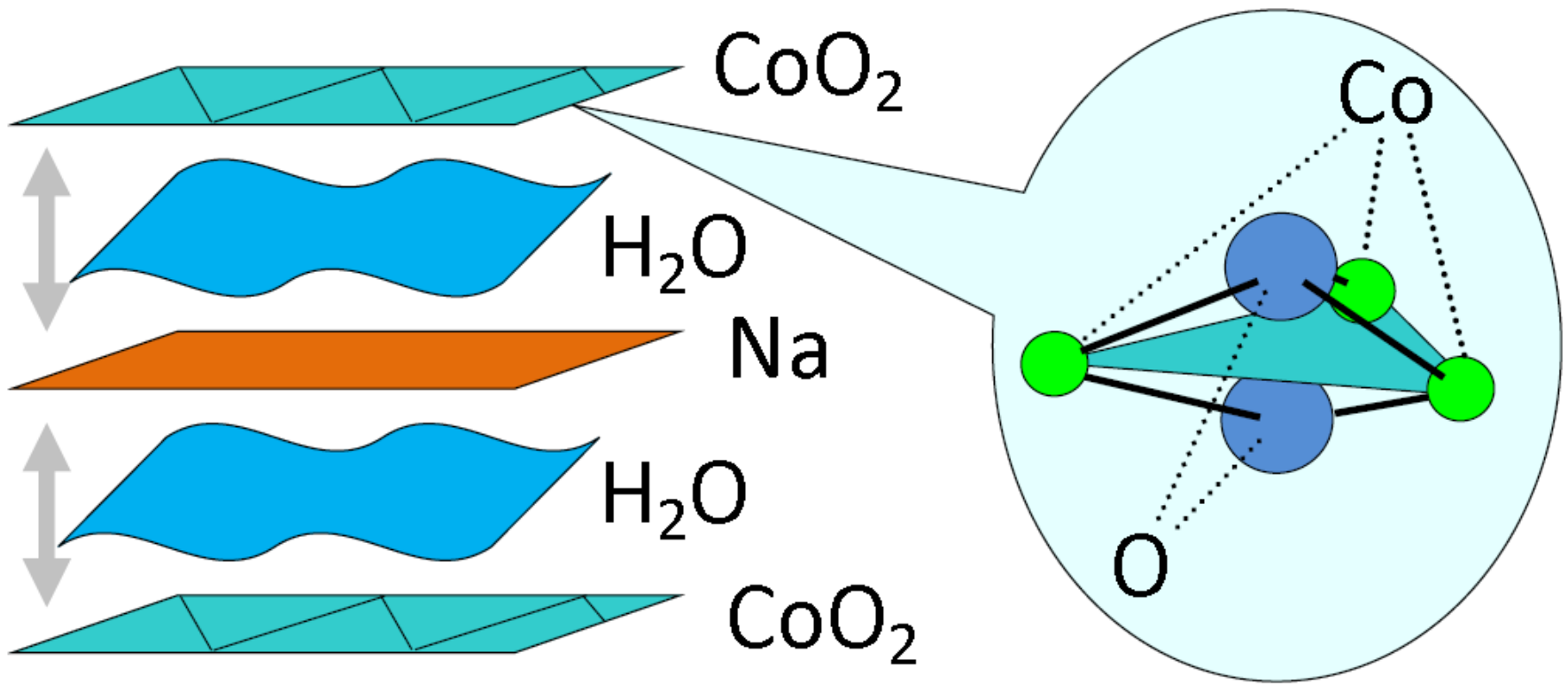

**Fig. 9** Schematics of the layer structure of hydrated sodium cobaltate $Na_{0.3}CoO_2 \cdot 1.4\ H_2O$ and atomic microstructure of the basic cobalt oxide triangular plaquets [177].

$RuSr_2GdCu_2O_8$ (Ru-1212) is a *ruthenate-cuprate hybrid* containing both $CuO_2$ and $RuO_2$ layers. It fits into the elucidated cuprate HTS layer structure scheme (see Fig. 2b) substituting the Ca of the canonical 1212-HTS structure (or the Y in YBCO, or the RE in RE-123, respectively) by Gd to render $CuO_2$/Gd/$CuO_2$ stacks, separated by a SrO "wrapping layer" from the $RuO_2$ layers as "charge reservoir layers". Like rare-earth borocarbides (see chapt. 7), Ru-1212 and some other closely related rutheno-cuprate compounds display ferromagnetism and superconductivity coexisting on a microscopic scale [173], with $T_{Curie} \sim 135$ K and $T_c$ up to 72 K for Ru-1212. The $CuO_2$/Gd/$CuO_2$ stacks are believed to be responsible for the superconductivity, whereas the (ferro)magnetic ordering arises from the $RuO_2$ layers. A clear intrinsic Josephson effect shows that the material acts as a natural superconductor-insulator-ferromagnet-insulator-superconductor superlattice [174]. This may be explained by electronic phase separation resulting in domains that are both SC and AF separated by ferromagnetic boundaries [175].

*Cobaltates* made in 2003 their entry into the SC oxide zoo. $T_c = 4.5$ K has been achieved in hydrated sodium cobaltate $Na_{0.3}CoO_2$•1.4 $H_2O$ [28]. Na provides here the doping. The intercalation of water increases the separation between the $CoO_2$ layers and seems to be essential for the onset of superconductivity: $Na_{0.3}CoO_2$•0.6 $H_2O$, with the same formal Co oxidation state but substantially less separation between the $CoO_2$ layers is not SC [176]. A major difference compared to cuprate HTS is the triangular lattice geometry of the $CoO_2$ layers (see Fig. 9) which introduces magnetic frustration into the Co spin lattice, in contrast to the square lattice geometry of Cu ions in cuprate HTS which favors unfrustrated AF spin orientation.

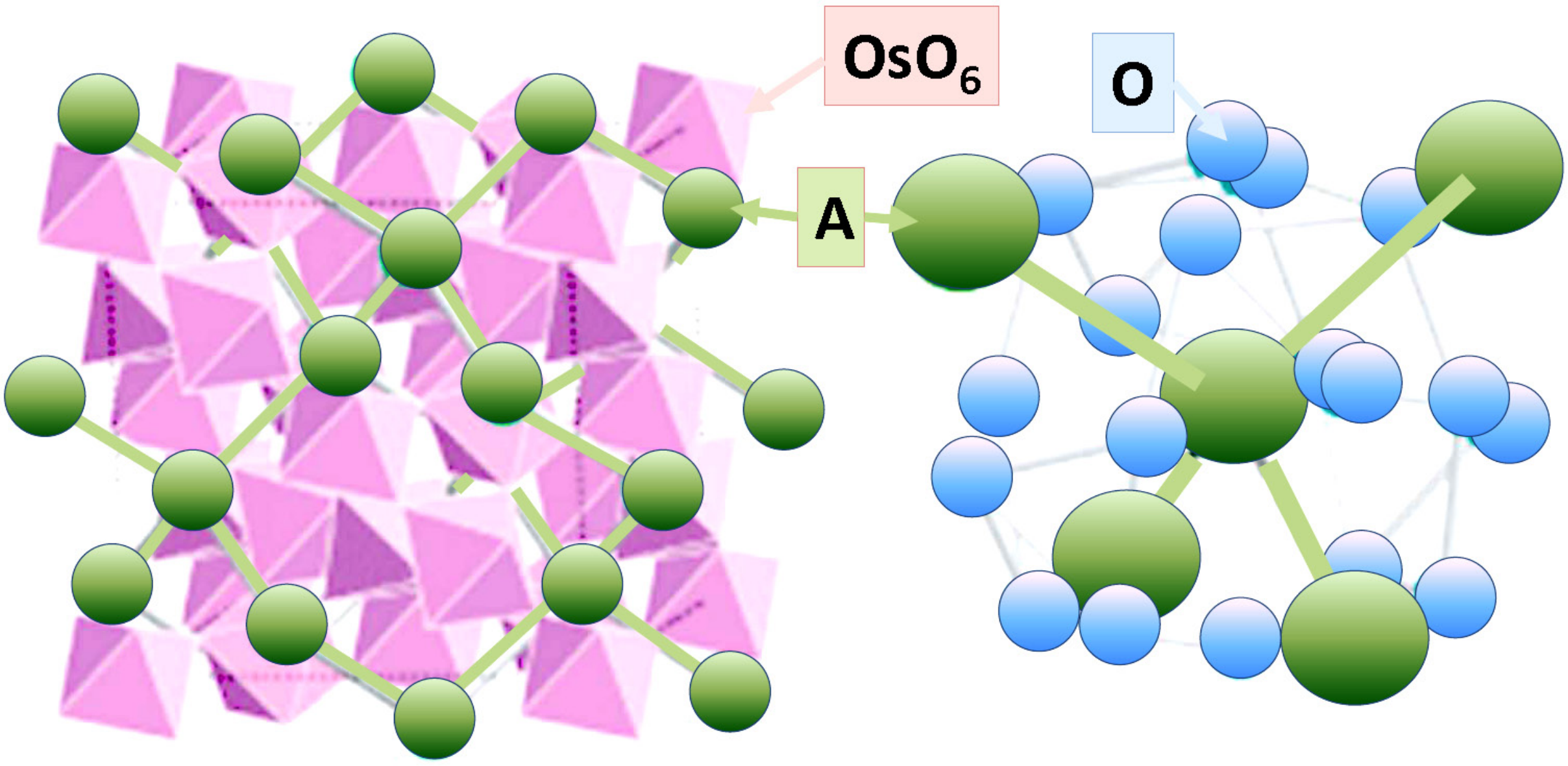

**Fig. 10** Crystal structure of the β-pyrochlore oxides $AOs_2O_6$. The A atom is located in an oversized atomic cage made of $OsO_6$ octahedra and can move with a large excursion along the four [111] directions pointing to the neighboring A atoms in adjacent cages [182].

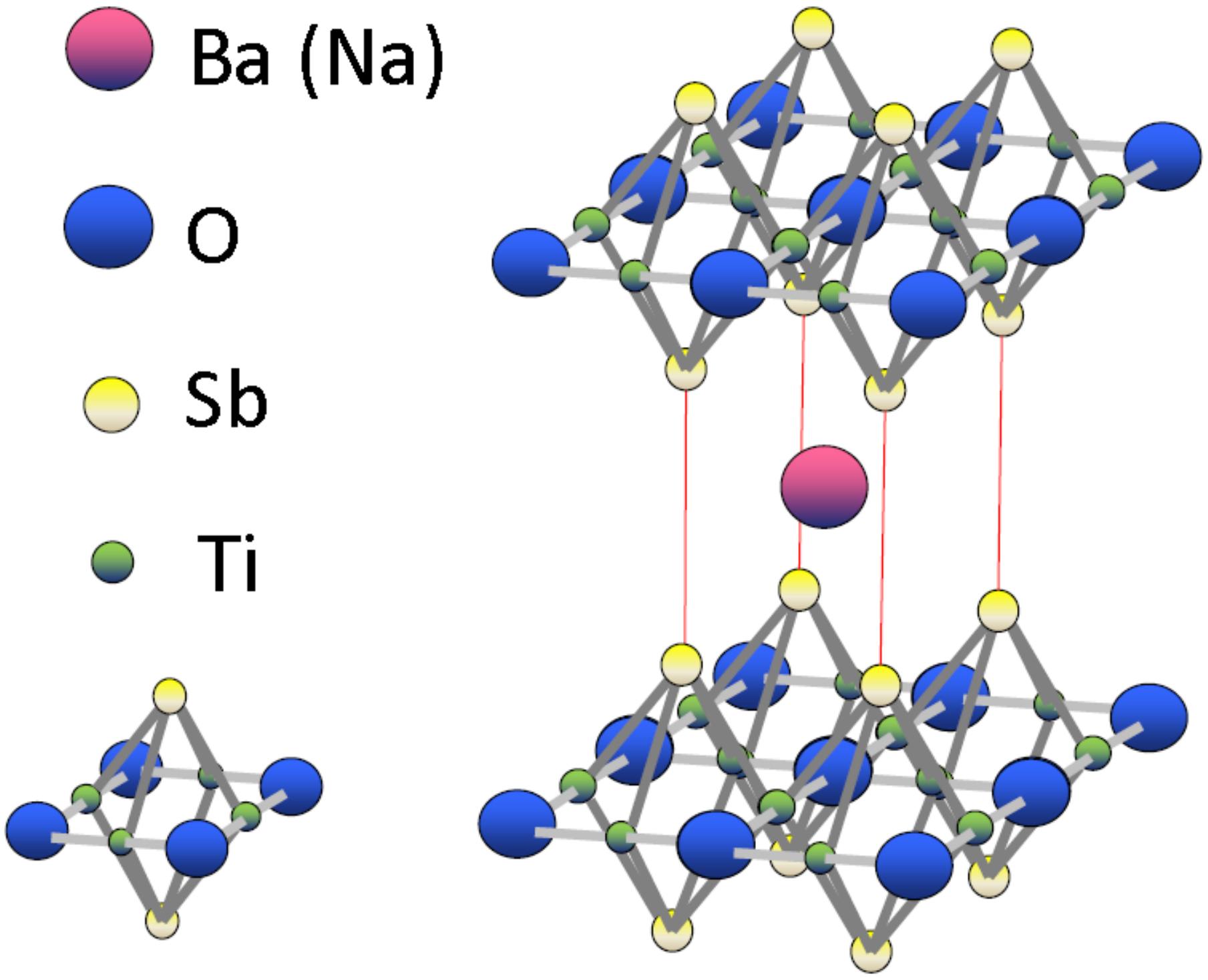

**Fig. 11** Crystal structure of $Ba_{1-x}Na_xTi_2Sb_2O$ [189].

*β-pyrochlore oxide superconductors* $AOs_2O_6$ with A = K, Rb, Cs and respective $T_c$ = 9.6 K, 6.3 K, 3.3 K [178] have a triangle-based crystal structure which is in principle even more subject to magnetic frustration [179]. The A ion sits here in a cage formed by the surrounding $OsO_6$ tetrahedra (see Fig. 10) [180]. Anomalous phonons are observed as an anharmonic oscillation ("rattling" motion) of the A ion cage [181]. Intriguingly, the rattling motion participates in the SC properties [182, 183]. In the three compounds $T_c$ rises with increasing magnitude of the rattling motion and electron-phonon coupling [184], while the density of state decreases [185], opposite to what is expected from BCS theory.

*Ti-based pnictide oxides* combine structural features of cuprate HTS and iron pnictides by featuring a $Ti_2O$ square lattice (anti structure to the $CuO_2$ planes) capped by pnictogen ions [186]. $BaTi_2Sb_2O$ ($T_c$ = 1 K) [187] and $Ba_{1-x}Na_xTi_2Sb_2O$ ($T_c$ = 5.5 K) [188] (see Fig. 11) are at present the only superconductive representatives of this materials family. Recent experiments suggest here in general an effective electron-phonon coupling with a strong tendency towards charge-density wave formation which is for these two particular cases sufficiently weak to allow for the superconductive transition [189].

## 3. IRON-BASED SUPERCONDUCTORS

The two-dimensional layer structure of iron-based superconductor families ("Fe-Sc") [38,190] bears close resemblance to the cuprate HTS structure: The transition element atoms (Fe/Cu) are arranged in a quadratic lattice and apparently provide the SC mechanism. Instead of the Cu-Cu bonding via O atoms sitting halfway between next-nearest Cu atoms in the cuprate HTS, in Fe-Sc the Fe-Fe bonding happens via tetrahedrally arranged P, As, Se or Te atoms above and underneath the Fe plane and does now affect the second-nearest Fe neighbouring atoms as well (see Fig. 12 left). The Fe atoms form thus with the (P/As/Se/Te) atoms a network of regular pyramids with alternating upward/downward orientation. For both SC families, optimum $T_c$ is observed for the most sysmmetric arrangement of these layer geometries, i.e., for flat $CuO_2$ layers [131] and for regular $Fe(P/As/Se/Te)_4$ tetrahedra [192].

The upward/downward orientation of the pyramids introduces an additional degree of freedom for the stacking of the Fe(P/As/Se/Te) layers: Except for the $BaFe_2As_2$ ("122") superconductor family where the neigbouring layers are in opposite orientation to provide a cage for the Ba atoms (see Fig. 12 top right), in most of the other Fe-Sc families the layers are "pyramidized" in the same orientation (see Fig. 12 bottom right): In "11" (e.g., FeSe) there is nothing between, in "111" (LiFeAs and NaFeAs) two alkali metal atom layers fill the "hollows" on each side of Fe(P/As/Se/Te) pyramid layers, in "1111" (e.g., LaFeAsO) an additional oxygen layer comes between these metal atom layers. In close analogy to the chemical pressure rationale applied successfully in cuprate HTS in 1987 to raise $T_c$ [83], the $T_c$ = 26 K of LaFeAsO was improved by replacing La by smaller rare-earth ions which resulted in the present record $T_c$ = 56 K in $Gd_{0.8}Th_{0.2}FeAsO$ [193]. The substitution of La with actinides to form ***A***nFeAsO presents an exciting opportunity to study the role of the rare earth ions. For ***A***n = Np, Pu [194,195], however, no superconducting transition has been detected down to T = 2 K.

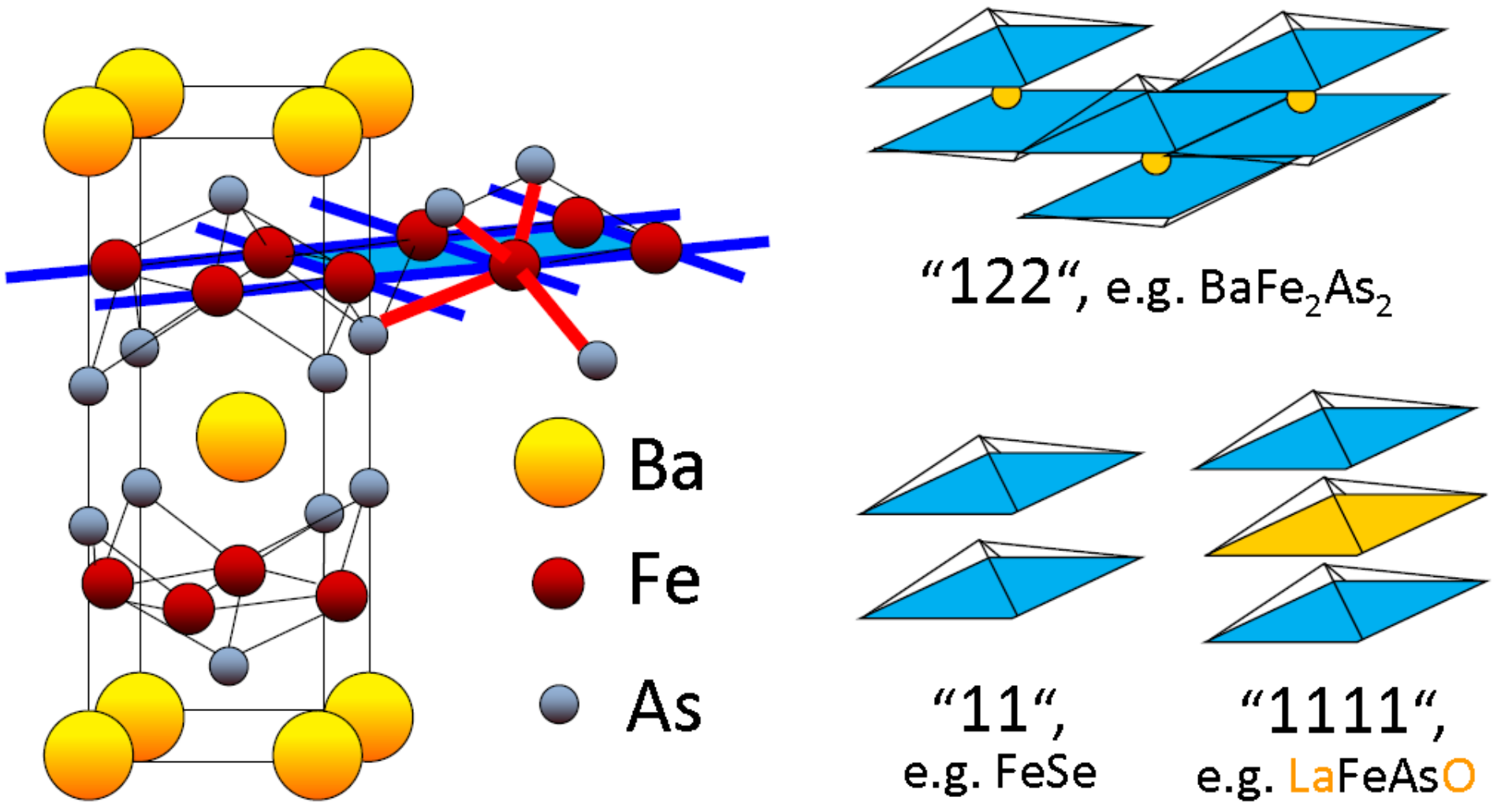

**Fig. 12** Left: Crystal structure of $BaFe_2As_2$ with the quadratic Fe lattice (blue) and tetrahedral As coordination of the Fe atoms (red)
Right: Schematic structure of the 122, 11 and 1111 Fe-Sc indicating the orientation of the Fe(P/As/Se/Te) pyramids (blue)

A huge difference between the two HTS families is the replacibility of the transition metal atoms: In cuprate HTS, 10 percent substitution of Cu atoms by Zn, the rightward neighbouring atom in the periodic table, suppresses superconductivity completely. In stark contrast, in $BaFe_{2-x}Co_xAs_2$ the introduction of Co into the Fe layers even introduces superconductivity by the concomitant electron doping, e.g., up to $T_c$ = 24 K for Co concentration x = 0.06 [196,197]: Fe-Sc apparently tolerate considerable disorder in the Fe planes. Another huge difference is that the undoped, "parent" compounds of cuprate HTS are antiferromagnetic insulators, whereas Fe-Sc derive from magnetic metal parent compounds [8]. The electronic correlations in the Fe-Sc are certainly much weaker than in cuprate HTS where the introduction of extra charges by doping apparently totally rearranges the electronic system. However, in spite of the normal metal state of the Fe-Sc parent compound, the scattering process of the conduction electrons in the Fe atoms involves novel electron-orbital selective correlation mechanisms which prevent an easy theoretical description [198].

Just like HTS, Fe-Sc are extreme type-II superconductors with $\lambda > 100$ nm [199] and $\xi \sim 1$ nm [202]. For cuprate HTS, the discrepancy between the distance d ~ 1 nm between the SC $(CuO_2/Ca/)_{n-1}CuO_2$ stacks [147] and the coherence length $\xi_c = 0.3$ nm perpendicular to the planes leads to the discussed materials-intrinsic weak-link behavior (with the particular exception of the Cu-HTS, e.g., YBCO due to the presumed proximity effect via the CuO chains). For Fe-Sc, the situation is a bit more benign with the reported d = 0.86 nm / $\xi_c$ = 0.6 nm for Nd-1111, d = 0.65 nm / $\xi_c$ = 1.5 nm for Ba-122 and d = 0.6 nm / $\xi_c$ = 0.6 nm for FeSe [202]. Therefore, at least for Ba-122 the application perspectives are really promising. The almost isotropic magnetic field behavior, e.g., for Ba-122 with $B_{c2}$ [$T$ = 0 K] estimates of ~ 50 T and 40 T for magnetic fields parallel and perpendicular to the planes, respectively [202], is not related to the supposed isotropic $s_{+/-}$ SC order parameter but stems from the fact that the Fe-Sc are apparently Pauli-limited: The Zeeman splitting of the electronic single-particle states makes it energetically favorable that the Cooper-pairs split into the constituent up- and down-spin states at fields below the "orbital limit" $B_{c2} = \Phi_0/(2\,\pi\,\xi^2)$ given by the magnetic flux quantum $\Phi_0$ and the SC coherence lengths perpendicular to the field.

The restrictions with respect to the crystalline alignment of neighbouring grains appear to be much less severe for Fe-Sc than for Cu-HTS: In both cases, for small misalignment angles $\alpha$ the critical current density $J_c$ is observed to remain more ore less constant up to a critical angle $\alpha_c$, followed by an exponential decrease $J_c \sim e^{-\alpha/\alpha 0}$ for larger $\alpha$. However, $\alpha_c \approx 10°$ and $\alpha_0 \approx 15°$ as recently reported values for Fe-Sc indicate a much less stringent texture requirement than for cuprate HTS ($\alpha_c \approx$ 3-5°, $\alpha_0 \approx$ 3° / 5°; see Fig. 7) [200]. Moreover, it is not yet clear if this granularity is already an intrinsic limit: Comparing these results from 2011 [200] with the state of the art in 2009 [201], further substantial improvement of the grain boundary $J_c$ transparency can be expected. The chemical complexity of Fe-Sc is not yet as well controlled as for cuprate HTS where it took a decade to arrive at sample qualities that allowed to distinguish between intrinsic and extrinsic effects [104]. By now, the only Fe-Sc system where the constitutional phase diagram has been determined is the binary Fe-Se system, i.e., the route to homogeneous and phase pure samples has not yet been clarified: Present bulk samples show many microstructural features that may result in granularity such as cracks, voids and grain boundaries with impurity phase inclusions [203]. All in all, Fe-Sc apparently still have plenty of room for materials improvement.

## 4. OTHER CHALCOGENIDE SUPERCONDUCTORS

*Chevrel phases* are Mo clusters of the type $A_xMo_6X_8$ with X = S or Se and $A_x$ an interstitial atom [204]. As type II superconductors with relatively high critical fields and temperatures ($SnMo_6S_8$: $T_c$ = 11.8 K, $H_{c2}$(T = 0 K) ~ 34 T; $PbMo_6S_8$: $T_c$ = 12.6 K, $H_{c2}$(T = 0 K) ~ 50 T; $Gd_{0.2}PbMo_6S_8$: $T_c$ = 14.3 K, $H_{c2}$(T = 0 K) ~ 60 T [205] ) these materials had been studied very actively with respect to wire application just before the discovery of cuprate HTS [206].

*Transition metal dichalcogenides* ("TMDs"; $MX_2$ with M = Nb, Ti, Mo, W… and X = S, Se or Te) are based on hexagonal transition metal layers sandwiched between two hexagonal chalcogenide layers. The various relative orientations of these layers give rise to different polytypes (s. Fig. 13 left). The weak van der Waals interactions between such neighbouring sandwiches allows easy slippage as well as easy cleavage along these planes (s. Fig. 13 right). Bulk superconductivity for 2*H*-$NbSe_2$ ($T_c$ ~ 7 K [207]) and 2*H*-$TaS_2$ ($T_c$ ~ 4 K [208]) is well-known for a long time. For 1*T*-$TiSe_2$ it was recently achieved by means of Cu-doping (up to $T_c$ = 4.15 K [209]) or pressure application (up to $T_c$ = 1.8 K [210]). Recently, new techniques for the application of high electric-fields enabled here a continuous doping charge adjustment to cover the full superconductive phase range with $T_c$ up to ~ 3 K [59]. For 2*H*-$MoS_2$ $T_c$ up to 10.8 K [211], for 2*H*-$MoSe_2$ up to ~ 7 K [212]) and for 2*H*-$WS_2$ up to ~ 7 K [212]) could be adjusted by electric-field effect. Charge density wave ("CDW") behavior is a competing instability of the electronic system common to all TMDs [59,213]. Whether the physical origin for this CDW instability is primarily of electronic or phononic origin is still under debate [214] and depends probably very much on the particular TMD material [215,216].

Recently, $Bi_4O_4S_3$ ($T_c$ = 4.5 K) [217] established a first representative of a new family of layered superconductors where *superconducting $BiS_2$ layers* are separated by various blocking layers [218,219]. The superconductive transition is apparently well described by conventional phonon-mediated pairing.

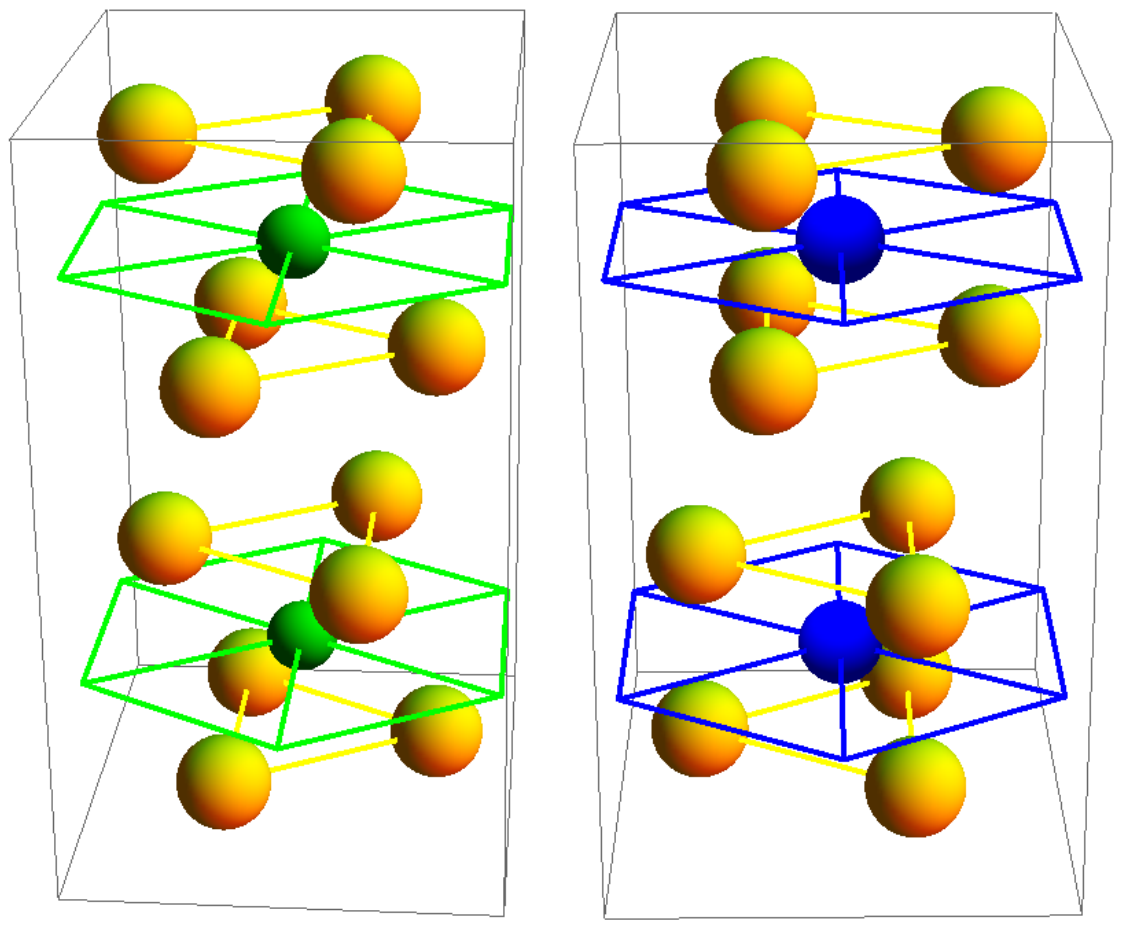
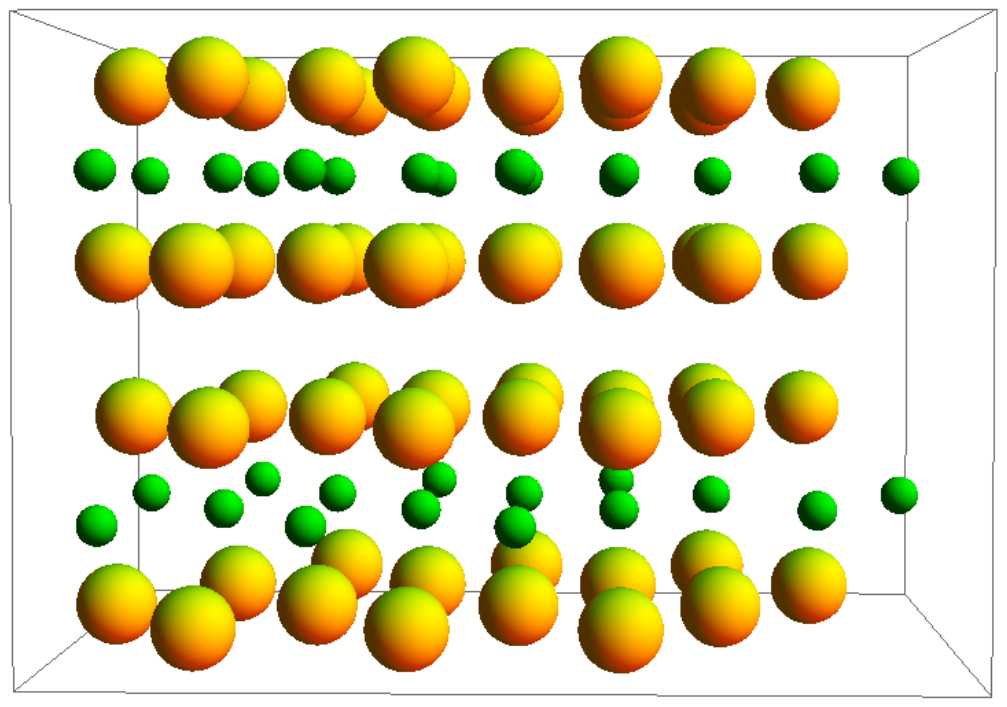

**Fig. 13** Left: Crystal structure of 1*T*-$TiSe_2$ (two unit cells) and 2*H*-$NbSe_2$ (one unit cell). Ti atoms are shown in green, Se atoms in yellow and Nb atoms in blue.

Right: 2d-layer structure of transition metal chalcogenides, e.g. 1*T*-$TiSe_2$.

## 5. HEAVY FERMION SUPERCONDUCTORS

Heavy-Fermion (HF) systems are stoichiometric lanthanide or actinide compounds whose qualitative low-temperature behavior in the normal state closely parallels the one well-known from simple metals. The key features are the specific heat which varies approximately linearly $C \sim \gamma T$, the magnetic susceptibility which approaches a temperature independent constant $\chi(0)$ and the electrical resistivity which increases quadratically with temperature $\rho(T) = \rho_0 + AT^2$. However, the coefficient $\gamma \sim 1$ J / mole K$^2$ as well as $\chi(0)$ are enhanced by a factor of $100-1000$ as compared to the values encountered in ordinary metals while the Sommerfeld-Wilson ratio $[\pi (k_B)^2 \chi(0)] / [3(\mu_B)^2 \gamma]$ is of order unity. The large enhancement of the specific heat is also reflected in the quadratic temperature coefficient $A$ of the resistivity $A \sim \gamma^2$. These features indicate that the normal state can be described in terms of a Fermi liquid. The excitations determining the low-temperature behavior correspond to heavy quasiparticles whose effective mass $m^*$ is strongly enhanced over the free electron mass m. The characteristic temperature $T^*$ which can be considered as a fictitious Fermi temperature or, alternatively, as an effective band width for the quasiparticles is of the order 10 – 100 K. Residual interactions among the heavy quasiparticles lead to instabilities of the normal Fermi liquid state. A hallmark of these systems is the competition or coexistence of various different cooperative phenomena which results in highly complex phase diagrams. Of particular interest are the SC phases which typically form at a critical temperature $T_c \leq 2$ K. $PuCoGa_5$ with $T_c \sim$ 18.5 K is up to now the only "high- $T_c$ " HF representative [220].

The discovery of superconductivity in $CeCu_2Si_2$ [11] (see Fig. 14) forced condensed-matter physicists to revise the generally accepted picture of the electrons occupying the inner shells of the atoms. Traditionally, the corresponding states were viewed as localized atomic-like orbitals which are populated according to Hund's rules in order to minimize the mutual Coulomb repulsion. This leads to the formation of local magnetic moments which tend to align and which are weakly coupled to the delocalized conduction electrons. The latter were viewed as "free" fermions which occupy coherent Bloch states formed by the valence orbitals of the atoms. An attractive residual interaction among the conduction electrons causes the normal metallic state to become unstable with respect to superconductivity. The Cooper pairs which characterize a SC phase are broken by magnetic centers. The damaging effect of 4f- and 5f-ions was well established by systematic studies of dilute alloys.

The key to the understanding of the origin and the nature of the unanticipated superconductivity in $CeCu_2Si_2$ [11] lies in a better appreciation of the physics of the highly unusual normal state out of which it forms. The characteristic features summarized above show that the magnetic degrees of freedom of the partially filled f-shells form a strongly correlated paramagnetic Fermi liquid with an effective Fermi energy of the order of 1-10 meV. The existence of a Fermi liquid state with heavy quasiparticles involving the f-degrees of freedom has been confirmed experimentally.

The SC phases of HF materials are characterized by BCS-type pair-correlations among the heavy quasiparticles of the normal state [221]. This picture is inferred from the fact that the discontinuities observed in various thermodynamic properties at the SC transition scale with the large effective masses of the normal state. The unusual energy and length scales in HF superconductors, however, lead to novel phenomena. First, the effective Fermi temperature typically exceeds the SC transition temperature only by one order of magnitude. This fact implies that standard weak-coupling theory which keeps only the leading contributions in an expansion with respect to the ratio $T_c / T^*$ can provide only qualitative results. Pronounced strong-coupling corrections are reflected in deviations from the universal behavior predicted for weak-coupling systems. Second, the small characteristic energy $k_B T^*$, or, equivalently, the narrow width of the quasiparticle bands implies a small value of the Fermi velocity $v_F$. As a result, the size of the Cooper pairs in the HF systems, i.e., the coherence length $\xi_0 = \frac{\hbar v_F}{k_B T_c}$ is much less than in a typical "ordinary" superconductor. Many of the systems can be considered as "clean" with the mean-free paths exceeding the coherence length. As a result, anisotropic pair states can form which are strongly suppressed by impurity scattering in "ordinary" superconductors with large coherence lengths.

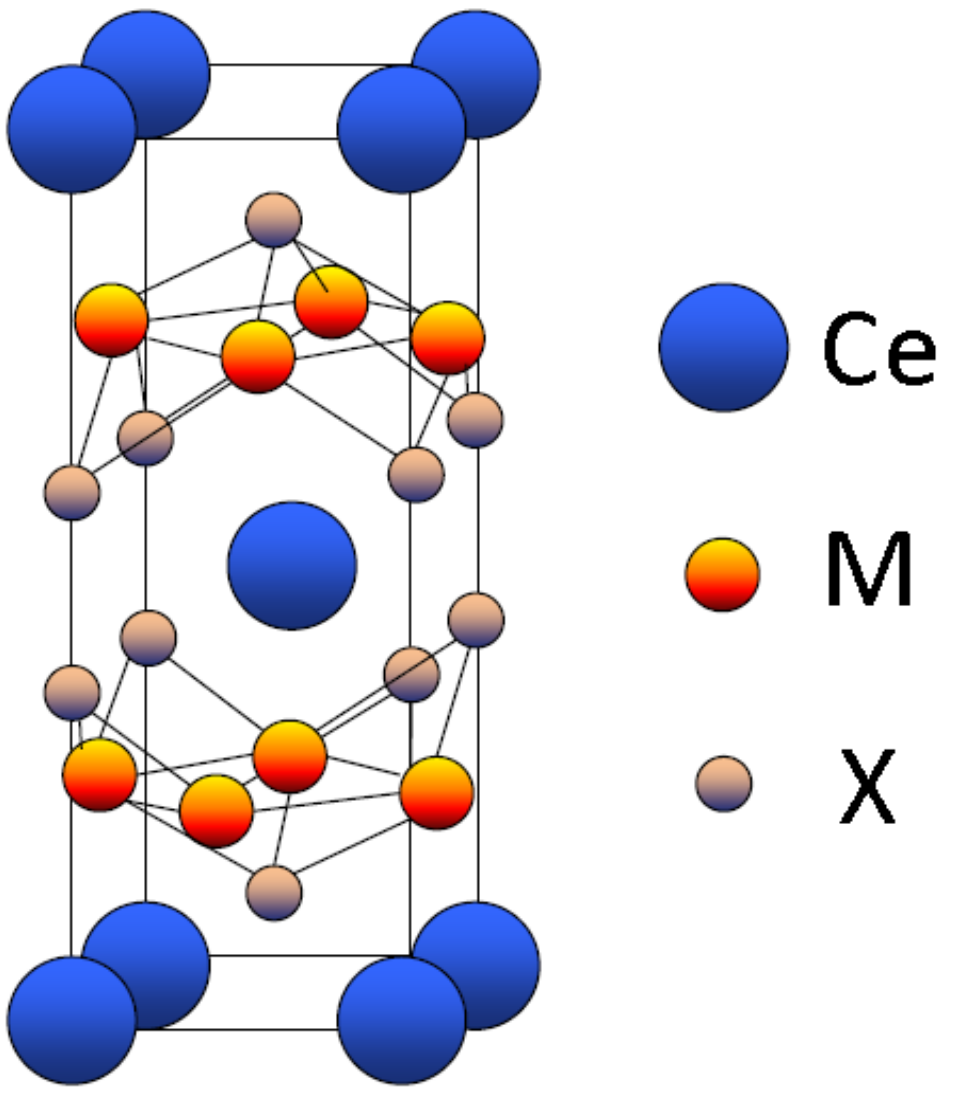

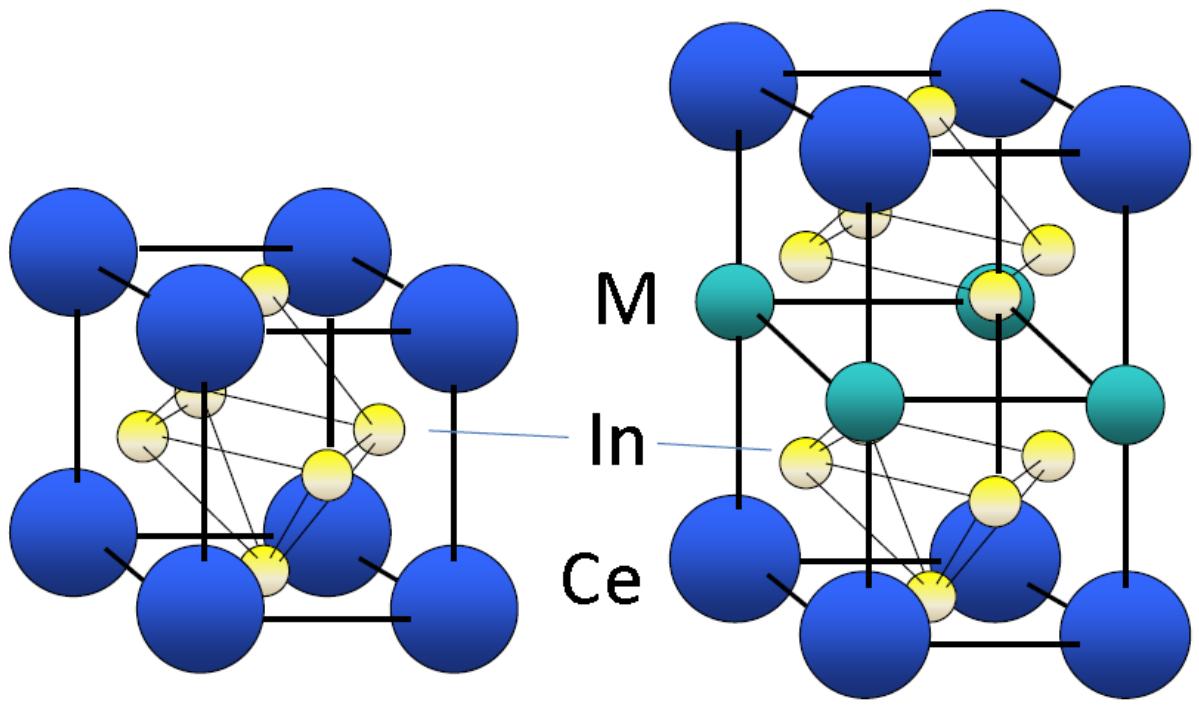

**Fig. 14** Conventional unit cell of $CeM_2X_2$ (M = Cu, Ni, Ru, Rh, Pd, Au, ...; X = Si, Ge) and $URu_2Si_2$.

**Fig. 15** Unit cell of $CeIn_3$ and $CeMIn_5$ (M = Co, Ir).

Although we know the correlations which characterize the SC state we are still far from a complete theory of superconductivity in these materials. Major questions are still open. Among the prominent ones is the question of which order parameters characterize the SC states and what is the origin of the attraction between the quasiparticles. The correct microscopic description of the interaction would yield, of course, the SC transition temperatures as well as the detailed form of the order parameter.

To describe ordered phases, the determination of the type and the symmetry of the order parameter is of central importance. The latter restricts the possible excitations in the ordered phases and hence determines the low-temperature properties. Order parameters given in terms of expectation values of physical observables like spin- and charge densities can be directly measured by x-ray or neutron diffraction. The magnetic phases of the lanthanide and actinide compounds are therefore rather well characterized. Superconductivity, however, corresponds to an off-diagonal long-range order parameter which is not directy observable. The usual procedure to determine the symmetry of the SC order parameter is to select plausible candidate states corresponding to irreducible representations of the symmetry group, calculate expected behavior of physical quantities and compare the predictions with experiment.

The occurrence of long-range order described by an order parameter is most frequently associated with spontaneous symmetry breaking. The simplest superconductors where only gauge invariance is broken are called conventional. In this case the SC state has the same symmetry as the underlying crystal. It should be noted that conventional is not a synonym for isotropic. A superconductor who has additional broken symmetries besides gauge symmetry, i.e., whose symmetry is lower than that of the underlying crystal is called unconventional. Considerable progress has been made recently in detecting unconventional order parameter symmetries. Angle-resolved studies of thermodynamic and transport properties in the vortex phase determine the position of order parameter nodes relative to the crystal axes.

Superconducting order parameters behave in many respects like two-fermion wave functions being represented by a 2 x 2 matrix in (pseudo-) spin space. The latter can be decomposed into a scalar antisymmetric (pseudo-) spin singlet and a multi-component (pseudo-) spin symmetric triplet contribution. The order parameter can be chosen as purely antisymmetric or purely symmetric provided either spin orbit interaction can be neglected or the crystal lattice has an inversion center. The majority of HF superconductors crystallizes in lattices with inversion center. There is, however, a restricted number of non-centrosymmetric HF superconductors where the strong spin-orbit interaction leads to a multitude of unique properties [222]. Highly promising systems are artificial superlattices consisting of the HF superconductor $CeCoIn_5$ (see Fig. 15) and its conventional metallic counterpart $YbCoIn_5$ [223,224,191].

In several systems, the order parameter has been shown to be unconventional. This was inferred mainly from anisotropies displayed by thermal transport in the presence of an external magnetic field. The most spectacular examples of unconventional superconductors are $UPt_3$ [225] and $PrOs_4Sb_{12}$ [226] where the split transition points to a multicomponent order parameter. In the majority of HF superconductors, the order parameter seems to be a complex scalar function.

The fact that HF superconductors can be considered as "clean" systems with the mean-free path exceeding the coherence length has rekindled speculations so as to find inhomogeneous superconductivity phases (Fulde-Ferrel-Larkin-Ovchinnikov states; “FFLO” / “LOFF”) at sufficiently low temperatures in sufficiently high magnetic magnetic fields [227, 228]. In fact, the order of the phase transition in an applied magnetic field has been found to change from second to first order with decreasing temperature – in agreement with long-standing theoretical predictions [229]. The claims that the HF superconductor $CeCoIn_5$ might show FFLO/LOFF states had to be revised [230]. Although there is clear evidence for a field-induced phase inside the superconducting state this phase does not exhibit the properties anticipated for a FFLO/LOFF state. The observed features rather suggest the emergence of antiferromagnetic order (see [231] and references therein).

An indispensable prerequisite to a microscopic theory of superconductivity is a microscopic theory of the normal state, of the quasiparticles and their interactions. The Landau theory does not make assumptions or predictions concerning the microscopic nature of the ground state and the low-lying excitations. It does not address the question how the latter emerge in an interacting electron system. During the past decade it became clear that there are different routes to heavy fermion behavior [232].

In Ce- and Yb-based compounds, the heavy quasiparticles with predominantly 4f-character arise through the Kondo effect in the periodic lattice. The Kondo picture for these heavy-fermion compounds is supported by the fact that the thermodynamic properties at low temperatures (e. g., the specific heat, the magnetic susceptibility) as well as the temperature-dependence of the spectroscopic data can be reproduced by an Anderson model. This picture has been confirmed in detail by deHaas-vanAlphen and photoemission studies [233]. The strongly renormalized quasiparticles can be described by a semiphenomenological ansatz, the Renormalized Band Method which yields realistic quasiparticle bands [234]. The emerging local Kondo screening and the formation of the coherent Fermi liquid state in $YbRh_2Si_2$ was convincingly demonstrated by high-resolution STM studies [235] and detailed transport measurements [236]. The nesting features of the calculated Fermi surfaces can serve as a useful guideline in the search for instabilities with respect to modulated structures [237, 238]. In particular, the calculations provide realistic models for the study of the competition/coexistence phase diagrams. Despite the efforts to implement modern many-body methods for strong correlations into realistic electronic structure calculations there is still no general concept for quantitative microscopic correlations. In particular, the subtle interplay between local and intersite effects continues to challenge theorists. The latter may lead to long-range order while the former favor the formation of a Fermi liquid state at low temperatures.

A microscopic picture for the heavy quasiparticles has finally emerged for the actinide compounds [232, 239]. Increasing experimental evidence points towards a dual character of the 5f-electrons with some of them being delocalized forming coherent bands while others stay localized reducing the Coulomb repulsion by forming multiplets. The "dual character" of the 5f electrons in actinides should be reflected in the line shapes of the core level spectra [240, 241]. This is confirmed by recent core level photoelectron spectra of U compounds which exhibit rich structures with wide yet systematic variation from compound to compound [242]. The hypothesis of the dual character is translated into a calculational scheme which reproduces both the Fermi surfaces and the effective masses determined by deHaas-vanAlphen experiments without adjustable parameter. The method yields a model for the residual interaction leading to various instabilities of the normal phase. The dual model should also provide insight into the mysterious hidden order phases of U compounds.

In Ce, Yb and actinide HF compounds, the coherent heavy quasiparticles are derived from the partially filled f-shells whose degrees of freedom have to be (partially) included into the Fermi surface. The situation is different for the Pr skutterudites where the quasiparticles are derived from the conduction states whose effective masses are strongly renormalized by low-energy excitations of the Pr 4f-shells [243]. From a microscopic point of view we are dealing with three different classes of HF SC materials.

It is generally agreed that the pairing interaction in HF superconductors is of electronic origin. Theoretical models usually involve the exchange of a boson. It is therefore convenient to classify the various mechanisms according to the boson which is exchanged [244]. The majority of models construct effective interactions based on the exchange of spin-fluctuations. Neither these models nor their refined variants which account for the internal orbital f-electron structure are able to properly predict the symmetry of the order parameter, e. g., in $UPt_3$. In particular, the calculations do not reproduce a multicomponent order parameter as stable solution. A model based on the exchange of weakly damped propagating magnetic excitons was suggested for U-based HF compounds [245,246]. First estimates of the transition temperature yield a value of the correct order of magnitude. Pairing due to intra-atomic excitations may also occur in the Pr-skutterudite HF superconductor. In this case, however, quadrupolar instead of magnetic excitons should be involved [247].

Superconductivity in HF compounds is usually found to coexist or to compete with various cooperative phenomena. Of particular importance in this context is itinerant antiferromagnetism as realized in a spin density wave (SDW). Since both order parameters appear in the itinerant quasiparticle system the observed behavior results from a subtle interplay between Fermi surface geometry and gap structures.

Many HF systems are on the verge of magnetic instability. By application of pressure these materials may be tuned in their normal states through a quantum critical point (QCP) from an antiferromagnet to a paramagnetic metal. The theoretical picture, however, is at present still rather controversial and contentious.

## 5A. Ce- AND Yb-BASED HF COMPOUNDS

The discovery of superconductivity in $CeCu_2Si_2$ ($T_c$ = 1.5 K; see Fig. 14) [11] initiated the rapid development of HF physics. For nearly two decades, this material was the only Ce-based heavy fermion superconductor at ambient pressure. In the past decade, superconductivity at ambient pressure was found in the Ce-based HFS $CeM_mIn_{3+2m}$ (M = Ir or Co; m = 0, 1) [248,249]. The most prominent member of this family is $CeCoIn_5$ which has a relatively high $T_c$ = 2.3 K (see Fig. 15) [250]. Of fundamental interest ist the discovery of HF superconductivity in $CePt_3Si$ ($T_c$ = 0.75 K; see Fig. 16) [251] which crystallizes in a lattice without inversion symmetry (for a recent review see [252] and references therein). β-$YbAl_4B$ ($T_c$ = 80 mK; see Fig. 17) [253] is the only Yb-based HF superconductor known to date. It remains a major challenge to give reasons for the apparent asymmetry between Ce-based and Yb-based HF systems, i.e., to explain why there is a great variety of Ce-based HF superconductors but only one weak Yb-based HF superconductor.

The SC phases in the lanthanide-based HF compounds are characterized by anisotropic order parameters which reflect the on-site repulsion introduced by the strong correlations of the partially filled 4f shells. The current experimental and theoretical research focuses on the question which factors determine the character of the low-temperature phases. The subtle interplay between local singlet-formation via Kondo effect and long-range magnetic order can be monitored experimentally in pressure studies where isostructural relatives of the HF superconductors are tuned from magnetic phases at ambient pressure to SC states, e. g., $CeCu_2Ge_2$ [254], $CePd_2Si_2$ [255,256] $CeNi_2Ge_2$ [257] and $CeRh_2Si_2$ [258], $CeSn_3$ [259] and $CeIn_3$ (see Fig. 15) [255]. Similar behavior is found in doping experiments where constituents of the metallic host are successively replaced. In general, a highly interesting type of competition between (anisotropic) superconductivity and magnetic order is encountered. In $CeCu_2Si_2$, the heavy Fermi liquid is unstable with respect to both superconductivity and a spin-density wave. The actual ground state realized in a sample depends sensitively on the composition of the sample [237]. The complicated interplay between magnetic excitations and superconductivity is reflected in INS spectra [238, 260] (for recent reviews see [261,262]).

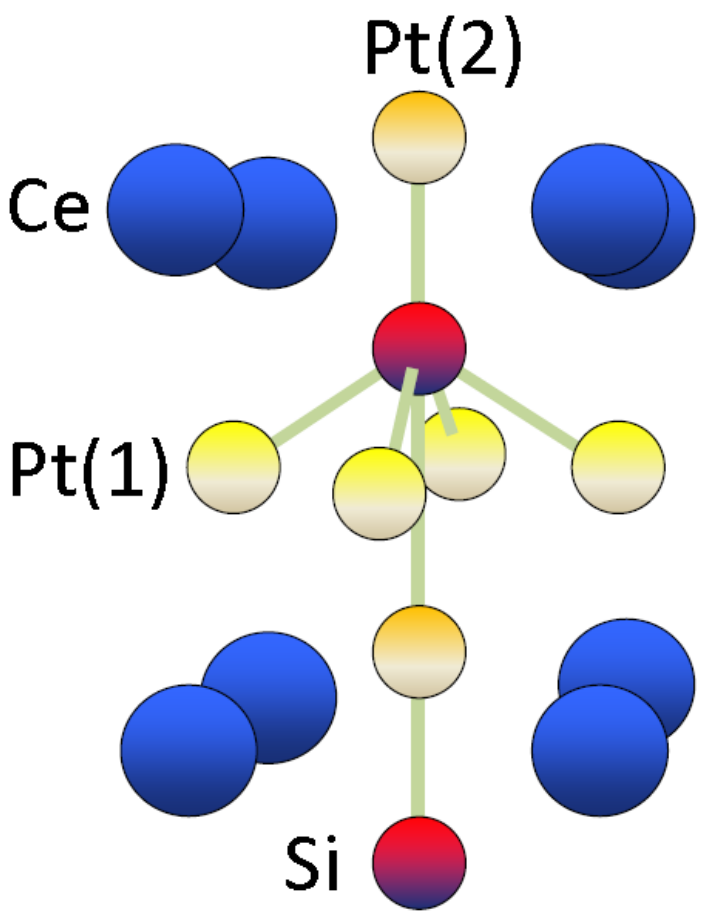

**Fig. 16** Crystal structure of $CePt_3Si$

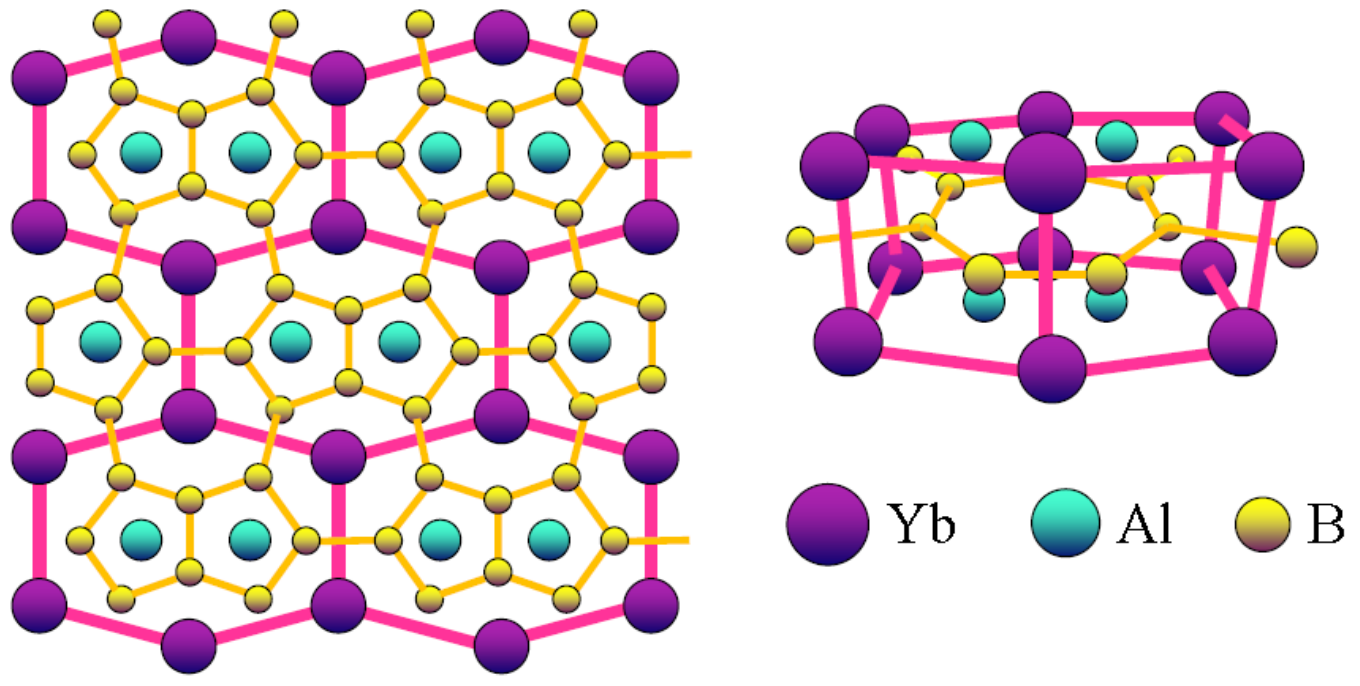

**Fig. 17** Crystal structure of β-$YbAl_4B$

## 5B. ACTINIDE-BASED HF COMPOUNDS

Heavy Fermion superconductivity is found more frequently in intermetallic actinide-compounds than in lanthanide-compounds. This may be related to the different nature of heavy quasiparticles in actinide-compounds where the 5f-electrons have a considerable, though orbitally dependent, degree of delocalization. The genuine Kondo mechanism is not appropriate for heavy quasiparticle formation as in lanthanide-compounds. This may lead to more pronounced delocalized spin fluctuations in U-compounds which mediate unconventional Cooper pair formation. Antiferromagnetic ("AF") order, mostly with small moments of the order $10^{-2}\ \mu_B$ is frequently found to envelop and coexist with the SC phase.

The hexagonal compound $UPt_3$ (see Fig. 18) [263] exhibits triplet pairing. It sticks out as the most interesting case of unconventional superconductivity with a multicomponent order parameter whose degeneracy is lifted by a symmetry-breaking field due to a small moment AF order. In contrast, in $UPd_2Al_3$ (see Fig. 19) [264] superconductivity coexists with large moment antiferromagnetism. Probably spin singlet pairing is realized. There is experimental evidence for a new kind of magnetic pairing mechanism mediated by propagating magnetic exciton modes. The sister compound $UNi_2Al_3$ [265] is an example of coexistence of large moment antiferromagnetism with a SC triplet order parameter. In $URu_2Si_2$ [266] the SC order parameter symmetry is still undetermined. The interest in this compound is focused more on the enveloping phase with a "hidden" order parameter presumably of quadrupolar type or an "unconventional" spin density wave (SDW). The oldest cubic U-HF superconductor $UBe_{13}$ [267] and its thorium alloy $U_{1-x}Th_xBe_{13}$ is also the most mysterious one. While for the pure system there is a single SC phase of yet unknown symmetry, in the small Th concentration range two distinct phases exist which may either correspond to two different SC order parameters or may be related to a coexistence of superconductivity with a SDW phase. In addition, in $UBe_{13}$ SC order appears in a state with clear non-Fermi liquid type anomalies. More recently the coexistence of ferromagnetism and superconductivity in $UGe_2$ [268] has been found. Due to the ferromagnetic polarization the triplet gap function contains only equal spin pairing.

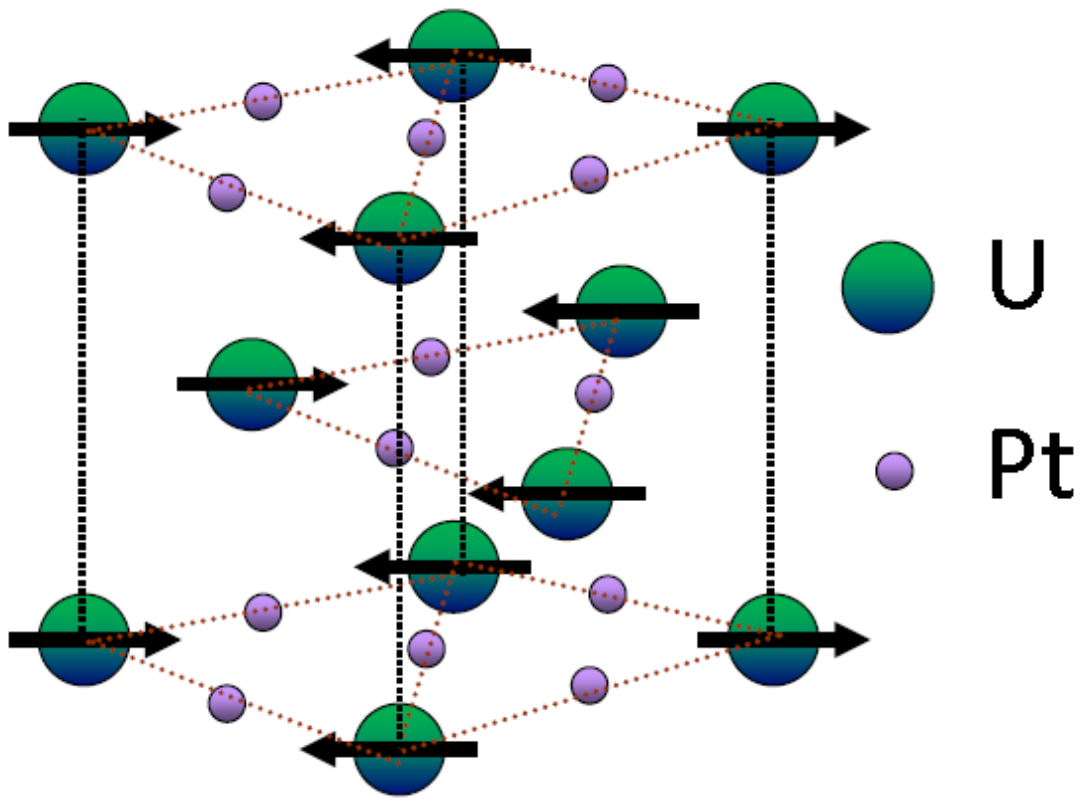

**Fig. 18** Crystal structure of $UPt_3$ (a = 0.5764 nm, c = 0.4884 nm) and AF magnetic structure ($T < T_N = 5.8$ K).

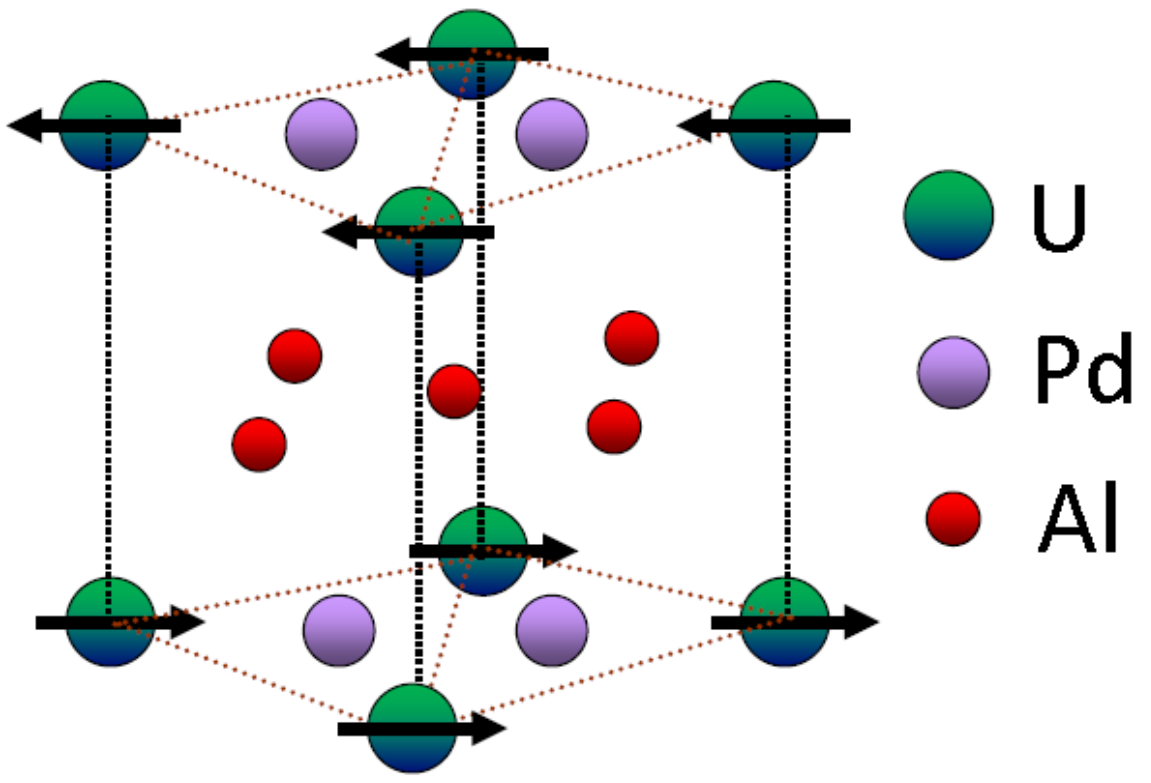

**Fig. 19** Conventional unit cell of $UPd_2Al_3$ (a = 0.5350 nm, c = 0.4185 nm) and simple AF magnetic structure.

The hexagonal heavy fermion compound $UPt_3$ (see Fig. 18) is the "flagship" of unconventional superconductivity, despite its rather low critical temperature of slightly less than 1 K. It is set aside from all other unconventional superconductors insofar as it exhibits two SC phase transitions. The exciting discovery of the split superconductive transitions in $UPt_3$ at $T_{c1}$ = 530 mK and $T_{c2}$ = 480 mK in specific heat measurements [225] has led to an enormous amount of experimental and theoretical work on $UPt_3$ [269]. The additional small moment antiferromagnetism observed in $UPt_3$ ($T_N$ = 5.8 K, $\mu$ = 0.035 $\mu_B$) plays a key role in the identification of the SC order parameter since the in-plane staggered magnetization acts as a symmetry breaking field (SBF) to the SC multicomponent order parameter. The SBF is believed to be responsible for the appearance of two SC transitions which otherwise would merge into one. Therefore one can identify three distinct SC phases. Despite the wealth of experimental results on $UPt_3$ there is no unequivocal consensus on the symmetry and node structure of the SC gap.

$UPd_2Al_3$ (see Fig. 19) [264] is a rather special case among the U-based HF superconductors. There is also AF order below $T_N$ = 14.3 K with almost atomic size local moments ($\mu$ = 0.85 $\mu_B$) in contrast to the small moments in other U-compounds. The AF order coexists with superconductivity below $T_c$ = 1.8 K. This suggests that in addition to the heavy itinerant quasiparticles nearly localized 5f-electrons should be present. They result from the dominating $5f^2$ configuration of the $U^{4+}$ ion [270]. This dual nature of 5f-electrons is even more obvious than in $UPt_3$ as is seen in various experiments. A direct confirmation of this dual nature of 5f-electrons in $UPd_2Al_3$ was obtained from inelastic neutron scattering (INS) [271] which found excitations that originate from local CEF transitions and disperse into bands of "magnetic excitons" due to intersite exchange. Complementary tunneling experiments with epitaxial $UPd_2Al_3$-$AlO_x$-Pb heterostructures probed the response of the itinerant quasiparticles and their SC gap. Strong coupling features in the tunneling density of states were reported [272] suggesting that the magnetic excitons identified in INS are the bosonic "glue" which binds the electrons to Cooper pairs [245]. This new mechanism for superconductivity is distinctly different from both the electron-phonon and spin fluctuation mechanism known so far. The pairing potential is mediated by a propagating boson (the magnetic exciton) as in the former case but depends on the spin state of conduction electrons as in the latter case.

The possibility of coexisting ferromagnetism and superconductivity was first considered by Ginzburg [273] who noted that this is only possible when the internal ferromagnetic field is smaller than the thermodynamic critical field of the superconductor. Such a condition is hardly ever fulfilled except immediately below the Curie temperature $T_C$ where coexistence has been found in a few superconductors with local moment ferromagnetism and $T_C < T_c$ such as $ErRh_4B_4$ and $HoMo_6S_8$. If the temperature drops further below $T_C$ the internal ferromagnetism molecular field rapidly becomes larger than $H_{c2}$ and superconductivity is destroyed. The reentrance of the normal state below $T_C$ has indeed been observed in the above compounds. The only compound known so far where local moment ferromagnetism coexists homogeneously with superconductivity for all temperatures below $T_c$ is the borocarbide compound $ErNi_2B_2C$ [274] (see chapter 8). The competition between ferromagnetism and superconductivity becomes more interesting if ferromagnetic order is due to itinerant electrons which also form the SC state. If the interaction slightly exceeds a critical value one has weak ferromagnetic order such as in $ZrZn_2$ with large longitudinal ferromagnetic spin fluctuations.

In this case p-wave superconductivity may actually be mediated by the exchange of ferromagnetic spin fluctuations and coexist with the small ferromagnetic moments [275]. p-wave superconductivity was predicted to exist in this case both on the ferromagnetic as well as on the paramagnetic side of this critical parameter region. The discovery of unconventional superconductivity under pressure in the itinerant ferromagnets $UGe_2$ [268] and later for URhGe [276] and the 3d ferromagnet $ZrZn_2$ [277] (more recent investigations showed that here only a surface layer probably with higher Zr content becomes SC [278]) under ambient pressure has opened this theoretical scenario to experimental investigation.

$URu_2Si_2$ (see Fig. 14), a “moderately heavy” fermion compound has mystified experimentalists and theorists alike with the discovery of AF order and another still unidentified ”hidden order” phase. In addition, the compound becomes a nodal superconductor below $T_c$ = 1.2 K [266, 279, 280]. The simple tetragonal AF order with wave vector in c-direction has tiny moments $\mu \sim 0.02\ \mu_B$ along c-axis [281, 282] which are of the same order as in $UPt_3$. However, unlike in $UPt_3$, very large thermodynamic effects, e. g., in specific heat, thermal expansion etc., occur which are hard to reconcile with the small ordered moments. These pronounced anomalies were interpreted as evidence for the presence of a second “hidden order” parameter, which cannot be seen in neutron or x-ray diffraction. In the itinerant models the order parameter is due to an unconventional pairing in the particle-hole channel leading to an unconventional SDW which has vanishing moment in the clean limit and also does not break time reversal invariance. The small staggered moments may then be induced in the condensate due to impurity scattering. Finally, in the dual models one assumes a localized singlet-singlet system in interaction with the itinerant electrons to cause induced moment magnetism with small moments but large anomalies. In all models it was previously taken for granted that both the primary “hidden” order parameter and AF order coexist homogeneously within the sample. However, hydrostatic and uniaxial pressure experiments [283, 284] have radically changed this view, showing that the order parameters exist in different parts of the sample volume; the tiny AF moment is not intrinsic but due to the small AF volume fraction under ambient pressure. Applying hydrostatic pressure or lowering the temperature increases the AF volume fraction and hence the ordered moment until it saturates at an atomic size moment of 0.4 $\mu_B$ per U atom. This means that the evolution of antiferromagnetism arises from the increase of AF volume with pressure rather than the increase of the ordered moment $\mu$ per U-atom. The character of the “hidden order” phase as well as its origin are still discussed rather controversially [285, 286, 287, 288, 289, 290].

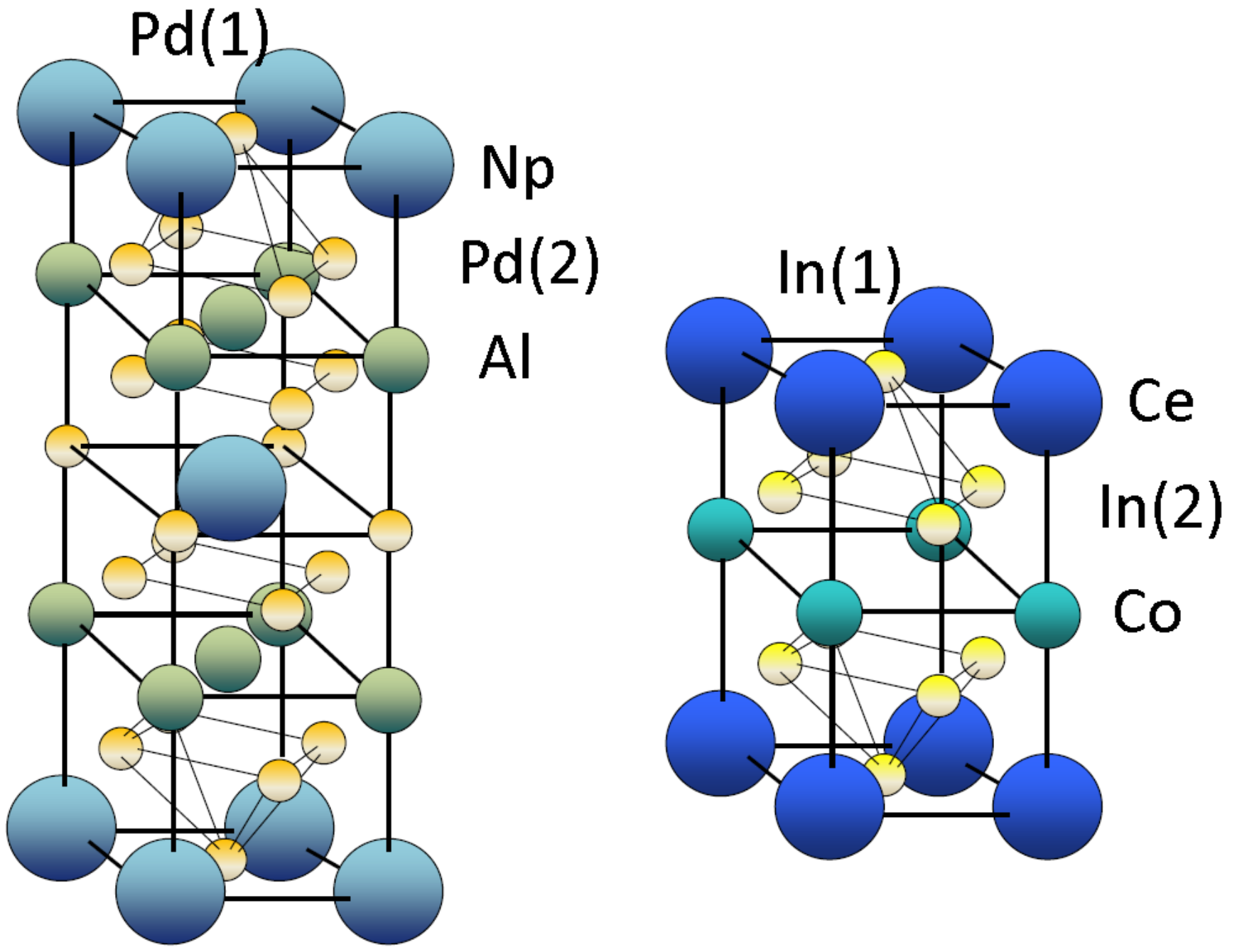

**Fig. 20** Structure of $NpPd_5Al_2$ (left panel) and $CeCoIn_5$ (right panel)

The cubic compound $UBe_{13}$ and $U_{1-x}Th_xBe_{13}$ was discovered rather early [267, 291] as a SC HF system. The U atom in this structure is embedded in an icosahedral cage of 12 Be-atoms [294]. A global understanding of the normal state and symmetry breaking in both SC and magnetic state is still elusive. Firstly, $UBe_{13}$ single crystals do not yet achieve the quality as, e. g., $UPt_3$ single crystals so that the symmetry of the anisotropic SC gap functions could not be identified yet. Furthermore, the Th-doped crystals $U_{1-x}Th_xBe_{13}$ show a perplexing variety of SC and possibly also magnetic phases whose microscopic origin and order parameter symmetries are not understood. Already in the normal state $UBe_{13}$ is a rather anomalous metal, e. g., non-Fermi-liquid behavior has been observed and attributed to a multichannel Kondo effect. The $T_c$ values for superconductivity, which occurs in the non-Fermi-liquid state, depend considerably on the type of sample. There are two classes with ‘high’ $T_c \sim 0.9$ K and ‘low’ $T_c \sim 0.75$ K which, however, are not much different in their impurity content.

The transuranium-based superconductors $PuCoGa_5$ ($T_c$ = 18.5 K) [220], $PuRhGa_5$ ($T_c$ = 8.7 K) [292] and $NpPd_5Al_2$($T_c$ = 4.9 K, see Fig. 20) [293] are all unconventional superconductors at ambient pressure with the highest transition temperatures $T_c$ among all the HF superconductors. The Pu-systems crystallize in the same structure as $CeCoIn_5$ (see Fig. 20) to which the structure of the Np-compound is closely related. The properties of the latter system are highly anisotropic.

## 5C. RARE-EARTH SKUTTERUDITES

The HF superconductor $PrOs_4Sb_{12}$ [226] is potentially of similar interest as $UPt_3$ because it represents the second example of multiphase superconductivity [295] with a critical temperature $T_c = 1.85$ K. The skutterudites $RT_4X_{12}$ (R = alkaline earth, rare earth or actinide; T = Fe, Ru or Os and X=P, As or Sb) show a cage structure where large voids formed by tilted $T_4X_{12}$ octahedrons can be filled with R atoms (see Fig. 21). They are, however, rather loosely bound and are therefore subject to large anharmonic oscillations ("rattling") in the cage. In addition, the presence of several equivalent equilibrium positions may give rise to tunneling split states. Both effects may lead to interesting low temperature elastic and transport phenomena, such as thermoelectric effects [296,297]. Depending on the cage-filling atom this large class of compounds displays also a great variety of interesting effects of strong electron correlation. Mixed valent and HF behavior, magnetic and quadrupolar order, non-Fermi liquid and Kondo insulating behavior have been found [226,297].

Studies of non-stoichiometric skutterudites $Pr(Os_{1-x}Ru_x)Sb_{12}$ [298] showed that the type of superconductivity changes at $x \sim 0.6$ where the transition temperature $T_c(x)$ has a minimum value of 0.75 K: For $x < 0.6$ one has an unconventional HF SC, for $x > 0.6$ conventional SC behavior, e.g., the $x = 1$ compound $PrRu_4Os_{12}$ with $T_c \sim 1$ K. In the HF multiphase superconductor $PrOs_4Sb_{12}$ ($x = 0$) the specific heat jump due to superconductivity is superposed on a Schottky anomaly due to the lowest CEF excitation. Nevertheless, its detailed analysis provides clear evidence for a split superconductivity transition [226,299] at $T_{c1} = 1.85$ K and $T_{c2} = 1.75$ K. The total jump of both transitions amounts to $\Delta_{SC}\, C/\gamma T_c \sim 3$ which considerably exceeds the BCS value 1.43 for a single transition. It also proves that the SC state is formed from the heavy quasiparticles that cause the enhanced γ-value of the electronic low-temperature specific heat contribution. A $T_c$-splitting of similar size also was clearly seen in thermal expansion measurements [300]. The two SC transitions are reminiscent of the split transition in $UPt_3$. There, a twofold orbitally degenerate SC state is split by weak AF order that reduces the hexagonal symmetry to an orthorhombic one. This also leads to two critical field curves in the B-T phase diagram. In $PrOs_4Sb_{12}$ no such symmetry breaking field exists.

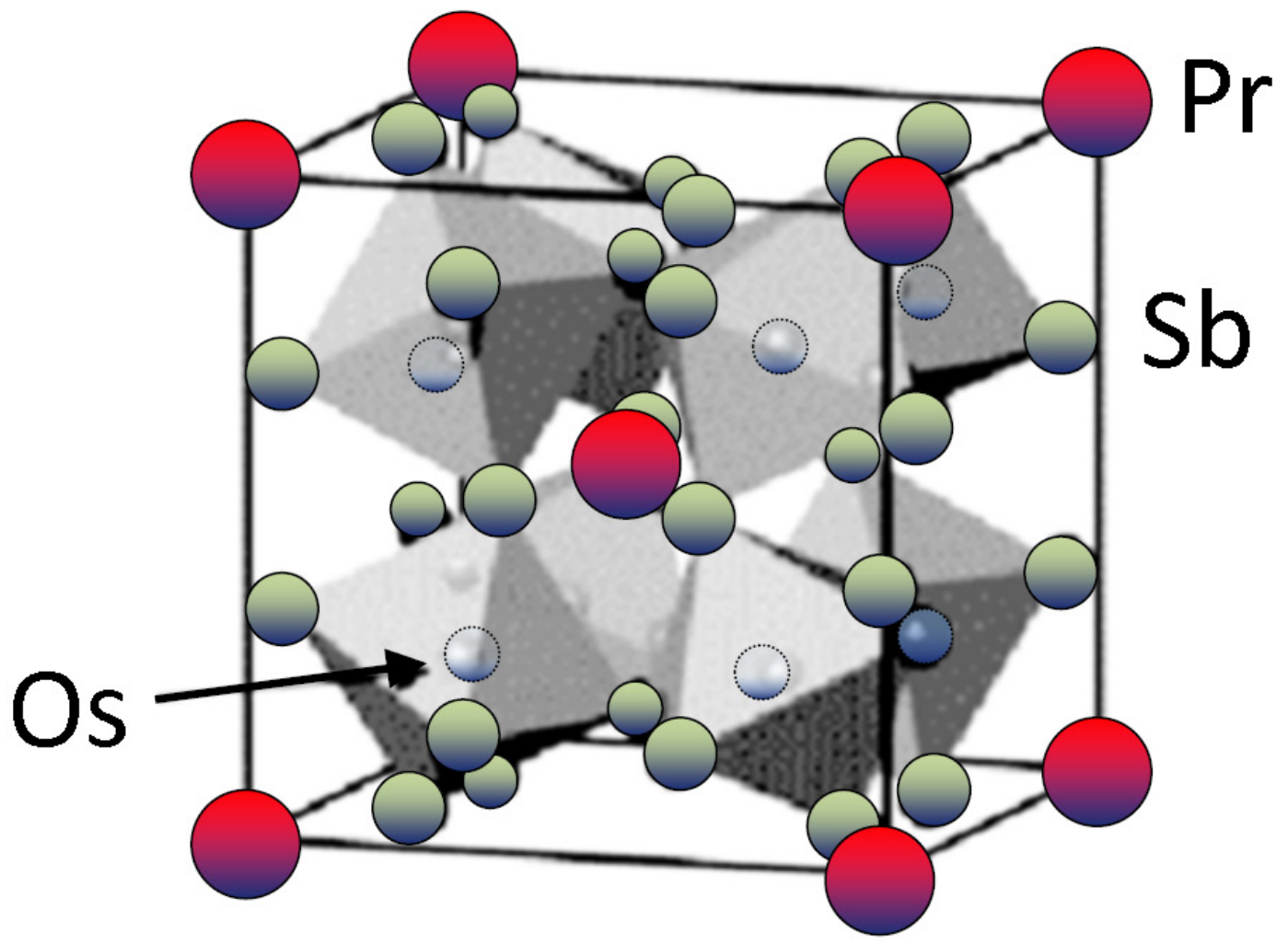

**Fig. 21** Cubic crystal structure of the filled skutterudite $RT_4X_{12}$. The T atoms are located in the center of the X octahedra. For $PrOs_4Sb_{12}$ the lattice constant is a = 0.93017 nm.

Up to now, various experiments gave inconclusive results on the question of the nature of gap anisotropy. As a preliminary conclusion, it seems clear that $PrOs_4Sb_{12}$ is a very unconventional multiphase HF superconductor of potentially the same interest as $UPt_3$. Since that heavy quasiparticles are presumably caused by coupling with virtual quadrupolar excitations from the nonmagnetic 5f ground state one is lead to speculate that superconductivity in $PrOs_4Sb_{12}$ might also result from an unprecedented pairing mechanism based on the exchange of quadrupolar fluctuations. At the moment, this quadrupolar superconductivity mechanism in $PrOs_4Sb_{12}$ as a third possibility for Cooper pair formation in HF compounds in addition to the spin-fluctuation and magnetic-exciton exchange mechanisms is still a conjecture.

## 6. NITRIDE SUPERCONDUCTORS

Nitrides were among the first discovered compound superconductors [301, 302]. NbN established in 1941 with $T_c$ = 16 K a new $T_c$ record value [303].

In 1994, in continuation of work on superconducting borocarbides (see chapter 8), superconductivity at $T_c$ = 12 - 13 K was observed in the structurally related $La_3Ni_2B_2N_3$[304]. Nitride halides MNX (M = Zr, Hf; X = Cl, Br, I) are insulators that contain hexagonal $X(MN)_2X$ layers which may be stacked in several polymorphic arrangements [305]. Chemical or electrochemical intercalation of alkali metals (Li, Na, K) into the van der Waals gaps between these layers, or removal of a small amount of halogen X, dopes electrons into the M d-band inducing superconductivity [306]. $T_c$ = 25.5 K has thus been achieved for $Li_{0.48}(THF)_yHfNCl$ containing cointercalated tetrahydrofuran solvent (THF) (see Fig. 22) [307, 308]. The intercalated MNX phases are very air-sensitive which has hampered study of their physical properties. Superconductivity with maximum $T_c$ = 16.3 K has recently been reported in $A_xTiNCl$ (A = Li, Na, K, Rb) [309], where the constituent layers have an orthorhombic structure that differs from that of the Zr and Hf based materials. This reveals that superconductivity is not specific to one underlying lattice symmetry in this family. Physical measurements show that the nitride halides are not conventional BCS superconductors [310].

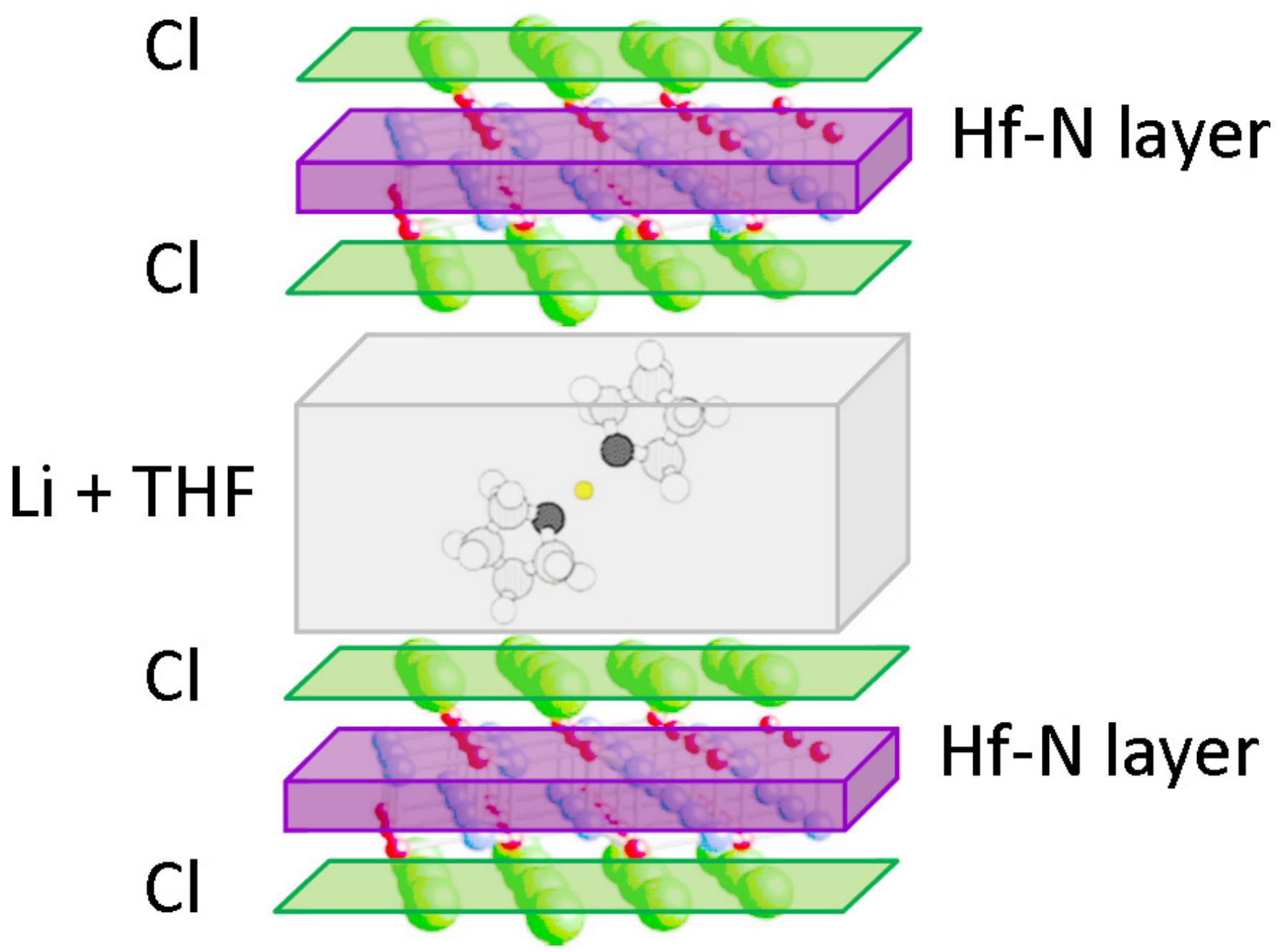

**Fig. 22** Schematic structural model for Li + THF cointercalated β-HfNCl with $T_c$ = 25.5 K. The superconducting Hf-N layers are "wrapped" by chloride layers. The Li + THF doping layers between the Hf-N-Cl sandwiches also adjust the distance between neighbouring superconducting Hf-N layers [308].

## 7. ORGANIC AND OTHER CARBON AND SILICON BASED SUPERCONDUCTORS

Carbides, e.g., NbC ($T_c$ = 12 K) [301,303] and MoC ($T_c$ = 14.3 K) [302,303], were among the first discovered compound superconductors and contended in these early days of superconductivity with nitrides and borides for the highest $T_c$. Theoretical speculations of superconductivity in organic compounds [13] were met for a long time with total disbelief from the experimental side, e.g., from B. Matthias. Things changed when immediately after Matthias's death in 1980 superconductivity was discovered below 0.9 K in the compound (TMTSF)$PF_6$ under a hydrostatic pressure of 12 kbar, with the organic molecule TMTSF (tetra-methyl-tetra-selenium-fulvalene; see Fig. 23) [14]. In 1991 it was found that compounds based on the "soccer ball" $C_{60}$ (see Fig. 24) become superconductors up to $T_c \sim 30$ K when doped with alkali atoms [311].

Meanwhile, a number of TMTSF-based superconductors with $T_c \sim 1$ K has been found, e.g., $(TMTSF)_2ClO_4$ which becomes SC at 1 K already under normal pressure conditions [312]. The organic molecules are stacked here on top of each other (see Fig. 23). The general chemical formula is $(TMTSF)_2X$ where X denotes an electron acceptor such as $PF_6$, $ClO_4$, $AsF_6$ or $TaF_6$. In the normal state, the TMTSF compounds have a relatively large electric conductivity along the stacks, but only a small conductivity perpendicular to the stacks, thus forming nearly-one-dimensional (normal) conductors. The TMTSF compounds are type-II superconductors with highly anisotropic properties. For example, in $(TMTSF)_2ClO_4$ along the stacks the Ginzburg-Landau coherence length is about 80 nm, whereas along the two perpendicular directions of the crystal axes it is about 35 nm and 2 nm, respectively. The latter value is of the same order of magnitude as the lattice constant along the c axis. Hence, the compound represents a nearly two-dimensional superconductor [313].

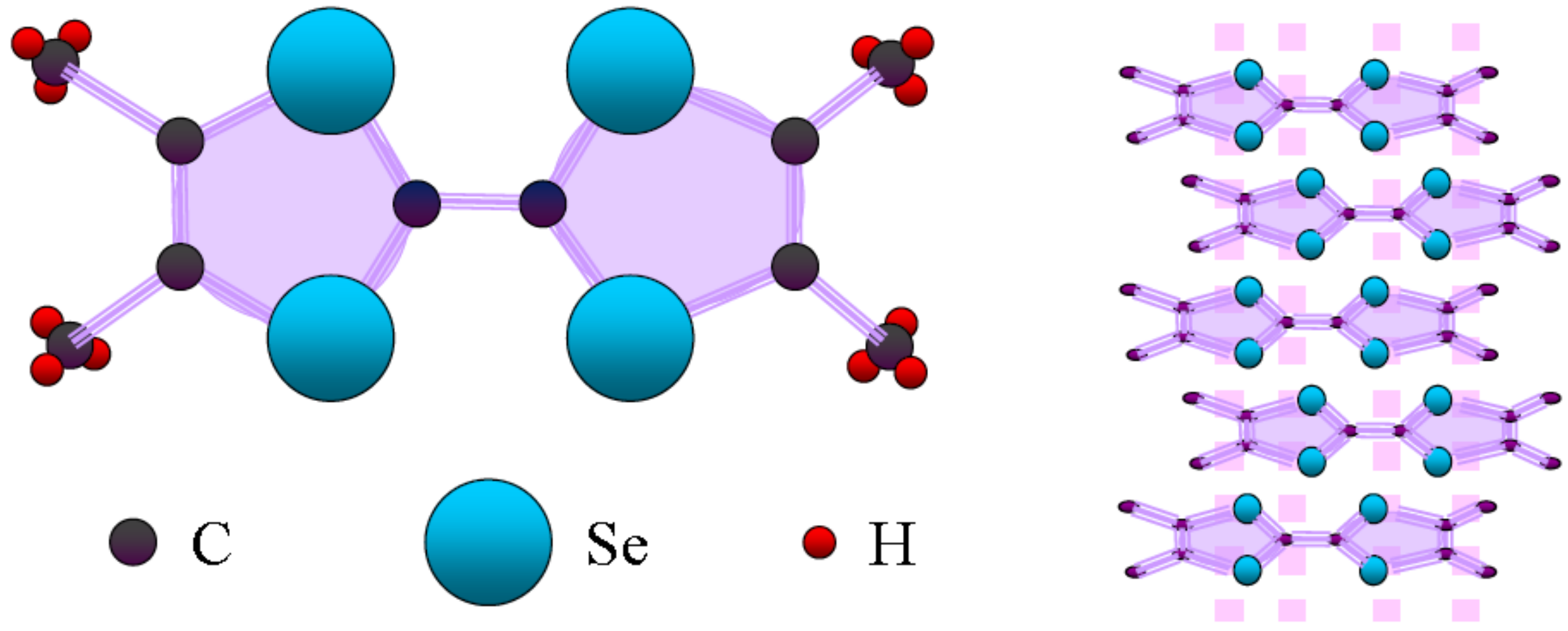

**Fig. 23** Structure of the organic molecule tetra-methyl-tetra-selenium-fulvalene (TMTSF) and stack arrangement of the molecules forming one-dimensional conduction channels

Another important class of organic superconductors, often exhibiting $T_c$ well above 1 K is based on the bis-ethylene-dithia-tetra-thiafulvalene molecule, abbreviated as "BEDT-TTF" or "ET". $(BEDT\text{-}TTF)_2Cu[N(CN)_2]Br$ becomes SC at 11.2 K [314], $(BEDT\text{-}TTF)_2Cu(NCS)_2$ at 10.4 K. The ET-compounds are also highly anisotropic. However, in contrast to the TMTSF-compounds, in the normal state they form two-dimensional layered structures with a large electric conductivity in two dimensions. Like the TMTSF based materials the ET-compounds are type-II superconductors as well, with very short out-of-plane coherence lengths. These compounds thus also represent SC layered structures making them in many respects similar to HTS. Like for HTS, the pairing mechanism of the organic superconductors is at present still unclear. At least some compounds appear to be d-wave superconductors. However, in the compound $(TMTSF)_2PF_6$ one may even deal with a spin-triplet superconductor [315].

The quasi-2D organic superconductors are prime candidates for exhibiting the long-sought FFLO/LOFF phases (for recent reviews see [316] and [228]. When the magnetic field is applied parallel to the conducting planes the orbital critical field is strongly enhanced and superconductivity is Pauli limited. First thermodynamic evidence for the formation of a FFLO/LOFF state was found in $\kappa$-$(BEDT\text{-}TTF)_2Cu(NCS)_2$ [317]. The angle-dependence of the formation of the FFLO/LOFF state was demonstrated in [318].

In 1994 superconductivity was found in *boron carbides* [319,320] (see the chapter 8), in 2004 in *diamond* with $T_c$ up to 4 K when doped with boron [3] and up to 11.4 K in thin films [321]. In silicon carbide bulk $T_c$ up to 1.4 K was achieved by doping with ~ 1 at % boron [322,323]. For silicon thin films a novel Gas Immersion Laser Doping technique [324] allows to increase the boron doping concentration homogenously up to 11 at %, far beyond the solubility limit of boron in silicon [325,326,327]. In spite of the modest $T_c$ of only up to 0.7 K the compatibility of this preparation technique with common silicon thin film technology is promising with respect to a link of superconducting and semiconducting electronics [327]. For yttrium carbide compounds $T_c$ as high as 18 K [328,329] was reported.

Superconductivity in a *graphite* intercalation compound was first observed in 1965 [330] on $KC_8$ which exhibits very low critical temperature $T_c$ = 0.14 K [331]. Later, several ternary graphite intercalation compounds revealed higher $T_c$ of 1.4 K for $KHgC_8$ [332] and 2.7 K for $KTl_{1.5}C_4$ [333]. Recently, the discovery of high critical temperatures in graphite intercalation compounds $YbC_6$ ($T_c$ =6.5K)[16], $CaC_6$ ($T_c$ =11.5 K) [334,335] and $Li_3Ca_2C_6$ ($T_c$ = 11.15 K) [336] has renewed the interest in this family of materials [337]. For *graphene,* the atomic monolayer modification of graphite, Li- or Ca-decoration is now been reported to induce superconductivity up to $T_c$ ~ 6 K [338,339]

*Fullerides* are metal-doped fullerenes. Compounds of the form $A_3C_{60}$ can become SC at surprizingly high temperatures [32,340,341]. By now, a number of SC fullerides based on the admixture of alkali atoms or of alkaline earth atoms are known. $Rb_3C_{60}$ has a value of $T_C$ of 29.5 K, the present record under pressure is held by $Cs_3C_{60}$ with $T_C$ = 40 K. The crystal structure of the fullerides is face centered cubic, with the alkali atoms occupying interstitial sites between the large $C_{60}$ molecules. Fullerides are BCS-like s-wave superconductors. Intramolecular $C_{60}$ phonons (see Fig. 24) seem to contribute the most important part of the pairing interactions [32]. However, for body-centered cubic A15-structured $Cs_3C_{60}$ (which is not SC at ambient pressure) an apparantly purely electronic transition to a SC state with $T_C$ up to 38 K can be induced by pressure, where the $T_c$ dependence on pressure can not be described by BCS theory in terms of the induced changes of anion packing density [342].

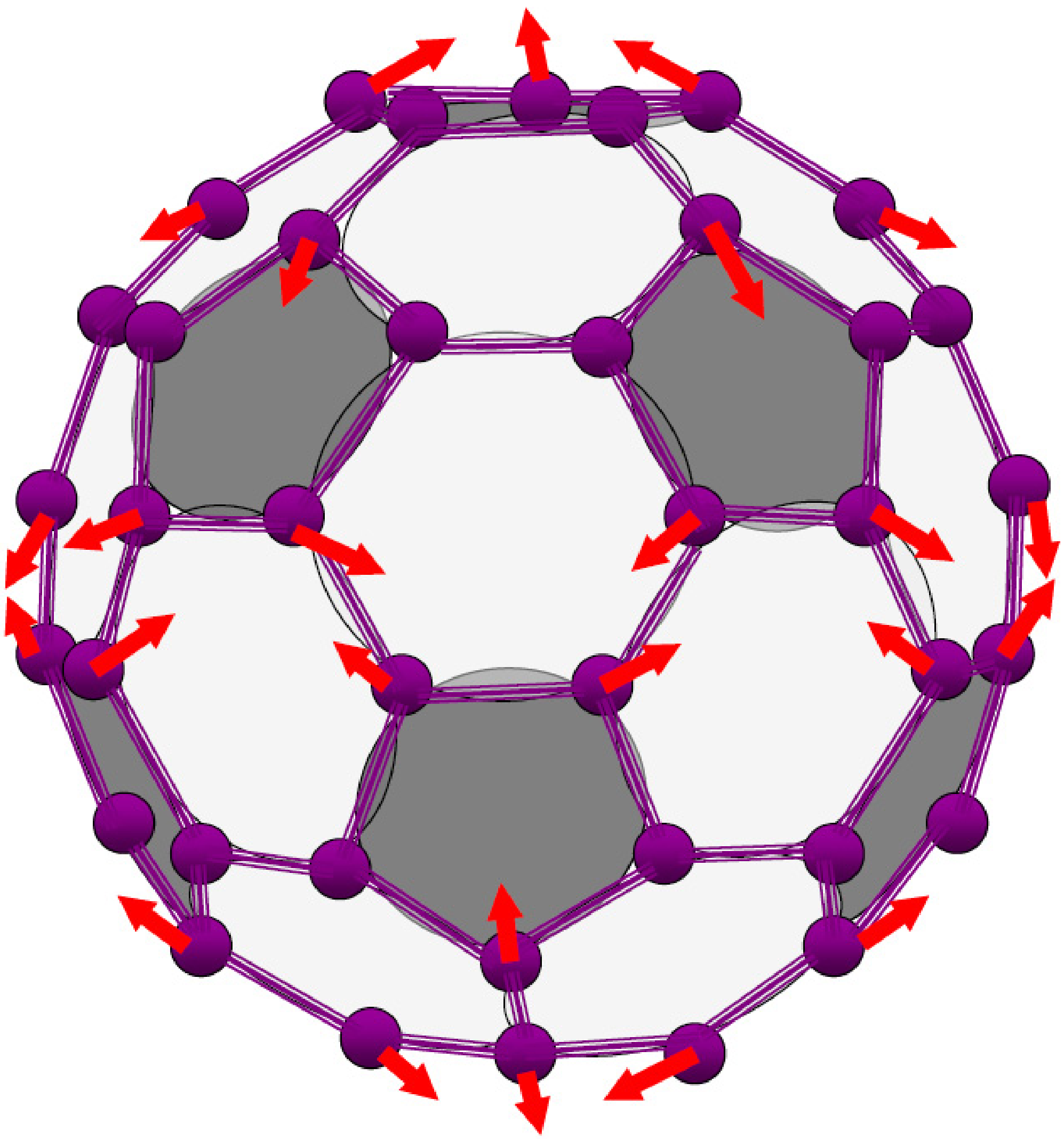

**Fig. 24** Structure of the $C_{60}$ molecule. The red arrows indicate one of the intramolecular phonon modes which are believed to be mainly responsible for the SC pairing [32].

Recent experiments on alkali-doped picene and dibenzopentacen, hydrocarbon molecules made up of an assembly of five and seven fused benzene rings, respectively, reported superconductivity up to $T_c$ of 18 K [343] and 33 K [344], respectively. A linear increase of $T_c$ with the number of constituent benzene rings is suspected. However, the fabrication process can not be controlled yet sufficiently to achieve single-phase preparation.

This holds true even more for *carbon natotubes* where long-standing speculations on superconductivity [345,346] have now been confirmed experimentally for the case of double-wall carbon natotubes (DWNT) with resistitively measured $T_c$ = 6.8 K [347].

## 8. BORIDES AND BOROCARBIDES

Rare-earth borocarbide superconductors have provided the first example of a homogeneous coexistence of superconductivity and ferromagnetism for all temperatures below $T_c$: The two antagonistic long-range orders are carried by different species of electrons that interact only weakly through contact exchange interaction leading to a small effect of the local moment molecular field on the SC conduction electrons. This allows a much better understanding of coexistence behavior as compared to the HF systems. Moreover, the nonmagnetic rare earth borocarbides have extremely large gap anisotropy ratios $\Delta_{max}/\Delta_{min} \geq 100$. Surely the standard electron-phonon mechanism has to be supplemented by something else, perhaps anisotropic Coulomb interactions to achieve this "quasi-unconventional" behavior in borocarbides.

The SC class of layered transition metal borocarbides $RNi_2B_2C$ (nonmagnetic R = Y, Lu, Sc; magnetic R = lanthanide elements in a $R^{3+}$ state; see Fig. 25) was discovered in 1994 [319,348,349,350]. The crystal structure consists of R C rock salt type planes separated by $Ni_2B_2$ layers built from $NiB_4$ tetrahedra and stacked along the c-axis. More general structures with more than one R C layer are possible [349]. The nonmagnetic borocarbides have relatively high $T_c$ values around 15 K. There is evidence that the SC mechanism is primarily of the electron-phonon type although this cannot explain the large anisotropy of the SC gap. At first sight the layered structure is similar to the HTS cuprates. However, unlike the copper oxide planes the $NiB_2$ planes show buckling. As a consequence, the electronic states at the Fermi level in the borocarbides do not have quasi-2-dimensional $d_{x^2-y^2}$ character and, therefore, have much weaker correlations excluding the possibility of AF spin-fluctuation mediated superconductivity.

The nonmagnetic borocarbides serve as a kind of reference point to separate the fascinating effects of AF and SC order parameter coupling in the magnetic $RNi_2B_2C$. However, the former have their own peculiarities, which are not yet completely understood. Foremost, despite their alleged electron-phonon nature, $LuNi_2B_2C$ and $YNi_2B_2C$ have strongly anisotropic gap functions and low energy quasiparticle states as is evident from specific heat and thermal conductivity. Furthermore, an anomalous upturn in $H_{c2}$ has been observed.

The magnetic $RNi_2B_2C$ are an excellent class of materials to study the effects of competition of magnetic order and superconductivity for the following reasons: The $T_c$ values are relatively high, and the $T_c / T_N$ ratio varies systematically across the R-series (see Fig. 26). Especially interesting are the cases of $RNi_2B_2C$ with R = Dy, Ho and Er where $T_c$ and $T_N$ (or $T_C$) are not too different, leading to strong competition of the magnetic and superconductivity order parameters. Furthermore, the SC condensate and magnetic moments are carried by different types of electrons, namely itinerant 3d-electrons for the $Ni_2B_2$ layers and localized $R^{3+}$ 4f-electrons for the R C layers, respectively. Finally, they are well separated and their coupling which is of the local exchange type can be treated in a controlled perturbative way, somewhat akin to the situation in the well known classes of Chevrel phase [351] and ternary compound [352] magnetic superconductors.

The AF molecular field establishes a periodic perturbation characterized by a length scale of the order of the Fermi wavelength. This implies that the spatial extent of the Cooper pairs extends over many periods of the alternating molecular field. The latter is therefore effectively averaged to zero and does not suppress superconductivity via an orbital effect. The system is invariant under the combined operation of time inversion followed by a translation with a lattice vector which allows to form Cooper pairs in a spin singlet state with vanishing (crystal) momentum in the AF lattice. This pair-state can be considered as a natural generalization of the pairing in time-reversed states encountered in usual non-magnetic superconductors.

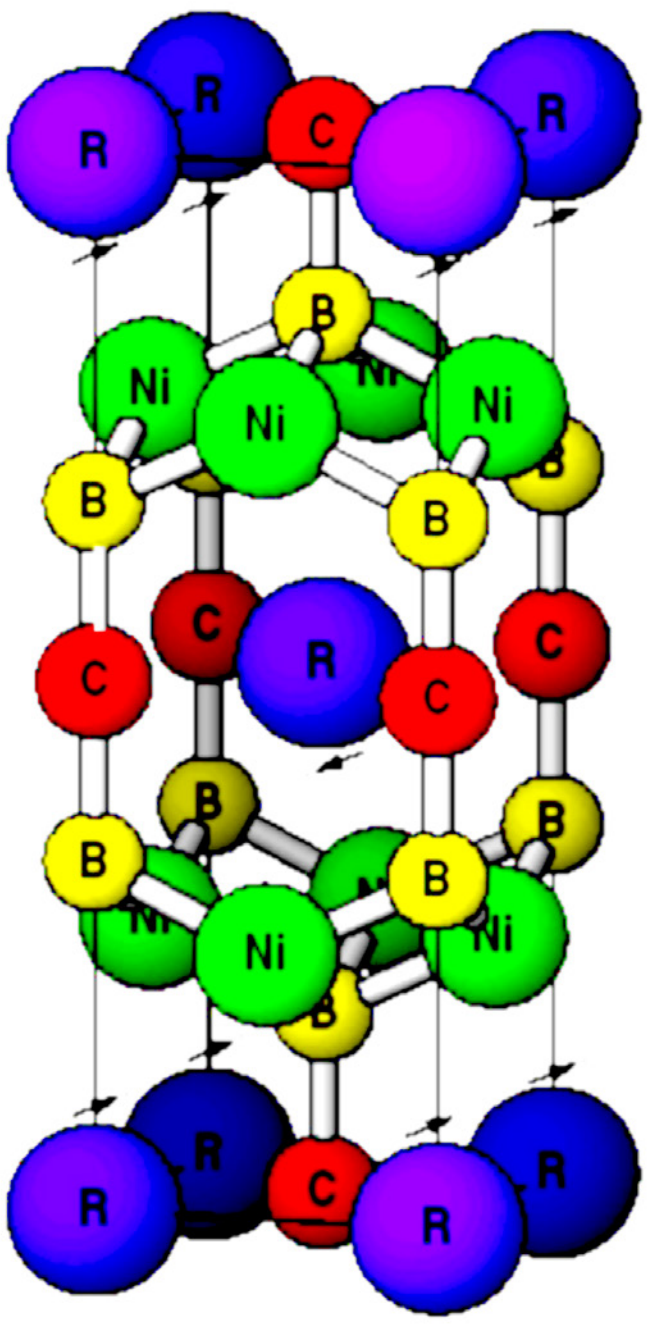

**Fig. 25** Tetragonal crystal structure of $RNi_2B_2C$ and low temperature AF magnetic structure with [110] easy axis as indicated by the arrows.

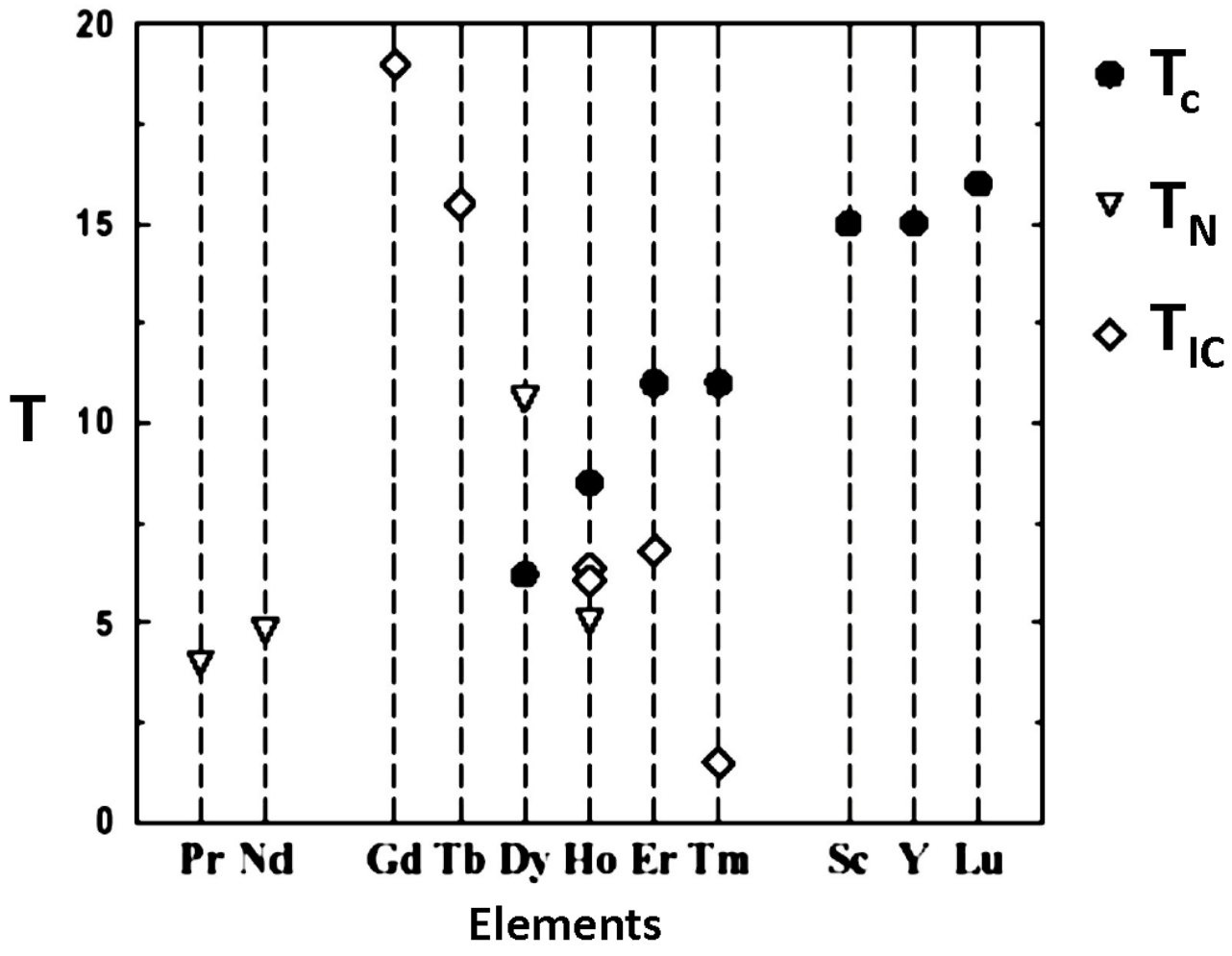

**Fig. 26** Magnetic ($T_{IC}$: incommensurate magnetic structure, $T_N$: simple AF structure) and SC ($T_c$) transition temperatures in Kelvin for the $RNi_2B_2C$ series.

The nonmagnetic $YNi_2B_2C$ and $LuNi_2B_2C$ compounds with comparatively high $T_c$ values of 16.5 K and 15.5 K serve as reference systems for the more difficult systems $RNi_2B_2C$ with both magnetic and SC phases. The electron-phonon nature of superconductivity in $YNi_2B_2C$ and $LuNi_2B_2C$ is inferred from a substantial s-wave character of the order parameter as witnessed by the appearance of a moderate Hebel-Slichter peak in the $^{13}C$ NMR relaxation rate [353].

On the other hand, the gap function is strongly anisotropic as can be seen both from temperature and field dependence of thermodynamic and transport quantities [349,354] indicating the presence of gap nodes [355]. More precisely, within experimental accuracy, there must be at least a gap anisotropy $\Delta_{max}/\Delta_{min} \geq 100$ [355]. For an electron-phonon superconductor this would be the largest anisotropy ever observed. This conjecture is also supported by the field dependence of the low temperature specific heat [356] and of the thermal conductivity along (001) [354]. Since in the latter case the heat current is perpendicular to the vortices this proves that quasiparticles must be present in the inter-vortex region. This is also required to explain the observation of dHvA oscillations far in the vortex phase. Experimental evidence therefore demands a nodal gap function for borocarbides. A s+g wave model [357] fulfills the requirements. In addition, it explains recent results on ultrasonic attenuation, which also confirmed the existence of gap nodes in the cubic plane [358].

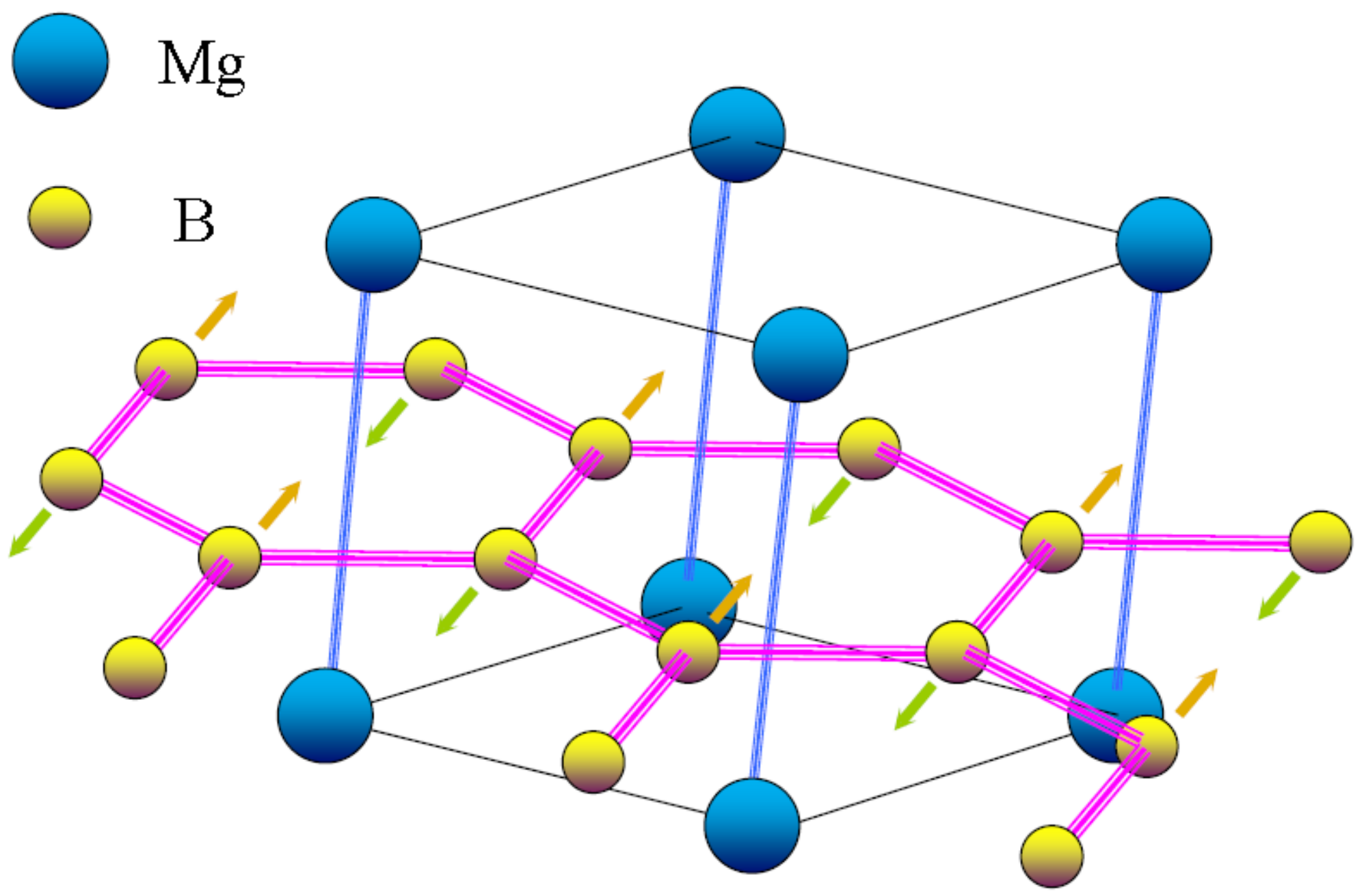

**Fig. 27** Hexagonal crystal structure of $MgB_2$. The arrows indicate the B-phonon mode which presumably introduces the strongest SC coupling.

The classical argument in favor of the electron-phonon mechanism is the observation of an isotope effect characterized by the isotope exponent for a specific atom with mass M as given by $\alpha_X = d\,[\ln T_c]/\,d\,[\ln M_X]$. However, this cannot be applied easily to complex layered superconductors, as is evident from the existence of an isotope effect in the probably nonphononic cuprate superconductors with nonoptimal doping. The boron isotope effect found in $YNi_2B_2C$ ($\alpha_B = 0.2$) and $LuNi_2B_2C$ ($\alpha_B = 0.11$) is much smaller than the BCS value $\alpha_X = 0.5$. A nonphononic origin stemming from the influence of boron on the charge density in the $B_2Ni_2$ layers has therefore been suggested [359]. The most direct support for phonon-mediated Cooper pairing comes from scanning tunneling spectroscopy ("STS")[360] which has shown the existence of a strong coupling signature in the tunneling DOS due to a soft optical phonon close to the Fermi surface nesting wave vector Q. The STS-derived electron-phonon coupling constant $\lambda = 0.5 - 0.8$ is compatible with the value $\lambda = 0.53$ obtained from resistivity data.

The discovery of superconductivity in $MgB_2$ (see Fig. 27) in early 2001 with $T_c \sim 40$ K, came as a huge surprise since this simple material was known since the early 1950s and had simply been missed in the systematic research for superconductivity [361,362]. The "materials preparation" of most of the research groups immediately after the discovery was to simply order $MgB_2$ powder from chemicals wholesale where it was commercially available in quantities of metric tons already for years!

Since no atomic d- or f-shells are involved in the conduction electron system of this binary compound of light elements Coulomb correlation do not play a role. The simple crystal structure consisting of graphite-like B-layers with intercalated Mg favors conduction along these layers and a respective superconductive and normal state anisotropy, but it does not introduce a reduction of the effective dimensionality, as in the case of organic superconductors due to the stacking of isolated aromatic rings. The coupling of the conduction electrons to a particular boron phonon mode (see Fig. 27) was identified right from the start as basic origin of superconductivity in $MgB_2$ [363, 364]. The observation of two energy gaps (at 1.8 and 6.8 meV [365, 366]) and the considerable superconductive anisotropy as large as 6–9 [367] challenged a more thorough theoretical investigation which explained these findings in terms of two-band superconductivity [36] on the basis of the large anharmonicity of the involved phonon mode and a refined treatment of its coupling with the different sheets of the electronic conduction band [365,368,369,370].

$MgB_2$ has many prerequisites for a future SC wire material for applications at operation temperatures up to ~ 20 K [67]:

- Low cost and abundant availability of the raw materials,
- high $T_c$ = 39 K,
- extreme type-II SC nature as a consequence of the coherence lengths $\xi$ ~ 4.4 nm and the penetration depth $\lambda$ ~ 132 nm in combination with the occurrence of suitable pinning centers which enable even at LHe temperature critical current densities that outperform those of commercial NbTi wire [371],
- transparency of grain boundaries to current,
- possibility of persistent current operation [372].
- The remarkably low normal state resistivity ρ(42 K) = 0.38 μΩ cm, comparable to that of copper wire and over a factor of 20 times lower than that of polycrystalline $Nb_3Sn$ [365], is extremely helpful with resepect to quench avoiding [367,373].

The most promising application perspective are coils for Magnetic Resonance Imaging (MRI) where a higher operation temperature would allow a much more flexible cryostat design and thus better accessibility of the imaging zone. $MgB_2$ is today the best candidate for the next entry in the list of technically mature and readily commercially available conductor materials.

## REFERENCES

[1] W. Buckel, R. Kleiner, *Supraleitung – Grundlagen und Anwendungen,* WILEY-VCH Verlag, Weinheim (2013);

W. Buckel, R. Kleiner, *Superconductivity - Fundamentals and Applications,* WILEY-VCH Verlag, Weinheim (2004)

[2] T. H. Geballe, Science **293 (**2001) 223

[3] E. A. Ekimov, V. A. Sidorov, E. D. Bauer, N. N. Mel'nik, N. J. Curro, J. D. Thompson, S. M. Stishov, Nature **428 (**2004) 542

[4] C. Buzea, T. Yamashita, Supercond. Sci. Techn. **14 (**2001) R115

[5] D. Goodstein, J. Goodstein, Phys. perspect. **2** (2000) 30

[6] R. Huebener, H. Lübbig, *Die Physikalisch-Technische Reichsanstalt*, Vieweg+Teubner-Verlag, Wiesbaden (2011)

[7] K. Fossheim, A. Sudbo, *Superconductivity - Physics and Applications,* John Wiley & Sons, Ltd. (2004)

[8] I. I. Mazin, Nature **464** (2010) 183

[9] J. Bardeen, L. N. Cooper, J. R. Schrieffer, Phys. Rev **108** (1957) 1175

[10] *BCS: 50 years*, ed. Leon N. Cooper and Dmitri Feldman, World Scientific Press, Singapore (2011)

[11] F. Steglich, J. Aarts, C. D. Bredl, W. Lieke, D. Meschede, W. Franz, H. Schäfer, Phys. Rev. Lett. **43** (1979) 1892

[12] C. Pfleiderer, Rev. Mod. Phys. **81** (2009) 1551

[13] W. A. Little, Phys. Rev. A **134** (1964) 1416

[14] D. Jérome, A. Mazaud, M. Ribault, K. Bechgaard, J. Phys. (Paris), Lett. **41** (1980) L95;

D. Jerome, A. Mazaud, M. Ribault,. K. Bechgaard, C. R. Hebd. Seances Acad. Sci. Ser. **B 290** (1980)

[15] G. Saito, H. Yamochi, T. Nakamura, T. Komatsu, M. Nakashima, H. Mori, K. Oshima, Physica B **169** (1991) 372

[16] J. Wosnitza, Crystals **2** (2012) 248

[17] N. D. Mermin, H. Wagner, Phys. Rev. Lett. **17** (1966) 1133 & 1307

[18] J. G. Bednorz, K. A. Müller, Z. Phys. B **64** (1986) 189

[19] C. W. Chu in [10]

[20] J. R. Schrieffer, D. J. Scalapino, J. W. Wilkins, Phys. Rev. Lett. **10** (1963) 336

[21] G. M. Eliashberg, Sov. Phys. JETP **11** (1960) 696

[22] S. N. Putilin, E. V. Antipov, A. M. Abakumov, M. G. Rozova, K. A. Lokshin, D. A. Pavlov, A. M. Balagurov, D. V. Sheptyakov, M. Marezio, Physica C **338** (2001) 52

[23] J. R. Gavaler, Appl. Phys. Lett. **23** (1973) 480

[24] W. L. McMillan, J. M. Rowell, Phys. Rev. Lett. **14** (1965) 108

[25] R. Zeyer, G. Zwicknagl, Z. Phys. B **78** (1990) 175

[26] B. Keimer, S. A. Kivelson, M. R. Norman, S. Uchida, J. Zaanen, Nature **518** (2015) 179

[27] Y. Maeno, T. M. Rice, M. Sigrist, Physics Today, January 2001, p. 42

[28] K. Takada, H. Sakurai, E. Takayama-Muromachi, F. Izumi, R. A. Dilanian, T. Sasaki, Nature **422** (2003) 53

[29] P. A. Lee, N. Nagaosa, X.-G. Wen, Rev. Mod. Phys. **78** (2006) 17

[30] T. A. Maier, D. Poilblanc, D. J. Scalapino, Phys. Rev. Lett. **100** (2008) 237001

[31] K. Tanigaki, T. W. Ebbesen, S. Saito, J. Mizuki, J. S. Tsai, Y. Kubo, S. Kuroshima, Nature **352** (1991) 222

[32] O. Gunnarsson, Rev. Mod. Phys. **69** (1997) 575

[33] P. C. Canfield, P. L. Gammel, D. J. Bishop, Physics Today, October 1998, p. 40

[34] J. Nagamatsu, N. Nakagawa, T. Muranaka, Y. Zenitani, J. Akimitsu, Nature **410** (2001) 63

[35] H. Suhl, B.T. Matthias, and L.R. Walker, Phys. Rev. Lett. **3** (1959) 552;
V. Moskalenko, Fiz. Met. Metalloved. **8** (1959) 503

[36] G. Binnig, A. Baratoff, H. E. Hoenig, J. G. Bednorz, Phys. Rev. Lett. **45** (1980) 1352

[37] J. Geerk, R. Schneider, G. Linker, A. G. Zaitsev, R. Heid, K.-P. Bohnen, H. v. Löhneysen, Phys. Rev. Lett. 94 (2005) 227005

[38] G. R. Stewart, Rev. Mod. Phys. **83** (2011) 1589

[39] A. V. Chubukov, P.J. Hirschfeld, Physics Today **68,** June 2015, 46

[40] P.J. Hirschfeld, Comptes Rendus Physique **17** (2016) 197

[41] Y. Kamihara, H. Hiramatsu, M. Hirano, R. Kawamura, H. Yanagi, T. Kamiya, H. Hosono, J. Am. Chem. Soc. **128** (2006) 10012

[42] Y. Kamihara, T. Watanabe, M. Hirano, H. Hosono, J. Am. Chem. Soc. **130** (2008) 3296

[43] Z.-A. Ren, G.-C. Che, X.-L. Dong, J. Yang, W. Lu, W. Yi, X.-L. Shen, Z.-Cai Li, L.-L. Sun, F. Zhou, Z.-X. Zhao, Europhys. Lett. **83** (2008) 17002

[44] F.-C. Hsu, J.-Y. Luo, K.-W. Yeh, T.-K. Chen, T.-W. Huang, P. M. Wu‡, Y.-C. Lee, Y.-L. Huang, Y.-Y. Chu, D.-C. Yan, M.-K. Wu, Proc. Natl. Acad. Sci U.S.A. **105** (2008) 14262

[45] J. F. Ge, Z. L. Liu, C. Liu, C. L. Gao, D. Qian, Q. K. Xue, Y. Liu, J.-F. Jia, Nature Mater. **14** (2015) 285

[46] J. Shiogai, Y. Ito, T. Mitsuhashi, T. Nojima, A. Tsukazaki, Nature Phys. **12** (2016) 42

[47] N. Reyren, S. Thiel, A.D. Caviglia, L.F. Kourkoutis, G. Hammerl, C. Richter, C.W. Schneider, T. Kopp, A.-S. Rüetschi, D. Jaccard, M. Gabay, D.A. Muller, J.-M. Triscone, J. Mannhart, Science **317** (2007) 1196

[48] A.D. Caviglia, S. Gariglio, N. Reyren, D. Jaccard, T. Schneider, M. Gabay, S. Thiel, G. Hammerl, J. Mannhart, J.-M. Triscone, Nature **456** (2008) 624

[49] A. P. Drozdov, M. I. Eremets, I. A. Troyan, V. Ksenofontov, S. I. Shylin, Nature **525** (2015) 73

[50] D. Duan, Y. Liu, F. Tian, D. Li, X. Huang, Z. Zhao, H. Yu, B. Liu, W. Tian, T. Cui, Sci. Rep. **4** (2014) 6968

[51] N. Bernstein, C. S. Hellberg, M. D. Johannes, I. I. Mazin, M. J. Mehl, Phys. Rev. B **91** (2015) 060511;

I. Errea, M. Calandra, C. J. Pickard, J. Nelson, R. J. Needs, Y. Li, H. Liu, Y. Zhang, Y. Ma, F. Mauri, Phys. Rev. Lett. **114** (2015) 157004;

J. A. Flores-Livas, A. Sanna, E. K. U. Gross, arXiv:1501.06336;

R. Akashi, M. Kawamura, S. Tsuneyuki, Y. Nomura, R. Arita, Phys. Rev. B **91** (2015) 224513

[52] A. Drozdov, M. I. Eremets, I. A. Troyan, arXiv:1508.06224;

J. A. Flores-Livas, M. Amsler, C. Heil, A. Sanna, L. Boeri, G. Profeta, C. Wolverton, S. Goedecker, E. K. U. Gross, Phys. Rev. B **93** (2016) 020508;

A. Shamp, T. Terpstra, T. Bi, Z. Falls, P. Avery, and E. Zurek, arXiv:1509.05455

[53] N. W. Ashcroft, Phys. Rev. Lett. **92** (2004) 187002

[54] R. Mankowsky, A. Subedi, M. Först, S. O. Mariager, M. Chollet, H. T. Lemke, J. S. Robinson, J. M. Glownia, M. P. Minitti, A. Frano, M. Fechner, N. A. Spaldin, T. Loew, B. Keimer, A. Georges, A. Cavalleri, Nature **516** (2014) 71;

W. Hu, S. Kaiser, D. Nicoletti, C. R. Hunt, I. Gierz, M. C. Hoffmann, M. Le Tacon, T. Loew, B. Keimer, A. Cavalleri, Nature Mater. **13** (2014) 705;

S. Kaiser, C. R. Hunt, D. Nicoletti, W. Hu, I. Gierz, H. Y. Liu, M. Le Tacon, T. Loew, D. Haug, B. Keimer, A. Cavalleri, Phys. Rev. B **89**, 184516 (2014)

[55] M. Först, A. Frano, S. Kaiser, R. Mankowsky, C. R. Hunt, J. J. Turner, G. L. Dakovski, M. P. Minitti, J. Robinson, T. Loew, M. Le Tacon, B. Keimer, J. P. Hill, A. Cavalleri, S. S. Dhesi, Phys. Rev. B **90** (2014) 184514

[56] D. Fausti, R. I. Tobey, N. Dean, S. Kaiser, A. Dienst, M. C. Hoffmann, S. Pyon, T. Takayama, H. Takagi, A. Cavalleri, Science **331** (2011) 189;

C. R. Hunt, D. Nicoletti, S. Kaiser, T. Takayama, H. Takagi, A. Cavalleri, Phys. Rev. B **91**, 020505 (2015)

[57] M. Först, R. I. Tobey, H. Bromberger, S. B. Wilkins, V. Khanna, A. D. Caviglia, Y.-D. Chuang, W. S. Lee, W. F. Schlotter, J. J. Turner, M. P. Minitti, O. Krupin, Z. J. Xu, J. S. Wen, G. D. Gu, S. S. Dhesi, A. Cavalleri, J. P. Hill, Phys. Rev. Lett. **112**, 157002 (2014)

[58] M. Mitrano, A. Cantaluppi, D. Nicoletti, S. Kaiser, A. Perucchi, S. Lupi, P. Di Pietro,D. Pontiroli, M. Riccò, S. R. Clark, D. Jaksch, A. Cavalleri, doi:10.1038/nature16522

[59] L. J. Li, E. C. T. O'Farrell, K. P. Loh, G. Eda, B. Özyilmaz, A. H. Castro Neto, Nature **529** (2016) 185

[60] T.G. Berlincourt, Cryogenics **27** (1987) 283

[61] R. Flükiger, C. Senatore, M. Cesaretti, F. Buta, D. Uglietti, B. Seeber, Supercond. Sci. Technol. **21** (2008) 054015

[62] http://fusionforenergy.europa.eu/mediacorner/newsview.aspx?content=819

[63] D. W. Hazelton, V. Selvamanickam, Proc. IEEE **97** (2007) 1831

[64] C. Senatore, M. Alessandrini, A. Lucarelli, R. Tediosi, D. Uglietti, Y. Iwasa, Supercond. Sci. Technol. **27** (2014) 1

[65] U. Trociewitz, M. Canassy, M. Hannion, D. Hilton, J. Jaroszynski, P. Noyes, Y. Viouchkov, H. Weijers, D. Larbalestier, Appl. Phys. Lett. **99** (2011) 202506

[66] S. Yoon, J. Kim, K. Cheon, H. Lee, S.Hahn, S. H. Moon, to be published in Supercond. Sci. Technol.

[67] Y. Hishinuma, A. Kikuchi, T. Takeuchi, S. Yamada, J. Phys.: Conf. Ser. 234 (2010) 022014

[68] C. James, M. Krishnan, B. Bures, T. Tajima, L. Civale, R. Edwards, J. Spradlin, H. Inoue, IEE Trans. Appl. Supercond. **23** (2013) 3500205

[69] S. K. Tolpygo, D. Amparo, Supercond. Sci. Technol. **23** (2010) 034024

[70] R. Hott, *High-Temperature Superconductivity 1*, Springer Verlag, Berlin (2005), p.9

[71] M. H. Devoret and R. J. Schoelkopf, Science **339** (2013) 1169;
D. Córcoles, E. Magesan, S. J. Srinivasan, A. W. Cross, M. Steffen, J. M. Gambetta, J. M. Chow, Nature Comm. **6** (2015) 6979

[72] M. Z. Hasan, C. L. Kane, Rev. Mod. Phys. **82** (2010) 3045;
X.-L. Qi, S.-C. Zhang, Rev. Mod. Phys. **83** (2011) 1057

[73] J. Alicea, Rep. Prog. Phys. **75** (2012) 076501

[74] C. Nayak, S. H. Simon, A. Stern, M.Freedman, S. Das Sarma, Rev. Mod. Phys. **80** (2008) 1083;
N. Read, Phys. Today **65** (2012) 38

[75] V. S. Pribiag, A. J. A. Beukman, F. Qu, M. C. Cassidy, C. Charpentier, W. Wegscheider, L. P. Kouwenhoven, Nature Nanotech. **10** (2015) 593

[76] D. R. Harshman, A. P. Mills, Jr., Phys. Rev. B **45** (1992) 10684

[77] C. W. Chu, J. Supercond. **12** (1999) 85

[78] C. W. Chu, IEEE Trans. Appl. Supercond. **7** (1997) 80

[79] H. Yamauchi, M. Karppinen, S. Tanaka, Physica C **263** (1996) 146

[80] H. Yamauchi, M. Karppinen, Supercond. Sci. Technol. **13** (2000) R33

[81] P. Bordet, S. LeFloch, C. Chaillout, F. Duc, M. F. Gorius, M. Perroux, J. J. Capponi, P. Toulemonde, J. L. Tholence, Physica C **276** (1997) 237

[82] N. L. Wu, Z. L. Du, Y. Y. Xue, I. Rusakova, D. K. Ross, L. Gao, Y. Cao, Y. Y. Sun, C. W. Chu, M. Hervieu, B. Raveau, Physica C **315** (1999) 227

[83] M. K. Wu, J. R. Ashburn, C. J. Torng, P. H. Hor, R. L. Meng, L. Gao, Z. J. Huang, Y. Q. Wang, C. W. Chu, Phys. Rev. Lett. **58** (1987) 908

[84] J. G. Lin, C. Y. Huang, Y. Y. Xue, C. W. Chu, X. W. Cao, J. C. Ho, Phys. Rev. B **51** (1995) 12900

[85] J. W. Chu, H. H. Feng, Y. Y. Sun, K. Matsuishi, Q. Xiong, C. W. Chu, *HTS Materials, Bulk Processing and Bulk Applications* - Proceedings of the 1992 TCSUH Workshop (Ed. C. W. Chu, W. K. Chu, P. H. Hor and K. Salama), World Scientific, Singapore (1992), p. 53; TCSUH preprint 92:043

[86] G. V. M. Williams, J. L. Tallon, Physica C **258** (1996) 41

[87] M. Guillaume, P. Allenspach, W. Henggeler, J. Mesot, B. Roessli, U. Staub, P. Fischer, A. Furrer, V. Trounov, J. Phys.: Condens. Matter **6** (1994) 7963

[88] M. Guillaume, P. Allenspach, J. Mesot, B. Roessli, U. Staub, P. Fischer, A. Furrer, Z. Phys. B **90** (1993) 13

[89] R. K. Williams; D. M. Kroeger, P. M. Martin, J. R. Mayotte, E. D. Specht, J. Brynestad, J. Appl. Phys. **76** (1994) 3673;

W. Zhang, K. Osamura, Physica C **190** (1992) 396;

G. F. Voronin, S. A. Degterov, Physica C **176** (1991) 387;

J. Karpinski, S. Rusiecki, B. Bucher, E. Kaldis, E. Jilek, Physica C **161** (1989) 618;

J. Karpinski, S. Rusiecki, B. Bucher, E. Kaldis, E. Jilek, Physica C **160** (1989) 449;

J. Karpinski, E. Kaldis, E. Jilek, S. Rusiecki, B. Bucher, Nature **336** (1988) 660

[90] P. Zoller, J. Glaser, A. Ehmann, C. Schultz, W. Wischert, S. Kemmler-Sack, T. Nissel, R. P. Huebener, Z. Phys. B **96** (1995) 505

[91] D. L. Feng, A. Damascelli, K. M. Shen, N. Motoyama, D. H. Lu, H. Eisaki, K. Shimizu, J.-I. Shimoyama, K. Kishio, N. Kaneko, M. Greven, G. D. Gu, X. J. Zhou, C. Kim, F. Ronning, N. P. Armitage, Z.-X. Shen, Phys. Rev. Lett. **88** (2002) 10700

[92] H. Eisaki, N. Kaneko, D. L. Feng, A. Damascelli, P. K. Mang, K. M. Shen, Z.-X. Shen, M. Greven, Phys. Rev. B **69** (2004) 064512

[93] S. Lösch, H. Budin, O. Eibl, M. Hartmann, T. Rentschler, M. Rygula, S. Kemmler-Sack, R. P. Huebener, Physica C **177** (1991) 271

[94] H. Yamauchi, T. Tamura, X.-J. Wu, S. Adachi, S. Tanaka, Jpn. J. Appl. Phys. **34** (1995) L349

[95] T. Tamura, S. Adachi, X.-J. Wu, T. Tatsuki, K. Tanabe, Physica C **277** (1997) 1

[96] A. Iyo, Y. Tanaka, Y. Ishiura, M. Tokumoto, K. Tokiwa, T. Watanabe, H. Ihara, Supercond. Sci. Technol. **14** (2001) 504

[97] A. Iyo, Y. Aizawa, Y. Tanaka, M. Tokumoto, K. Tokiwa, T. Watanabe, H. Ihara, Physica C **357-360** (2001) 324

[98] D. Tristan Jover, R. J. Wijngaarden, R. Griessen, E. M. Haines, J. L. Tallon, R. S. Liu, Phys. Rev. B **54** (1996) 10175

[99] Z. Y. Chen, Z. Z. Sheng, Y. Q. Tang,Y. F. Li, L. M. Wang, D. O. Pederson, Supercond. Sci. Technol. **6** (1993) 261

[100] E. V. Antipov, A. M. Abakumov, S. N. Putilin, Supercond. Sci. Technol. **15** (2002) R31

[101] C. Acha, S. M. Loureiro, C. Chaillout, J. L. Tholence, J. J. Capponi, M. Marezio, M. Nunez-Regueiro, Solid State Comm. **102** (1997) 1

[102] T. Tatsuki, A. Tokiwa-Yamamoto, A. Fukuoka, T. Tamura, X.-J. Wu, Y. Moriwaki, R. Usami, S. Adachi, K. Tanabe, S. Tanaka, Jpn. J. Appl. Phys. **35** (1996) L205

[103] E. Stangl, S. Proyer, M. Borz, B. Hellebrand, D. Bäuerle, Physica C **256** (1996) 245

[104] A. Erb, E. Walker, R. Flükiger, Physica C **245** (1995) 245;

A. Erb, E. Walker, J.-Y. Genoud, R. Flükiger, Physica C **282-287** (1997) 89

[105] H. Ihara, Physica C **364-365** (2001) 289

[106] P. W. Klamut, B. Dabrowski, S. M. Mini, M. Maxwell, S. Kolesnik, J. Mais, A. Shengelaya, R. Khasanov I. Savic, H. Keller, T. Graber, J. Gebhardt, P. J. Viccaro, Y. Xiao, Physica C **364-365** (2001) 313;

B. Lorenz, R. L. Meng, J. Cmaidalka, Y. S. Wang, J. Lenzi, Y. Y. Xue, C. W. Chu, Physica C **363** (2001) 251

[107] T. Kawashima, Y. Matsui, E. Takayama-Muromachi, Physica C **254** (1995) 131

[108] I. Bozovic, G. Logvenov, I. Belca, B. Narimbetov, I. Sveklo, Phys. Rev. Lett. **89** (2002) 107001

[109] S. Karimoto, H. Yamamoto, T. Greibe, M. Naito, Jpn. J. Appl. Phys. **40** (2001) L127

[110] M. Naito, M. Hepp, Jpn. J. Appl. Phys. **39** (2000) L485

[111] L. Alff, S. Meyer, S. Kleefisch, U. Schoop, A. Marx, H. Sato, M. Naito, R. Gross, Phys. Rev. Lett. **83** (1999) 2644

[112] C. Q. Jin, Y. S. Yao, S. C. Liu, W. L. Zhou, W. K. Wang, Appl. Phys. Lett. **62** (1993) 3037

[113] A. Iyo, Y. Tanaka, M. Tokumoto, H. Ihara, Physica C **366** (2001) 43

[114] C. U. Jung, J. Y. Kim, S. M. Lee, M.-S. Kim, Y. Yao, S. Y. Lee, S.-I. Lee, D. H. Ha Physica C **364-365** (2001) 225

[115] M. Al-Mamouri, P. P. Edwards, C. Greaves, M. Slaski, Nature **369** (1994) 382

[116] T. Kawashima, Y. Matsui, E. Takayama-Muromachi, Physica C **257** (1996) 313

[117] Z. Hiroi, N. Kobayashi, M. Takano, Nature **371** (1994) 139

[118] L. Gao, Y. Y. Xue, F. Chen, Q. Xiong, R. L. Meng, D. Ramirez, C. W. Chu, J. H. Eggert, H. K. Mao, Phys. Rev. B **50** (1994) 4260

[119] P. Majewski, J. Mater. Res. **15** (2000) 854

[120] M. P. Siegal, E. L. Venturini, B. Morosin, T. L. Aselage, J. Mater. Res. **12** (1997) 2825

[121] J. J. Capponi, J. L. Tholence, C. Chaillout, M. Marezio, P. Bordet, J. Chenavas, S. M. Loureiro, E. V. Antipov, E. Kopnine, M. F. Gorius, M. Nunez-Regueiro, B. Souletie, P. Radaelli, F. Gerhards, Physica C **235-240** (1994) 146

[122] T. Ito, H. Suematsu, M. Karppinen, H. Yamauchi, Physica C **308** (1998) 198

[123] J. Leggett, Phys. Rev. Lett. **83** (1999) 392

[124] K. Kuroda, K. Abe, S. Segawa, J.-G. Wen, H. Unoki, N. Koshizuka, Physica C **275** (1997) 311

[125] J. P. Attfield, A. L. Kharlanov, J. A. McAllister, Nature **394** (1998) 157

[126] M. W. Pieper, F. Wiekhorst, T. Wolf, Phys. Rev. B **62** (2000) 1392;

A. J. Markwardsen, A. T. Boothroyd, B. Buck, G. J. McIntyre, Th. Wolf, J. Magn. Magn. Mater. **177-181** (1998) 502

[127] K. Oka, Z. Zou, J. Ye, Physica C **300** (1998) 200;

Z. Zou, J. Ye, K. Oka, Y. Nishihara, Phys. Rev. Lett. **80** (1998) 1074

[128] J. Ye, Z. Zou, A. Matsushita, K. Oka, Y. Nishihara, T. Matsumoto, Phys. Rev. B **58** (1998) 619

[129] Y. S. Lee, F. C. Chou, A. Tewary, M. A. Kastner, S. H. Lee, R. J. Birgeneau, Phys. Rev. B **69** (2004) 020502

[130] P. K. Mang, S. Larochelle, A. Mehta, O. P. Vajk, A. S. Erickson, L. Lu, W. J. L. Buyers, A. F. Marshall, K. Prokes, M. Greven, Phys. Rev. **B** 70 (2004) 094507

[131] O. Chmaissem, J. D. Jorgensen, S. Short, A. Knizhnik, Y. Eckstein, H. Shaked, Nature **397** (1999) 45

[132] F. Izumi, J. D. Jorgensen, Y. Shimakawa, Y. Kubo, T. Manako, S. Pei, T. Matsumoto, R. L. Hitterman, Y. Kanke, Physica C **193** (1992) 426

[133] Z. F. Ren, J. H. Wang, D. J. Miller, Appl. Phys. Lett. **69** (1996) 1798

[134] A. Yamato, W. Z. Hu, F. Izumi, S. Tajima, Physica C **351** (2001) 329

[135] J. Orenstein, A. J. Millis, Science **288** (2000) 468

[136] Y. Onose, Y. Taguchi, K. Ishizaka, Y. Tokura, Phys. Rev. Lett. **87** (2001) 217001

[137] J. L. Tallon, J. W. Loram, Physica C **349** (2001) 53

[138] C. M. Varma, Phys. Rev. B **55** (1997) 14554

[139] L. Alff, Y. Krockenberger, B. Welter, M. Schonecke, R. Gross, D. Manske, M. Naito, Nature **422** (2003) 698

[140] Y. Ando, A. N. Lavrov, S. Komiya, K. Segawa, X. F. Sun, Phys. Rev. Lett. **87** (2001) 017001

[141] P. W. Anderson, Science **235** (1987) 1196

[142] T. H. Geballe, B. Y. Moyzhes, Physica C **341-348** (2000) 1821

[143] D. H. Lu, D. L. Feng, N. P. Armitage, K. M. Shen, A. Damascelli, C. Kim, F. Ronning, Z.-X. Shen, D. A. Bonn, R. Liang, W. N. Hardy, A. I. Rykov, S. Tajima, Phys. Rev. Lett. **86** (2001) 4370

[144] E. H. Brandt, Phys. Rev. B **64** (2001) 024505

[145] G. Grasso, R. Flükiger, Supercond. Sci. Technol. **10** (1997) 223

[146] J. Sato, K. Ohata, M. Okada, K. Tanaka, H. Kitaguchi, H. Kumakura, T. Kiyoshi, H. Wada, K. Togano, Physica C **357-360** (2001) 1111;

M. Okada, Supercond. Sci. Technol. **13** (2000) 29

[147] A. K. Saxena, *High-Temperature Superconductors,* Springer-Verlag Berlin-Heidelberg (2010, 2012)

[148] N. P. Plakida, *High-Temperature Superconductivity,* Springer-Verlag Berlin-Heidelberg (1995)

[149] D. Larbalestier, A. Gurevich, D. M. Feldmann, A. Polyanskii, Nature **414** (2001) 368

[150] H. Hilgenkamp, J. Mannhart, Rev. Mod. Phys. **74** (2002) 485

R. Gross, L. Alff, A. Beck, O. M. Froehlich, D. Koelle, A. Marx, IEEE Trans. Appl. Supercond. **7** (1997) 2929

[151] R. Kleiner, F. Steinmeyer, G. Kunkel, P. Müller, Phys. Rev. Lett. **68** (1992) 2394

[152] U. Welp, K. Kadowaki, R. Kleiner, Nature Photonics **7** (2013) 702

[153] E. H. Brandt, Rep. Prog. Phys. **58** (1995) 1465; arXiv:cond-mat/9506003

[154] K. Tanaka, A. Iyo, Y. Tanaka, K. Tokiwa, M. Tokumoto, M. Ariyama, T. Tsukamoto, T. Watanabe, H. Ihara, Physica B **284-288** (2000) 1081;
T. Watanabe, S. Miyashita, N. Ichioka, K. Tokiwa, K. Tanaka, A. Iyo, Y. Tanaka, H. Ihara,Physica B **284-288** (2000) 1075

[155] C. C. Tsuei, J. R. Kirtley, Rev. Mod. Phys. **72** (2000) 969;
C. C. Tsuei, J. R. Kirtley, Physica C **367** (2002) 1

[156] H. Hilgenkamp, J. Mannhart, Rev. Mod. Phys. **74** (2002) 485

[157] H. Hilgenkamp, J. Mannhart, B. Mayer, Phys. Rev. B **53** (1996) 14586;
P. A. Nilsson, Z. G. Ivanov, H. K. Olsson, D. Winkler, T. Claeson, E. A. Stepantsov, A. Ya. Tzalenchuk, J. Appl. Phys. **75** (1994) 7972

[158] J. Betouras, R. Joynt, Physica C **250** (1995) 256

[159] J. Alarco, E. Olsson, Phys. Rev. B **52** (1995) 13625

[160] D. Agassi, D. K. Christen, S. J. Pennycook, Appl. Phys. Lett. **81** (2002) 2803

[161] H. Hilgenkamp, J. Mannhart, Appl. Phys. Lett. **73** (1998) 265

[162] S. Graser, P. J. Hirschfeld, T. Kopp, R. Gutser, B. M. Andersen, J. Mannhart, Nature Phys. **6** (2010) 609

[163] K. Heine, J. Tenbrink, M. Thöner, Appl. Phys. Lett. 55 (1989) 2441

[164] S. Honjo, T. Mimura, Y. Kitoh, Y. Noguchi, T. Masuda, H. Yumura, M. Watanabe, M. Ikeuchi, H. Yaguchi, IEEE Trans. Appl. Supercond. **21** (2011) 967

[165] Y. Yamada, Ma. Mogi, K. Sato, SEI Technical Review **65** (2007) 51

[166] A. W. Sleight, J. L. Gillson, P. E. Bierstedt, Solid State Commun. **17** (1975) 27

[167] M. Merz, N. Nücker, S. Schuppler, D. Arena, J. Dvorak, Y. U. Idzerda, S. N. Ustinovich, A. G. Soldatov, S. V. Shiryaev, S. N. Barilo, Europhys. Lett. **72** (2005) 275

[168] L. F. Mattheiss, E. M. Gyorgy, D. W. Johnson Jr., Phys. Rev. B **37** (1988) 3745

[169] E. S. Hellman, E. H. Hartford, Jr., Phys. Rev. B **52** (1995) 6822

[170] R. J. Cava, B. Batlogg, J. J. Krajewski, R. Farrow, L. W. Rupp, Jr., A. E. White, K. Short, W. F. Peck, T. Kometani, Nature **332** (1988) 814

[171] S. Pei, J. D. Jorgensen, B. Dabrowski, D. G. Hinks, D. R. Richards, A. W. Mitchell, J. M. Newsam, S. K. Sinha, D. Vaknin, A. J. Jacobson, Phys. Rev. B **41** (1990) 4126

[172] L. A. Klinkova, M. Uchida, Y. Matsui, V. I. Nikolaichik, N. V. Barkovskii, Phys. Rev. B **67** (2003) 140501

[173] C. Bernhard, J. L. Tallon, Ch. Niedermayer, Th. Blasius, A. Golnik, E. Brücher, R. K. Kremer, D. R. Noakes, C. E. Stronack, E. J. Asnaldo, Phys. Rev. B **59** (1999) 14099

[174] T. Nachtrab, D. Koelle, R. Kleiner, C. Bernhard, C. T. Lin, Phys. Rev. Lett. **92** (2004) 117001

[175] C. W. Chu, B. Lorenz, R. L. Meng, Y. Y. Xue, *Frontiers in Superconducting Materials,* Springer Verlag, Berlin (2005), p.331

[176] M. L. Foo, R. E. Schaak, V. L. Miller, T. Klimczuk, N. S. Rogado, Y. Wang, G. C. Lau, C. Craley, H. W. Zandbergen, N. P. Ong, R. J. Cava, Solid State Commun. **127** (2003) 33

[177] J. W. Lynn, Q. Huang, C. M. Brown, V. L. Miller, M. L. Foo, R. E. Schaak, C. Y. Jones, E. A. Mackey, R. J. Cava, Phys. Rev. B **68** (2003) 214516

[178] S. Yonezawa, Z. Muraoka, Y. Matsushita, Z. Hiroi, J. Phys.: Condensed Matter **16** (2004) L9;

S. Yonezawa, Z. Muraoka, Z. Hiroi, cond-mat/0404220;

S. Yonezawa, Y. Muraoka, Y. Matsushita, Z. Hiroi, J. Phys. Soc. Jpn. **73** (2004) 819;

Z. Hiroi, S. Yonezawa, Y. Muraoka, cond-mat/0402006

[179] N. P. Ong, R. J. Cava, Science **305** (2004) 52

[180] G. Schuck, S. M. Kazakov, K. Rogacki, N. D. Zhigadlo, J. Karpinski, Phys. Rev. B **73** (2006) 144506

[181] J. Kunes, T. Jeong T, W. E. Pickett, Phys. Rev. B **70** (2004) 174510

[182] Y. Nagao , J. Yamaura, H. Ogusu, Y. Okamoto, Z. Hiroi, J. Phys. Soc. Jpn. **78** (2009) 064702

[183] K. Hattori, H. Tsunetsugu, Phys. Rev. B **81** (2010) 134503

[184] Z. Hiroi, J. Yamaura, S. Yonezawa, H- Harima, Physica C **460-462** (2007) 20

[185] R. Sainz, A. J. Freeman, Phys. Rev. B **72** (2005) 024522

[186] B. Lorenz, A. M. Guloy, P. C. W. Chu, Int. J. Mod. Phys. B **28** (2014) 1430011

[187] T. Yajima, K. Nakano, F. Takeiri, T. Ono, Y. Hosokoshi, Y. Matsushita, J. Hester, Y. Kobayashi, and H. Kageyama, J. Phys. Soc. Jpn. **81** (2012) 103706

[188] P. Doan, M. Gooch, Z. J. Tang, B. Lorenz, A. Moller, J. Tapp, P. C. W. Chu, A. M. Guloy, J. Am. Chem. Soc. **134** (2012) 16520

[189] Q. Song, Y. J. Yan, Z. R. Ye, M. Q. Ren, D. F. Xu, S. Y. Tan, X. H. Niu, B. P. Xie, T. Zhang, R. Peng, H. C. Xu, J. Jiang, D. L. Feng. Phys. Rev. B **93** (2016) 024508

[190] D. C. Johnston, Adv. Phys. **59** (2010) 803;

D. Mandrus, A. S. Sefat, M. A. McGuire, B. C. Sales, Chem. Mater. **22** (2010) 715

[191] D. Maruyama, M. Sigrist, Y. Yanase, J. Phys. Soc. Jpn. **81** (2012) 034702

[192] C.-H. Lee, A. Iyo, H. Eisaki, H. Kito, M. T. Fernandez-Diaz, T. Ito, K. Kihou, H. Matsuhata, M. Braden, and K. Yamada, J. Phys. Soc. Jpn. **77** (2008) 083704

[193] C. Wang, L. Li, S. Chi, Z. Zhu, Z. Ren, Y. Li, Y. Wang, X. Lin, Y. Luo, S. Jiang, X. Xu, G. Cao, Z. Xu, EPL **83** (2008) 67006

[194] T. Klimczuk, H. C. Walker, R. Springell, A. B. Shick, A. H. Hill, P. Gaczynski, K. Gofryk, S. A. J. Kimber, C. Ritter, E. Colineau, J.-C. Griveau, D. Bouëxière, R. Eloirdi, R. J. Cava, R. Caciuffo, Phys. Rev. B **85** (2012) 174506

[195] T. Klimczuk, A. B. Shick, R. Springell, H. C. Walker, A. H. Hill, E. Colineau, J.-C. Griveau, D. Bouëxière, R. Eloirdi, R. Caciuffo, Phys. Rev. B **86** (2012) 174510

[196] F. Hardy, P. Burger, T. Wolf, R. A. Fisher, P. Schweiss, P. Adelmann, R. Heid, R. Fromknecht, R. Eder, D. Ernst, H. v. Löhneysen, C. Meingast, Europhys. Lett. **91** (2010) 47008

[197] A. S. Sefat, R. Jin, M. A. McGuire, B. C. Sales, D. J. Singh, D. Mandrus, Phys. Rev. Lett. **101** (2008) 117004

[198] N. Lanata, H. U. R. Strand, G. Giovanetti, B. Hellsing, L. de' Medici, M. Capone, Phys. Rev. B **87** (2013) 045122

[199] R. Prozorov, V. G. Kogan, Rep. Prog. Phys. **74** (2011) 124505

[200] T. Katase, Y. Ishimaru, A. Tsukamoto, H. Hiramatsu, T. Kamiya, K. Tanabe, H. Hosono, Nature Comm. **1419** (2011)

[201] S. Lee, J. Jiang, J. D. Weiss, C. M. Folkman, C. W. Bark, C. Tarantini, A. Xu, D. Abraimov, A. Polyanskii, C. T. Nelson, Y. Zhang, S. H. Baek, H. W. Jang, A. Yamamoto, F. Kametani, X. Q. Pan, E. E. Hellstrom, A. Gurevich, C. B. Eom, D. C. Larbalestier, Appl. Phys. Lett. **95** (2009) 212505

[202] M. Putti, I. Pallecchi, E. Bellingeri, M. R. Cimberle, M. Tropeano, C. Ferdeghini, A. Palenzona, C. Tarantini, A. Yamamoto, J. Jiang, J. Jaroszynski, F. Kametani, D. Abraimov, A. Polyanskii, J. D. Weiss, E. E. Hellstrom, A. Gurevich, D. C. Larbalestier, R. Jin, B. C. Sales, A. S. Sefat, M. A. McGuire, D. Mandrus, P. Cheng, Y. Jia, H. H. Wen, S. Lee, C. B. Eom, Supercond. Sci. Technol. **23** (2010) 034003

[203] T. Katase, H. Hiramatsu, H. Yanagi, T. Kamiya, M. Hirano, H. Hosono, Solid State Commun. **149** (2009) 2121;

L. Wang, Y. Ma, Q. Wang, K. Li, X. Zhang, Y. Qi, Z. Gao, X. Zhang, D. Wang, C. Yao, C. Wang Appl. Phys. Lett. **98** (2011) 222504;

M. Rotter, M. Tegel, I. Schellenberg, F. M. Schappacher, R. Pöttgen, J. Deisenhofer, A. Günther, F. Schrettle, A. Loidl, D. Johrendt, New J. Phys. **11** (2009) 025014

[204] R. Chevrel, M. Hirrien, M. Sergent, Polyhedron **5** (1986) 87

[205] Ø. Fischer, Appl. Phys. **16** (1978) 1

[206] G. Rimikis, W. Goldacker, W. Specking, R. Flükiger, IEEE Trans. Magn. **27** (1991) 1116

[207] E. Revolinsky, G.A. Spiering, D.J. Beerntsen, J. Phys. Chem. Solids **26** (1965) 1029

[208] R. E. Schwall, G. R. Stewart, T. H. Geballe, J. Low Temp. Phys. **22** (1976) 557

[209] E. Morosan, H. W. Zandbergen, B. S. Dennis, J. W. G. Bos, Y. Onose, T. Klimczuk, A. P. Ramirez, N. P. Ong, R. J. Cava, Nature Phys. **2** (2006) 544

[210] A. F. Kusmartseva, B. Sipos, H. Berger, L. Forró, E. Tutiš, Phys. Rev. Lett. **103** (2009) 236401

[211] J. T. Ye, Y. J. Zhang, R. Akashi, M. S. Bahramy, R. Arita, Y. Iwasa, Science **338** (2012) 1193

[212] W. Shi, J. Ye, Y. Zhang, R. Suzuki, M. Yoshida, J. Miyazaki, N. Inoue, Y. Saito, Y. Iwasa, Sci. Rep. **5** (2015) 12534

[213] D. J. Rahn, S. Hellmann, M. Kalläne, C. Sohrt, T. K. Kim, L. Kipp, K. Rossnagel, Phys. Rev. B **85** (2012) 224532

[214] M. D. Johannes, I. I. Mazin, Phys. Rev. B **77** (2008) 165135

[215] M. Porer, U. Leierseder, J.-M. Menard, H. Dachraoui, L. Mouchliadis, I. E. Perakis, U. Heinzmann, J. Demsar, K. Rossnagel, R. Huber, Nature Mater. **13** (2014) 857

[216] F. Weber, S. Rosenkranz, J.-P. Castellan, R. Osborn, R. Hott, R. Heid, K.-P. Bohnen, T. Egami, A. H. Said, D. Reznik, Phys. Rev. Lett. **107** (2011) 107403

[217] Y. Mizuguchi, H. Fujihisa, Y. Gotoh, K. Suzuki, H. Usui, K. Kuroki, S. Demura, Y. Takano, H. Izawa, and O. Miura, Phys. Rev. B **86** (2012) 220510

[218] T. Hiroi, J. Kajitani, A. Omachi, O. Miura, Y. Mizuguchi, J. Phys. Soc. Jpn. **84** (2015) 024723

[219] D. Yazici, I. Jeon, B.D. White, M.B. Maple, Physica C **514** (2015) 218

[220] J. L. Sarrao, L. A. Morales, J. D. Thompson, B. L. Scott, G. R. Stewart, F. Wastin, J. Rebizant, P. Boulet, E. Colineau, G. H. Lander, Nature **420** (2002) 297

[221] L. Taillefer, G. G. Lonzarich, Phys. Rev. Lett. **60** (1988) 1570;
H. Aoki, S. Uji, A. K. Albessard, Y. Onuki, Phys. Rev. Lett. **71** (1993) 2110;
F. S. Tautz, S. R. Julian, G. J. McMullen, G. G. Lonzarich, Physica B **206-207** (1995) 29

[222] *Non-Centrosymmetric Superconductors,* ed. E. Bauer and M. Sigrist, Lecture Notes in Physics, Vol. 847, Springer, Heidelberg (2012)

[223] H. Shishido, T. Shibauchi, K. Yasu, T. Kato, H. Kontani, T. Terashima,Y. Matsuda, Science **327** (2010) 980

[224] Y. Mizukami, H. Shishido, T. Shibauchi, M. Shimozawa, S. Yasumoto, D. Watanabe, M. Yamashita, H. Ikeda,T. Terashima, H. Kontani, Y. Matsuda, Nature Phys. **7** (2011) 849

[225] R. A. Fisher, S. Kim, B. F. Woodfield, N. E. Phillips, L. Taillefer, K. Hasselbach, J. Flouquet, A. L. Giorgi, J. L. Smith, Phys. Rev. Lett. **62** (1989) 1411

[226] E. D. Bauer, N. A. Frederick, P.-C. Ho, V. S. Zapf, M. B. Maple, Phys. Rev. B **65** (2002) 100506

[227] T. Watanabe, Y. Kasahara, K. Izawa, T. Sakakibara, C. vanderBeek, T. Hanaguri, H. Shishido, R. Settai, Y. Onuki, Y. Matsuda, Phys. Rev. B **70** (2004) 020506;
For a review on the LOFF / FFLO state see
R. Casalbuoni, G. Nardulli, Rev. Mod. Phys. **76** (2004) 263

[228] G. Zwicknagl, J. Wosnitza, in [10]

[229] A. Bianchi, R. Movshovich, C. Capan, A. Lacerda, P. G. Pagliuso, J. L. Sarrao, Phys. Rev. Lett. **91** (2003) 187004

[230] M. Kenzelmann, T. Strässle, C. Niedermayer, M. Sigrist, B. Padmanabhan, M. Zolliker, A. D. Bianchi, R. Movshovich, E. D. Bauer, J. L. Sarrao, J. D. Thompson, Science **321** (2008) 1652

[231] Y. Tokiwa, E. D. Bauer, P. Gegenwart, Phys. Rev. Lett. **109** (2012) 116402

[232] P. Fulde, P. Thalmeier, G. Zwicknagl, *Strongly Correlated Electrons,* Solid State Physics, Advances in Research and Applications, Vol. 60 (2006) p. 1

[233] J. D. Denlinger, G.-H. Gweon, J. W. Allen, C. G. Olson, M. B. Maple, J. L. Sarrao, P. E. Armstrong, Z. Fisk, H. Yamagami, J. Electron. Spectrosc. Relat. Phenom**. 117&118** (2001) 347;

F. Reinert, D. Ehm, S. Schmidt, G. Nicolay, S. Hüfner, J. Kroha, O. Trovarelli, C. Geibel, Phys. Rev. Lett. **87** (2001) 106401

[234] G. Zwicknagl, Adv. Phys. **41** (1992) 203

[235] S. Ernst, S. Kirchner, C. Krellner, C. Geibel, G. Zwicknagl, F. Steglich, S. Wirth, Nature **474** (2011) 362

[236] H. Pfau, R. Daou, S. Lausberg, H. R. Naren, M. Brando, S. Friedemann, S. Wirth, T. Westerkamp, U. Stockert, P. Gegenwart, C. Krellner, C. Geibel, G. Zwicknagl, and F. Steglich, Phys. Rev. Lett **110** (2013) 256403

[237] O. Stockert, E. Faulhaber, G. Zwicknagl, N. Stüßer, H. S. Jeevan, M. Deppe, R. Borth, R. Küchler, M. Loewenhaupt, C. Geibel, F. Steglich, Phys. Rev. Lett. **92** (2004) 136401

[238] I. Eremin, G. Zwicknagl, P.Thalmeier, P. Fulde, Phys. Rev. Lett. **101** (2008) 187001

[239] G. Zwicknagl, A. N. Yaresko, P. Fulde, Phys. Rev. B **63** (2002) 081103;

G. Zwicknagl, A. N. Yaresko, P. Fulde, Phys. Rev. B **68** (2003) 052508

[240] G. Zwicknagl, Mater. Res. Soc. Symp. Proc. **1444** (2012)

[241] G. Zwicknagl, Phys. Staus. Solidi B **250** (2013) 634

[242] S. Fujimori, T. Ohkochi, I. Kawasaki, A. Yasui, Y. Takeda, T. Okane, Y. Saitoh, A. Fujimori, H. Yamagami, Y. Haga, E. Yamamoto, Y. Tokiwa, S. Ikeda, T. Sugai, H. Ohkuni,N. Kimura, Y. OnukiJ. Phys. Soc. Jpn. **81** (2011) 014703

[243] G. Zwicknagl, P. Thalmeier, and P. Fulde, Phys. Rev. **B 79** (2009) 115132

[244] For a review see
P. Thalmeier, G. Zwicknagl, *Handbook on the Physics and Chemistry of Rare Earths*, Vol. 34, chapter "*Unconventional Superconductivity and Magnetism in Lanthanide and Actinide Intermetallic Compounds*"; arXiv:cond-mat/0312540

[245] N. K. Sato, N. Aso, K. Miyake, R. Shiina, P. Thalmeier, G. Varelogiannis, C. Geibel, F. Steglich, P. Fulde, T. Komatsubara, Nature **410** (2001) 340

[246] P. McHale, P. Thalmeier, P. Fulde, Phys. Rev. B **70** (2004) 014513

[247] P. Thalmeier, G. Zwicknagl, G. Sparn, F. Steglich, *Frontiers in Superconducting Materials,* Springer Verlag, Berlin (2005), p.109

[248] J. D. Thompson, R. Movshovich, Z. Fisk, F. Bouquet, N. J. Curro, R. A. Fisher, P. C. Hammel, H. Hegger, M. F. Hundley, M. Jaime, P. G. Pagliuso, C. Petrovic, N. E. Phillips, J. L. Sarrao, J. Magn. Magn. Mat. **226-230** (2001) 5

[249] C. Petrovic, R. Movshovich, M. Jaime, P. G. Pagliuso, M. F. Hundley, J. L. Sarrao, Z. Fisk, J. D. Thompson, Europhys. Lett. **53** (2001) 354

[250] C. Petrovic, P. G. Pagliuso, M. F. Hundley, R. Movshovich, J. D. Sarrao, J. D. Thompson, Z. Fisk, P. Monthoux, J. Phys.: Condens. Matter **13** (2001) L337

[251] E. Bauer, G Hilscher, H Michor, Ch Paul, EW Scheidt, A Gribanov, Y Seropegin, H Noël, M Sigrist, and P Rogl, Phys. Rev. Lett. **92** (2004) 027003

[252] N. Kimura, I. Bonalde in [222]

[253] S. Nakatsuji, K. Kuga, Y. Machida, T. Tayama, T. Sakakibara, Y. Karaki, H. Ishimoto, S. Yonezawa, Y. Maeno, E. Pearson, G. G. Lonzarich, L. Balicas, H. Lee, and Z. Fisk, Nature Phys. **4** (2008) 603

[254] D. Jaccard, K. Behnia, J. Sierro, Phys. Lett. A **163** (1992) 475

[255] F. M. Grosche, S. R. Julian, N. D. Mathur, G. G. Lonzarich, Physica B **223&224** (1996) 50

[256] N. D. Mathur, F. M. Grosche, S. R. Julian, I. R. Walker, D. M. Freye, R. K. W. Haselwimmer, G. G. Lonzarich, Nature **394** (1998) 39

[257] F. M. Grosche, P. Agarwal S. R. Julian, N. J. Wilson, R. K. W. Haselwimmer, S. J. S. Lister, N. D. Mathur, F. V. Carter, S. S. Saxena, G. G. Lonzarich, J. Phys. Condens. Matter. **12** (2000) L533

[258] R. Movshovich, T. Graf, D. Mandrus, J. D. Thompson, J. L. Smith, Z. Fisk, Phys. Rev. B **53** (1996) 8241

[259] I. R. Walker, F. M. Grosche, D. M. Freye, G. G. Lonzarich, Physica C **282** (1997) 303

[260] O. Stockert, J. Arndt, E. Faulhaber, C. Geibel, H. S. Jeevan, S. Kirchner, M. Loewenhaupt, K. Schmalzl, W. Schmidt, Q. Si, F. Steglich, Nature Phys. **7** (2011) 119

[261] F. Steglich, J. Arndt, O. Stockert, S. Friedemann, M. Brand , C. Klingner, C. Krellner, C. Geibel, S.Wirth, S. Kirchner , Q. Si, J. Phys.: Condens. Matter **24** (2012) 294201

[262] F. Steglich, O. Stockert, S. Wirth, C. Geibel, H. Q. Yuan, S. Kirchner, Q. Si, arXiv:1208.3684

[263] G. R. Stewart, Z. Fisk, J. O. Willis, J. L. Smith, Phys. Rev. Lett. **52** (1984) 679

[264] C. Geibel, C. Schank, S. Thies, H. Kitazawa, C. D. Bredl, A. Böhm, M. Rau, A. Grauel,R. Caspary, R. Helfrich, U. Ahlheim, G. Weber, F. Steglich, Z. Phys. B **84** (1991) 1

[265] C. Geibel, S. Thies, D. Kczorowski, A. Mehner, A. Grauel, B. Seidel, U. Ahlheim, R. Helfrich, K. Petersen, C. D. Bredl, F. Steglich, Z. Phys. **83** (1991) 305

[266] T. T. M. Palstra, A. A. Menovsky, J. van den Berg, A. J. Dirkmaat, P. H. Res, G. J. Nieuwenhuys, J. A. Mydosh, Phys. Rev. Lett. **55** (1985) 2727

[267] H. R. Ott, H. Rudigier, Z. Fisk, J. L. Smith, Phys. Rev. Lett. **80** (1983) 1595

[268] S. S. Saxena, P. Agarwal, K. Ahllan, F. M. Grosche, R. W. K. Haselwimmer, M. J. Steiner, E. Pugh, I. R. Walker, S. R. Julian, P. Monthoux, G. G. Lonzarich, A. Huxley, I. Shelkin, D. Braithwaite, J. Flouquet, Nature **406** (2000) 587

[269] R. Joynt, L. Taillefer, Rev. Mod. Phys. **74** (2002) 237

[270] A. Grauel, A. Böhm, H. Fischer, C. Geibel, R. Köhler, R. Modler, C. Schank, F. Steglich, G. Weber, Phys. Rev. B **46** (1992) 5818

[271] T. E. Mason, G. Aeppli, Matematisk-fysiske Meddelelser 45 (1997) 231

[272] M. Jourdan, M. Huth, H. Adrian, Nature **398** (1999) 47

[273] V. Ginzburg, JETP **4** (1957) 153

[274] P. C. Canfield, S. L. Bud'ko, Physica C **262** (1996) 249

[275] D. Fay, J. Appel, Phys. Rev. B **22** (1980) 3173

[276] D. Aoki, A. Huxley, E. Ressouche, D. Braithwaite, J. Flouquet, J.-P. Brison, E. Lhotel, Nature **413** (2001) 613

[277] C. Pfleiderer, M. Uhlarz, S. M. Hayden, R. Vollmer, H. V. Löhneysen, N. R. Bernhoeft, G. G. Lonzarich, Nature **412** (2001) 58

[278] E. A. Yelland, S. M. Hayden, S. J. C. Yates, C. Pfleiderer, M. Uhlarz, R. Vollmer, H. v. Löhneysen, N. R. Bernhoeft, R. P. Smith S. S. Saxena, N. Kimura, Phys. Rev. B 72 (2005) 214523

[279] W. Schlabitz, J. Bauman, B. Politt, U. Rauchschwalbe, H. M. Mayer, U. Ahlheim, C. D. Bredl, Z. Phys. B **52** (1986) 171

[280] A. de Visser, F. E. Kayzel, A. A. Menovsky, J. J. M. Franse, J. van der Berg, G. J. Nieuwenhuys, Phys. Rev. B **34** (1986) 8168

[281] C. Broholm, J. K. Kjems, W. J. L. Buyers, P. Matthews, T. T. M. Palstra, A. A. Menovsky, J. A. Mydosh, Phys. Rev. Lett. **58** (1987) 1467

[282] M. B. Walker, W. J. L. Buyers, Z. Tun, W. Que, A. A. Menovsky, J. D. Garrett, Phys. Rev. Lett. **71** (1993) 2630

[283] H. Amitsuka, M. Sato, N. Metoki, M. Yokohama, K. Kuwahara, T. Sakakibara, H. Morimoto, S. Kawarazaki, Y. Miyako, J. A. Mydosh, Phys. Rev. Lett. **83** (1999) 5114

[284] H. Amitsuka, M. Yokoyama, S. Miyazaki, K. Tenya, T. Sakakibara, W. Higemoto, K. Nagamine, K. Matsuda, Y. Kohori, T. Kohara, Physica B **312-313** (2002) 390

[285] A. R. Schmidt, M. H. Hamidian, P. Wahl, F. Meier, A. V. Balatsky, J. D. Garrett, T. J. Williams, G. M. Luke, J. C. Davis, Nature **465** (2010) 570

[286] P. Aynajian, E. H. da Silva Netoa, C. V. Parkera, Y. Huang, A. Pasupathy, J. Mydosh, A. Yazdani, PNAS **107** (2010) 10383

[287] T. Yuan, J. Figgins, D. K. Morr, Phys. Rev B **86** (2012) 035129

[288] T. Shibauchi, Y. Matsuda, Physica **C 481** (2012) 229

[289] T. Das, Sci. Rep. **2** (2012) 596

[290] A. Pourret, A. Palacio-Morales, S. Kramer, L. Malone, M. Nardone, D. Aoki, G. Knebel, J. Flouquet, J. Phys. Soc. Jpn. **82** (2013) 034706

[291] H. R. Ott, H. Rudigier, T. M. Rice, K. Ueda, Z. Fisk, J. L. Smith, Phys. Rev. Lett. **82** (1984) 1915

[292] F. Wastin, P. Boulet, J. Rebizant, E. Colineau, G. H. Lander, J. Phys.: Condens. Matter **15** (2003) S2279

[293] D. Aoki, Y. Haga, T. D. Matsuda, N. Tateiwa, S. Ikeda, Y. Homma, H. Sakai, Y. Shiokawa, E. Yamamoto, A. Nakamura, R. Settai, Y. Ōnuki, J. Phys. Soc. Jpn **76** (2007) 063701

[294] M. W. McElfresh, J. H. Hall, R. R. Ryan, J. L. Smith, Z. Fisk, Acta Cryst. C **46** (1990) 1579

[295] K. Izawa, Y. Nakajima, J. Goryo, Y. Matsuda, S. Osaki, H. Sugawara, H. Sato, P. Thalmeier, K. Maki, Phys. Rev. Lett. **90** (2003) 117001

[296] B. C. Sales, D. Mandrus, R. K. Williams, Science **272** (1996) 1325

[297] B. C. Sales, *Handbook on the Physics and Chemistry of the Rare Earths*, Vol. 33, Elsevier (2003) chapter 211, p. 1

[298] N. A. Frederick, T. D. Do, P.-C. Ho, N. P. Butch, V. S. Zapf, M. B. Maple, Phys. Rev. B **69** (2004) 024523

[299] R. Vollmer, A. Faißt, C. Pfleiderer, H. v. Löhneysen, E. D. Bauer, P.-C. Ho, V. Zapf, M. B. Maple, Phys. Rev. Lett. **90** (2003) 5700

[300] N. Oeschler, P. Gegenwart, F. Steglich, N. A. Frederick, E. D. Bauer, M. B. Maple, Acta Phys. Pol. B **34** (2003) 959

[301] W. Meissner, H. Franz, Z. Phys. **65** (1930) 30

[302] B. T. Matthias, J. K. Hulm, Phys. Rev. **87** (1952) 799

[303] A. Mourachkine, *Room-Temperature Superconductivity*, Cambridge International Science Publishing, Cambridge (2004)

[304] R. J. Cava, H. W. Zandbergen, B. Batlogg, H. Eisaki, H. Takagi, J. J. Krajewski, W. F. Peck Jr., E. M. Gyorgy, S. Uchida, Nature **372** (1994) 245

[305] J. P. Attfield, J. Mater. Chem. **21** (2011) 4756

[306] S. Yamanaka, Annu. Rev. Mater. Sci. **30** (2000) 53;
S. Yamanaka, J. Mater. Chem. **20** (2010) 2922

[307] S. Yamanaka, K. Hotehama, H. Kawaji, Nature 392 (1998) 580

[308] Y. Kasahara, K. Kuroki, S. Yamanaka, Y. Taguchi, Physica C **514** (2015) 354

[309] S. Yamanaka, T. Yasunaga, K. Yamaguchi, M. Tagawa, J. Mater. Chem. **19** (2009) 2573

[310] Y. Taguchi, A. Kitora, Y. Iwasa, Phys. Rev. Lett. **97** (2006) 107001

[311] A. Hebard, M. J. Rosseinsky, R. C. Haddon, D. W. Murphy, S. H. Glarum, T. T. M. Palstra, A. P. Ramirez, A. R. Kortam, Nature **350** (1991) 600;
R. M. Fleming, A. P. Ramirez, M. J. Rosseinsky, D. W. Murphy, R. C. Haddon, S. M. Zahurak, A. V. Markhija, Nature **352** (1991) 787

[312] K. Bechgaard, K. Carneiro, M. Olsen, R B. Rasmussen, C. B. Jacobsen, Phys. Rev. Lett. **46** (1981) 852

[313] P. A. Mansky, G. Danner, P. M. Chaikin, Phys. Rev. B **52** (1995) 7554

[314] G. Saito, H. Yamochi, T. Nakamura, T. Komatsu, M. Nakashima, H. Mori, K. Oshima, Physica B **169** (1991) 372

[315] I. J. Lee, S. E. Brown, W. G. Clark, M. J. Strouse, M. J. Naughton, W. Kang, P. M. Chaikin, Phys. Rev. Lett. **88** (2002) 017004

[316] R. Beyer, J. Wosnitza, Low Temp. Phys. **39** (2013) 225

[317] R. Lortz, Y. Wang, A. Demuer, P. H. M. Böttger, B. Bergk, G. Zwicknagl, Y. Nakazawa, J. Wosnitza, Phys. Rev. Lett. **99** (2007 )187002

[318] R. Beyer, B. Bergk, S. Yasin, J. A. Schlueter, and J. Wosnitza, Phys. Rev. Lett. **109** (2012) 027003

[319] R. J. Cava, H. Takagi, H. W. Zandbergen, J. J. Krajewski, W. F. Peck, Jr., T. Siegrist, B. Batlogg, R. B. van Dover, R. J. Felder, K. Mizuhashi, J. O. Lee, H. Eisaki, S. Uchida, Nature **367** (1994) 252

[320] J. Cava, H. Takagi, B. Batlogg, H. W. Zandbergen, J. J. Krajewski, W. F. Peck Jr., R. B. van Dover, R. J. Felder, K. Mizuhashi, J. O. Lee, H. Eisaki, S. A. Carter, S. Uchida, Nature **367** (1994) 146

[321] Y. Takano, Science and Technology of Advanced Materials 7 (2006) S1

[322] Z.-A.Ren, J. Kato, T. Muranaka, J. Akimitsu, M. Kriener, Y. Maeno, JPSP **76** (2007) 103710

[323] M. Kriener, Y. Maeno, T. Oguchi, Z.-A. Ren, J. Kato, T. Muranaka, J. Akimitsu, Phys. Rev. B **78** (2008) 024517

[324] F. Chiodi, A. Grockowiak, J.E. Duvauchelle, F. Fossard, F. Lefloch,T. Klein, C. Marcenat, D. Débarre, Appl. Surf. Sci. **302** (2014) 209

[325] E. Bustarret, C. Marcenat, P. Achatz, J. Kacmarcik, F. Lévy, A. Huxley, L. Ortéga, E. Bourgeois, X. Blase, D. Débarre, J. Boulmer , Nature **444** (2006) 465

[326] X. Blase, E. Bustarret, C. Chapelier, T. Klein, C. Marcenat, Nature Mater. **8** (2009) 375

[327] J. E. Duvauchelle, A. Francheteau, C. Marcenat, F. Chiodi, D. Débarre, K. Hasselbach, J. R. Kirtley, and F. Lefloch, Appl. Phys. Lett. **107** (2015) 072601

[328] G. Amano, S. Akutagawa, T. Muranaka, Y. Zenitani, J. Akimitsu, J. Phys. Soc. Jpn. **73** (2004) 530

[329] T. Nakane, T. Mochiku, H. Kito, J. Itoh, M. Nagao, H. Kumakura, Y. Takano, Appl. Phys. Lett. **84** (2004) 2859

[330] N. B. Hannay, T. H. Geballe, B. T. Matthias, K. Andres, P. Schmidt, D. Mac Nair, Phys. Rev. Lett. **14** (1965) 225

[331] Y. Koike, S. Tanuma, H. Suematsu, K. Higuchi, J. Phys. Chem. Solids **41** (1980) 1111

[332] L. A. Pendrys, R. Wachnik, F. L. Vogel, P. Lagrange, G. Furdin, M. El Makrini, A. Hérold, Solid State Commun. **38** (1981) 677

[333] R. A. Wachnik, L. A. Pendrys, F. L. Vogel, P. Lagrange, Solid State Commun. **43** (1982) 5

[334] T. E. Weller, M. Ellerby, S. S. Saxena, R. P. Smith, N. T. Skipper, Nature Phys. **1** (2005) 39

[335] N. Emery, C. Hérold, M. d'Astuto, V. Garcia, C. Bellin, J. F. Marêché, P. Lagrange, G. Loupias, Phys. Rev. Lett. **95** (2005) 087003

[336] N. Emery, C. Hérold, J. F. Marêché, C. Bellouard, G. Loupias, P. Lagrange, J. Solid State Chem. **179** (2006) 1289

[337] N. Emery, C. Hérold, J.-F. Marêché, P. Lagrange, Sci. Technol. Adv. Mater. 9 (2008) 44102

[338] B. M. Ludbrook, G. Levy, P. Niggea, M. Zonno, M. Schneider, D. J. Dvorak, C. N. Veenstraa, S. Zhdanovich, D. Wonga, P. Dosanjh, C. Straßer, A. Stöhr, S. Forti, C. R. Ast, U. Starke, A. Damascellia, PNAS **112** (2015) 11795

[339] J. Chapman, Y. Su, C. A. Howard, D. Kundys, A. Grigorenko, F. Guinea, A. K.Geim, I.V. Grigorieva, R. R. Nair, arXiv:1508.06931

[340] C. H. Pennington, V. A. Stenger, Rev. Mod. Phys. **68** (1996) 855

[341] V. Buntar, H. W. Weber, Supercond. Sci. Technol. **9** (1996), 599

[342] Y. Takabayashi, A. Y. Ganin, P. Jegli, D. Arčon, T. Takano, Y. Iwasa, Y. Ohishi, M. Takata, N. Takeshita, K. Prassides, M J. Rosseinsky, Science **323** (2009) 1585

[343] R. Mitsuhashi, Y. Suzuki, Y. Yamanari, H. Mitamura, T. Kambe, N. Ikeda, H. Okamoto, A. Fujiwara, M. Yamaji, N. Kawasaki, Y. Maniwa, Y. Kubozono, Nature **464** (2010) 76

[344] M. Xue, T. Cao, D. Wang, Y. Wu, H. Yang, X. Dong, Ju. He, F. Li, G. F. Chen, Scientific Reports **2**:389

[345] Z. K. Tang, L. Zhang, N. Wang, X. X. Zhang, G. W. Wen, G. D. Li, J. N. Wang, C. T. Chan, P. Sheng, Science **292** (2001) 2462

[346] M. Bockrath, Nature Phys. **2** (2006) 155

[347] W. Shi, Z. Wang, Q. Zhang, Y. Zheng, C. Ieong, M. He, R. Lortz,Y. Cai, N. Wang, T. Zhang, H. Zhang, Z. Tang, P. Sheng, H. Muramatsu,Y. Ahm Kim, M. Endo, P. T. Araujo, M. S. Dresselhaus, Scientific Reports 2:625

[348] R. Nagarajan, C. Mazumdar, Z. Hossain, S. K. Dhar, K. V. Gopalakrishnan, L. C. Gupta, C. Godart, B. D. Padalia, R. Vijayaraghavan, Phys. Rev. Lett. **72** (1994) 274

[349] G. Hilscher, H. Michor, *Studies in High Temperature Superconductors*, Vol. 28, Nova Science Publishers (1999) pp. 241–286

[350] K.-H. Müller, V. N. Narozhnyi, Rep. Prog. Phys. **64** (2001) 943

[351] O. Fischer, *Magnetic Superconductors*, Vol. 5, North Holland, Amsterdam (1990)

[352] *Topics in Current Physics* (Eds. O. Fischer & M. B. Maple), Vol. 32 and 34 Springer Verlag, Berlin (1982)

[353] K. Mizuno, T. Saito, H. Fudo, K. Koyama, K. Endo, H. Deguchi, Physica B **259-261** (1999) 594

[354] E. Boaknin, R. W. Hill, C. Proust, C. Lupien, L. Taillefer, P. C. Canfield, Phys. Rev. Lett. **87** (2001) 237001

[355] K. Izawa, K. Kamata, Y. Nakajima, Y. Matsuda, T. Watanabe, M. Nohara, H. Takagi, P. Thalmeier, K. Maki, Phys. Rev. Lett. **89** (2002) 137006

[356] M. Nohara, M., Isshiki, H., Takagi, R. Cava, J. Phys. Soc. Jpn. **66** (1997) 1888

[357] K. Maki, P. Thalmeier, H. Won, Phys. Rev. B **65** (2002) 140502

[358] T. Watanabe, M. Nohara, T. Hanaguri, H. Takagi, Phys. Rev. Lett. **92** (2004) 147002

[359] S. -L. Drechsler, H. Rosner, S. V. Shulga, I. Opahle, H. Eschrig, J. Freudenberger, G. Fuchs, K. Nenkov, K. -H. Müller, H. Bitterlich, W. Löser, G. Behr, D. Lipp, A. Gladun, Physica C **364-365** (2001) 31

[360] H. Suderow, P. Martinez-Samper, N. Luchier, J. P. Brison, S. Vieira, P. C. Canfield, Phys. Rev. B. **64** (2003) 020503

[361] P. Grant, Nature **411** (2001) 532

[362] A. M. Campbell, Science **292** (2001) 65

[363] J. Kortus, I. I. Mazin, K. D. Belashchenko, V. P. Antropov, L. L. Boyer, Phys. Rev. Lett. **86** (2001) 4656

[364] K.-P. Bohnen, R. Heid, B. Renker, Phys. Rev. Lett. **86** (2001) 5771

[365] H. J. Choi, D. Roundy, H. Sun, M. L. Cohen, S. G. Louie, Nature **418** (2002) 758

[366] P. C. Canfield, G. Crabtree, Phys. Today **56** (March 2003) 34

[367] S. L. Bud'ko, P. C. Canfield, V. G. Kogan, Physica C **382** (2002) 85

[368] Y. Kong, O. V. Dolgov, O. Jepsen, O. K. Andersen, Phys. Rev. B **64** (2001) 020501

[369] I. I. Mazin, O. K. Andersen, O. Jepsen, A. A. Golubov , O. V. Dolgov, J. Kortus, Phys. Rev. B **69** (2004) 056501

[370] H. J. Choi, D. Roundy, H. Sun, M. L. Cohen, S. G. Louie, Phys. Rev. B **69** (2004) 056502

[371] J. H. Kim, S. Oh, Y.-U. Heo, S. Hata, H. Kumakura, A. Matsumoto, M. Mitsuhara, S. Choi, Y. Shimada, M. Maeda, J. L. MacManus-Driscoll, S. X. Dou, NPG Asia Materials **4** (2012) e3

[372] X. H. Li, L. Y. Ye, Z. S. Gao, D. L. Wang, L. Z. Lin, G. M. Zhang, S. T. Dai, Y. W. Ma, L. Y. Xiao, J. Phys.: Conf. Ser. **234** (2010) 022020

[373] M. Tomsic, M. Rindfleisch, J. Yue, K. McFadden, J. Phillips, M. D. Sumption, M. Bhatia, S. Bohnenstiehl, E. W. Collings, Int. J. Appl. Ceram. Technol. **4** (2007) 250